\documentclass[prd,preprintnumbers,aps,preprint,amsmath,amssymb,superscriptaddress,floatfix,nofootinbib,unsortedaddress]{revtex4}
\pdfoutput=1

\usepackage{amssymb, bm, amsmath, graphicx, physymb}
\usepackage[colorlinks=true,linktocpage=true,linkcolor=blue,citecolor=blue]{hyperref}
\usepackage[usenames,dvipsnames]{color}
\usepackage{slashed}
\usepackage[utf8]{inputenc}
\usepackage[T1]{fontenc}
\usepackage{enumerate}
\usepackage{amsfonts}
\usepackage{mathtools}
\usepackage{latexsym}
\usepackage{hyperref}
\usepackage{epstopdf} 
\usepackage{bbm}
\usepackage{soul}

\newcommand{\half}{\frac{1}{2}}
\newcommand{\thalf}{\tfrac{1}{2}}
\newcommand{\ord}[1]{\mathcal{O} \left( #1 \right)}
\newcommand{\tr}{\mathrm{tr}}

\newcommand{\tvec}[1]{\vec{#1}_\bot}

\renewcommand\a{\alpha}

\renewcommand\d{\delta}

\renewcommand\l{\lambda}
\renewcommand\r{\rho}

\newcommand\e{\epsilon}
\newcommand\g{\gamma}

\newcommand\m{\mu}

\newcommand\p{\pi}

\newcommand\s{\sigma}

\renewcommand\O{\Omega}

\newcommand\G{\Gamma}

\newcommand{\eq}[1]{Eq.~(\ref{#1})}

\renewcommand{\vec}{\boldsymbol}
\renewcommand{\part}{{\rm part}}

\newcommand{\com}[1]{{\sf\color[rgb]{0,0.6,0}[#1]}}

\newcommand{\be}{\begin{equation}}
\newcommand{\ee}{\end{equation}}
\newcommand{\bes}{\begin{subequations}}
\newcommand{\ees}{\end{subequations}}
\newcommand{\bea}{\begin{eqnarray}}
\newcommand{\eea}{\end{eqnarray}}

\newcommand{\pa}{\partial}

\renewcommand{\tr}{\textrm{tr}}

\begin{document}

\title{Ab Initio Coupling of Jets to Collective Flow \\ in the Opacity Expansion Approach}

\author{Andrey V. Sadofyev}
\email[Email: ]{andrey.sadofyev@usc.es}
\affiliation{Instituto Galego de F{\'{i}}sica de Altas Enerx{\'{i}}as,  Universidade de Santiago de Compostela, Santiago de Compostela 15782,  Spain}
\affiliation{Institute for Theoretical and Experimental Physics, Moscow 117218, Russia}
\author{Matthew D. Sievert}
\email[Email: ]{msievert@nmsu.edu}
\affiliation{Department of Physics, New Mexico State University, Las Cruces, NM 88003, USA}
\author{Ivan Vitev}
\email[Email: ]{ivitev@lanl.gov}
\affiliation{Theoretical Division, Los Alamos National Laboratory, Los Alamos, NM 87545, USA}

\begin{abstract}
    We calculate the leading corrections to jet momentum broadening and medium-induced branching that arise from  the velocity of a moving medium at first order in opacity. These results advance our knowledge of jet quenching and demonstrate how it couples to the collective flow of the quark-gluon plasma in heavy-ion collisions.  We also compute the leading corrections to jet momentum broadening due to transverse gradients of temperature and density.  We find that the velocity effects lead to both anisotropic transverse momentum diffusion proportional to the medium velocity and anisotropic medium-induced radiation emitted preferentially in the direction of the flow. We isolate the relevant sub-eikonal corrections by working with jets composed of scalar particles with arbitrary color factors interacting with the medium by scalar QCD. Appropriate substitution of the color factors and light-front wave functions allow us to immediately apply the results to a range of processes including $q \rightarrow q g$ branching in real QCD. The resulting general expressions can be directly coupled to hydrodynamic simulations on an event-by-event basis to study the correlations between jet quenching and the dynamics of various forms of nuclear matter. 
\end{abstract}

\date{\today}
\maketitle

\tableofcontents
\newpage

\section{Introduction}
\label{sec:Intro}

In ultra-relativistic heavy-ion collisions a new extreme state of nuclear matter is created far from equilibrium, with an energy density so high that the quarks and gluons normally confined within hadrons are liberated into a continuous medium~\cite{Shuryak:1980tp}.  After thermalization, when the medium is sufficiently close to equilibrium to be characterized by local thermodynamic quantities, it is known as the quark-gluon plasma (QGP).  The QGP exhibits pronounced collectivity~\cite{Adams:2005dq,Adcox:2004mh,Aamodt:2010pa}, behaving as a nearly ideal fluid~\cite{Gyulassy:2004zy}; while its constituents are deconfined, they are also strongly interacting and far from being ``free.''  The subsequent expansion and cooling of these QGP droplets are well described by relativistic viscous hydrodynamics \cite{Romatschke:2009im,Gale:2013da}, and when the energy density in a given fluid element falls low enough, the nuclear matter becomes confined once again into a gas of hadrons.  The multi-phase evolution of this hot and dense medium is especially interesting because of its sensitivity to the fundamental mechanisms of color confinement, the distributions of color charge and momentum inside the initial colliding nuclei, and the origin of hadronic matter (for a review see Ref.~\cite{Busza:2018rrf}).

The ``soft'' particles originating from the QGP after hadronization have characteristic transverse momenta close to the temperature of the medium and are characterized by their strong final-state collectivity~\cite{Andronic:2005yp,Torrieri:2004zz}.  On the other hand, the ``hard probes'' of the QGP -- highly energetic particles and jets -- originate from the initial hard scatterings in the nuclear collision. While the information encoded in the distributions of soft particles entangles different stages of medium evolution, the production of jets across a range of transverse momenta $p_\bot$ and rapidities $y$ provides time resolution of the initial dynamics. By interacting with the medium in a path-length-dependent way, jets lose energy and have their substructure modified, providing an ``X-ray'' of the nuclear matter.  This concept is the core idea of the jet tomography \cite{Vitev:2002pf}.

Hadrons and jets provide an important tomographic tool as well in the cold nuclear matter probed in deep inelastic scattering events at the electron-ion collider (EIC) \cite{Li:2020zbk,Li:2020rqj,Li:2020wyc, Arratia:2020nxw}.  Part of the  mission of the EIC is to perform tomographic imaging of the partonic content of protons and nuclei, including the distribution of those partons in space and in momentum.  The energy loss and modification of jets in this cold nuclear medium is sensitive to these distributions, and in this context the intrinsic orbital motion of partons plays an analogous role to the collective flow in the QGP.

The energy loss of an energetic parton in QCD is dominated by soft gluon bremsstrahlung -- the radiation of small quanta of energy while propagating through nuclear matter~\cite{Gyulassy:1993hr}. The emission of multiple soft gluons leads to the energetic leading parton dissipating a considerable fraction of its initial energy. All first-principles descriptions of this process are based on perturbative techniques in Quantum Chromodynamics (QCD) which treat the interaction of the jet with the medium (quasi-)particles through $t$-channel gluon exchanges (see e.g. \cite{Baier:1996sk, Zakharov:1996fv, Zakharov:1997uu, Baier:1998kq, Gyulassy:1999zd, Gyulassy:2000fs, Gyulassy:2000er,Wang:2001ifa,Zhang:2003wk,Arnold:2002ja,Djordjevic:2003zk,Feal:2018sml,Andres:2020vxs}). These approaches agree at the level of the fundamental perturbative processes, differing primarily in assumptions about the characteristic length scales such as the mean free path of the parton, the size of the matter, and the gluon emission/splitting coherence length.  Further differences are related to the description of the type of nuclear medium -- quark-gluon plasma or cold nuclear matter.

The perturbative QCD (pQCD) description of parton energy loss is usually treated with simplifying assumptions which make the calculation tractable.  These include the eikonal approximation, corresponding to the limit of an infinitely energetic leading parton.
The approximation (correlated with the eikonal approximation) that the medium scattering centers can be treated as static is often applied. Furthermore, the limit in which the initial hard-scattering point is well-separated in space from the subsequent rescatterings is usually taken as well. Finally, particular choices (usually equivalent within the aforementioned approximations) of the limits of integration for the soft gluon bremsstrahlung are made.
Taken together, these approximations allow one to neglect the collisional energy transfer and in-medium source recoil in a single collision, with the jet-medium interaction being governed purely by Glauber gluon exchange carrying only transverse momentum~\cite{Ovanesyan:2011xy,Kang:2016ofv,Fickinger:2013xwa}.  Under these approximations, the medium velocity is decoupled from the jet energy loss, with such correlations neglected as sub-eikonal corrections.  Moreover, the thermodynamic medium parameters (such as the momentum broadening per unit length $\hat{q}$) are commonly treated as constant in the directions transverse to the jet axis, while in realistic heavy-ion collisions these quantities can be expected to vary considerably across the jet's evolution.  Thus, while jet production and substructure are in principle amenable to a full-fledged program of jet tomography, realizing this goal in practice will require significant effort to relax these simplifying assumptions.

An example of efforts in this direction is the use of a hard thermal loop potential~\cite{Djordjevic:2008iz} which relaxes the recoilless approximation. The effects of medium expansion and dilution (due to Bjorken and transverse expansion) were considered early on~\cite{Gyulassy:2000gk,Gyulassy:2001kr}. Some transverse flow effects on the energy loss were also discussed on the basis of purely kinematic arguments, see e.g. \cite{Baier:2006pt,Liu:2006he,Renk:2006sx}. Still there is no dynamical treatment of the coupling of a moving medium, nor are real gradient corrections computed from first principles. Motivated by phenomenological considerations, a model that incorporates an additional  momentum transfer in the scattering potential  was proposed in \cite{Armesto:2004pt,Armesto:2004vz}. It was later established that  the apparent elongation of away-side particle correlations is in fact a superposition of symmetric minijets and long-range flow harmonics of the underlying bulk matter. As we will show below, the motion of the scattering centers of the medium translates into different effects on the broadening of jets and soft gluon radiation.  More recently, it has been shown that gradient corrections to the eikonal jet-medium coupling can be used for tomographic purposes as well \cite{He:2020iow}.

Some conceptual progress in this direction was achieved in strongly-interacting holographic models of the jet-medium interaction, see e.g. \cite{Lekaveckas:2013lha, Rajagopal:2015roa, Sadofyev:2015hxa,Casalderrey-Solana:2015vaa, Casalderrey-Solana:2016jvj, Li:2016bbh, Brewer:2017dwd, Brewer:2017fqy, Brewer:2018mpk, Reiten:2019fta, Arefeva:2020jvo}.  While not directly connected to first-principles pQCD, the holographic duality is a powerful tool allowing one to relate some 3+1-dimensional conformal field theories in the strong-coupling regime with a higher-dimensional gravitational theory \cite{Maldacena:1997re, Gubser:1998bc, Witten:1998qj}. This description is particularly well-suited to the computation of transport parameters in theories with known duals. While the exact dual of QCD is still unknown, these studies give valuable information about the features of strongly-coupled theories similar to QCD in the non-perturbative regime.  For example, in Ref.~\cite{Lekaveckas:2013lha}, it was shown that within this framework the drag force of the medium on the jet can be determined within the same gradient expansion which underlies the (traditional) hydrodynamic description of the medium itself.  Notably, the gradient corrections to the drag force were non-negligible and were necessary to bring the theoretical rate of energy loss in line with realistic simulations within the same model \cite{Chesler:2013urd}.

In pursuit of the goal of full-fledged jet tomography in heavy-ion collisions, we consider the analog of the holographic studies of Refs.~\cite{Lekaveckas:2013lha, Rajagopal:2015roa, Sadofyev:2015hxa, Reiten:2019fta} within the pQCD framework of the Gyulassy-Levai-Vitev (GLV) opacity expansion formalism.  We will treat explicitly the sub-eikonal motion of the medium scattering centers and compute the corresponding modification of the collisional and radiative processes within the opacity expansion. We will also assume that there is an order of time scales ensuring that the sources can be considered to be moving with a constant velocity during a single interaction but their velocity and thermodynamic parameters of the matter may change between different scattering centers. In doing so, we directly obtain a coupling between the motion of the medium and the pattern of medium-induced radiation, taking a significant step toward the eventual goal of full-fledged jet tomography.

This work is organized as follows. In Section~\ref{sec:Formalism} we set up the problem of jet propagation in a moving medium and use scalar QCD as the underlying field theory. The accuracy of our derivation and underlying assumptions are discussed.  The effect of the medium's velocity on the distribution of jets is derived, and we further obtain the leading linear gradient corrections in Section~\ref{sec:Broadening}. We illustrate these effects with moments of the final-state momentum  distribution.  Section~\ref{sec:Rad} shows the  calculation of medium-induced branching processes in scalar QCD for general kinematics of the splitting and the effect of medium motion. We further derive the soft parton emission limit. We conclude in Section~\ref{sec:concl}, where we also discuss what aspects of this  calculation are applicable to the future electron-ion collider.  All of the technical details of our derivation are documented extensively in the appendixes. In Appendix~\ref{app:Regge} we show that the unpolarized $2 \rightarrow 2$ scattering of partons by $t$-channel gluon exchange is universal (independent of species) even after including the first sub-eikonal corrections. Appendix~\ref{app:snglBorn} presents the derivation of the single Born scattering diagrams with velocity and gradient corrections. The corresponding calculation for double Born diagrams is documented in Appendix~\ref{app:dblBorn}.  Lastly, the derivation of the in-medium splitting amplitudes with single Born and  double Born interactions is given in Appendix~\ref{sec:BRsnglBorn} and  Appendix~\ref{sec:BRdblBorn}, respectively.

\section{Theoretical Formalism}
\label{sec:Formalism}
%

%
\subsection{Jet-Medium Interactions in Regge Kinematics}
%

The theory of medium modification of jets in QCD is fundamentally based on a large separation of scales between the jet with energy $E$ and the medium, whose momentum scales are characterized by the temperature $T$ or Debye mass $\mu = g T$.  The jet kinematics $E \gg \mu$ are an essential starting point of all perturbative descriptions of jet quenching, which leads to the interactions of the jet with the medium taking a particular form.  Specifically, the multiple scatterings which the jet undergoes as it propagates through the medium take place in \textit{Regge kinematics} (not to be confused with the pre-QCD ``Regge trajectories''), in which the center-of-mass energy $s \sim M E$ between the jet and the medium particles is much larger than the typical momentum transfer $|t| \sim \mu^2$; that is, $|t / s| \ll 1$.  (Here $M$ is the mass of a particle in the medium and the jet energy $E$ is evaluated in the medium rest frame.)  The leading-power behavior as $s \rightarrow \infty$, or equivalently as the jet energy $E \rightarrow \infty$, is referred to as the \textit{eikonal approximation}.  Whether one works at eikonal accuracy or includes the contributions of sub-eikonal, power-suppressed corrections in $\frac{p_\bot}{E} \sim \frac{\mu}{E}$, the relevant physics for the propagation of high-energy partons through a medium is the small-angle, forward scattering of Regge kinematics.  These purely kinematic features are common to any field-theoretic description of high-energy jets losing energy and being modified by the interaction with the medium.

While there are different schemes for implementing perturbative jet-medium physics, in this paper we work within the framework of the opacity expansion.  The opacity expansion corresponds to a perturbative order-by-order expansion in the typical number of scatterings a high-energy parton will undergo in the medium before escaping.  If the length of the medium is $L$ and the mean free path is $\lambda = (\rho \, \sigma_0)^{-1}$, with $\rho$ the number density of particles in the medium and $\sigma_0$ the total $2 \rightarrow 2$ elastic cross section for the jet to scatter on those particles, then the expected number of scatterings in the medium is $L / \lambda$, which is referred to as the opacity.  The perturbative opacity expansion, in the form which we employ here, was pioneered by Gyulassy, Levai, and Vitev (GLV) \cite{Gyulassy:1999zd,Gyulassy:2000er,Gyulassy:2000fs}.  Other approaches fundamentally based on the same perturbative ingredients also exist, notably the path integral formalism of Baier, Dokshitzer, Mueller, Peigne, Schiff, and Zakharov (BDMPS or BDMPS-Z) \cite{Zakharov:1996fv,Zakharov:1997uu,Baier:1996sk,Baier:1998kq} which formally resums the opacity series into a path integral and then makes subsequent approximations suitable for a very dense medium.  These two formulations built out of elementary QCD ingredients are consistent and compatible, providing useful analytic insight into the physics of jet-medium interaction in complementary regimes.  Additionally, interesting recent work \cite{Mehtar-Tani:2019tvy,Andres:2020vxs}  has also explored the viability of a hybrid description that captures elements of both approaches.

Because of the large separation of scales $E \gg \mu$ between the jet and the medium, the formalism of jet-medium interactions can be expressed quite generally at the Lagrangian level in terms of the interaction of the jet with a background field \cite{Idilbi:2008vm,Ovanesyan:2011xy}.  At eikonal accuracy, the background field modifies the jet in two essential ways: by exchanging transverse momentum between the jet and the medium, and by imprinting various quantum phases on the scattering amplitude of the jet.  The accumulation of transverse momentum by the jet as it propagates through the medium leads to the phenomenon of jet broadening characterized by $\hat{q}$, the typical transverse momentum squared acquired per unit length.  Likewise the accumulation of quantum phases leads to a highly nontrivial modification of the soft gluon radiation pattern known as the non-Abelian Landau-Pomeranchuk-Migdal (LPM) effect.  These modifications of jet observables by the medium are the basis for the program of jet tomography, whereby one attempts to deconvolute the jet observables to extract the interferometric information they carry about the microscopic content of the medium.

%
\begin{figure}[t]
    \centering
	\includegraphics[width=0.7\textwidth]{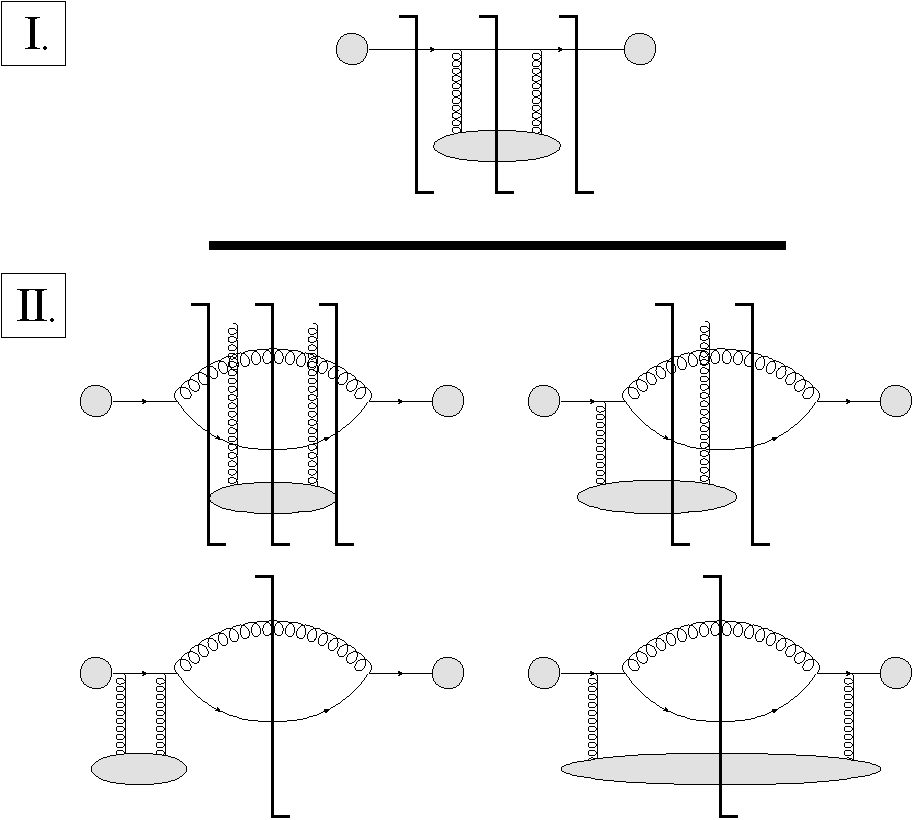}
	\caption{
	    Diagrams contributing to jet broadening (I.) and medium-induced radiation (II.) of a quark jet in QCD at first order in opacity.  Thick vertical lines denote all possible final-state cuts, and vertical gluons left dangling denote all possible attachments to partons in the jet at that point.  Asymmetric diagrams have mirror conjugates which are not drawn explicitly.
	} 
	\label{f:usualdiags}
\end{figure}
%

We illustrate in Fig.~\ref{f:usualdiags} the diagrammatic contributions to both jet broadening and medium-induced radiation in QCD at first order in the opacity expansion.  These perturbative contributions sum over all possible single scatterings of the jet with the medium, and the resulting integral over the position of the single scattering effectively iterates this kernel $n = L/\lambda$ times across the medium of length $L$.  Higher-order corrections to the diagrams shown in Fig.~\ref{f:usualdiags} systematically include more correlated scatterings or branchings \cite{Fickinger:2013xwa,Casalderrey-Solana:2015vaa} in the kernel.  For the medium-induced radiation, this kernel always includes the vacuum-like elementary branching vertex, together with the associated medium scattering in all possible ways.  Because of this, the medium-induced radiation kernel can be expressed in terms of the light-front wave functions (LFWF) of the elementary branching \cite{Brodsky:1997de}, which (when squared) are simply proportional to the Altarelli-Parisi vacuum splitting functions \cite{Sievert:2018imd}.  The appearance of LFWF as elementary ingredients of the medium-modified splitting functions is natural, since the diagrams of Fig.~\ref{f:usualdiags} can be computed equally well in light-front (time-ordered) perturbation theory by collecting the on-shell poles of all intermediate propagators.

%
\subsection{Scalar Approximation to Unpolarized QCD}
%

In Regge kinematics, the interaction of the jet with the medium depends on the species of partons involved in only a limited number of ways.  The $g \rightarrow g g$ QCD process for instance  (see Fig.~\ref{f:Born_Rad_A}), in which the parent gluon scatters in the medium before splitting in the final state, differs from the corresponding $g \rightarrow q \bar{q}$ process only by the replacement of the LFWF.  The elastic scattering of quarks versus gluons certainly differs by an overall color factor owing to the different color representations of the partons, but it is well known that the kinematic dependence of the elastic scattering cross section is identical for $q q \rightarrow q q$, $q g \rightarrow q g$, and $g g \rightarrow g g$ in the eikonal limit \cite{Gyulassy:1993hr}.  Indeed, it has long been known that the kinematic dependence of elastic scattering in Regge kinematics depends only on the spin of the exchanged parton, not the species of partons being scattered.  For these reasons, it is a simple matter to transform the eikonal calculations of jet broadening and medium-induced branching from one partonic process to another, simply by replacing the associated color factors and LFWF; this mapping was explicitly used in \cite{Sievert:2019cwq} and validated against previous direct calculations in the literature.  This universality has been exploited since the early days of jet quenching theory to remove unessential details of the calculation, such as replacing quarks with fundamental-representation scalars in scalar QCD in the high-energy (Regge) limit in the original GLV papers if only soft gluon emission is considered \cite{Gyulassy:1999zd}.

Because the main focus of this paper is the inclusion of velocity-dependent sub-eikonal corrections to the known jet broadening and medium-induced radiation results, we again wish to employ the simplification to equivalent scalar theories.  Doing so will help us to isolate the relevant $\sim\ord{\frac{\bot}{E}}$ kinematic corrections to the usual results, where $\bot$ is one of the characteristic transverse scales such as $|\vec{p}_\bot|$ or $\m$. However, one may rightly question whether the elastic scattering processes remain universal beyond the eikonal limit.  Certainly the exact $2 \rightarrow 2$ elastic cross sections for various partonic species differ in their kinematic dependence, so at some order in the sub-eikonal corrections $\ord{ \left( \frac{\bot}{E} \right)^n }$ this universality must break down.  Whether that universality continues to hold at the accuracy $\sim\ord{\frac{\bot}{E}}$ we wish to keep in order to capture the leading sub-eikonal velocity corrections is not obvious \textit{a priori}, so we verify it explicitly in Appendix~\ref{app:Regge}.

Based on the universality of Regge scattering and the interchangeability of the light-front wave functions, we will hereafter replace the quarks and gluons of QCD with scalar particles, both in the jet and in the medium.  We can choose the color representation of the various scalars to be fundamental or adjoint, and we can choose the scalar fields to be either real or complex.  We will consider two cases: either the medium particles are real scalar fields in the adjoint representation, or they are complex scalar fields in the fundamental representation.  The choice of either real adjoint scalars or complex fundamental scalars for the jet allows us to mimic the color structures of real QCD without the additional complexity of the numerator algebra.  In our case the real adjoint scalars are analogous to scalar gluons, and the complex fundamental scalars are analogous to scalar quarks and antiquarks.  

These choices can be encoded at the level of the Lagrangians describing the interactions of the medium particles among themselves and of the jet with the medium.  For the medium particles we use a massive scalar field $\Phi$ chosen to correspond to scalar gluons or scalar (anti)quarks as
\begin{subequations}    \label{e:Lmed}
\begin{align}
    \mathcal{L}_{med} &= \half 
    \Big( \partial_\mu \Phi^a - i g A_\mu^b (t_\mathrm{adj}^b)_{ac} \Phi^c \Big)
    \Big( \partial^\mu \Phi^a - i g A^{\mu d} (t_\mathrm{adj}^d)_{ae} \Phi^e \Big)
    \notag \\ &\hspace{1cm}
    - \half M^2 \Phi^a \Phi^a - \frac{1}{4} F_{\mu \nu}^a F^{a \mu \nu} \: ,
    \\
    \mathcal{L}_{med} &= 
    \Big( \partial_\mu \Phi_i - i g A_\mu^a (t_\mathrm{fund}^a)_{ij} \Phi_j \Big)^\dagger
    \Big( \partial^\mu \Phi^i - i g A^{\mu b} (t_\mathrm{fund}^b)_{ik} \Phi_k \Big)
    \notag \\ &\hspace{1cm}
    - M^2 \Phi_i^\dagger \Phi_i - \frac{1}{4} F_{\mu \nu}^a F^{a \mu \nu} \: ,
\end{align}
\end{subequations}
respectively.  Here $t^a_\mathrm{fund}$ and $t^a_\mathrm{adj}$ are the generators of $SU(N_c)$ in the fundamental or adjoint representations, respectively.  For our purposes these two cases will differ only by the representation of the generator $t^a$, so we will leave it arbitrary and evaluate specific color factors only at the end of the calculation.  The purpose of introducing the Lagrangian \eqref{e:Lmed} for the medium particles is to calculate the leading correction to the external potential $A^{\mu a}_\mathrm{ext}$ they generate describing the interaction with the jet, which we detail in Sec.~\ref{sec:potl} below.

For the jet particles, we similarly introduce a massless scalar field $\phi$ corresponding to either scalar gluons or scalar (anti)quarks; the choice for the jet partons $\phi$ need not be the same as for the medium partons $\Phi$.  The Lagrangians describing the interaction of the scalar gluons or scalar (anti)quarks in the jet with the external field produced by the medium are
\begin{subequations}    \label{e:Ljet}
\begin{align}   
    \mathcal{L}_{jet} &= \half (\partial_\mu \phi^a)(\partial^\mu \phi^a) 
    - i g A_\mathrm{ext}^{\mu a} \: (\partial^\mu \phi^b) \, (t^a_\mathrm{adj})_{b c} \, \phi^c \:,
    \\ 
    \mathcal{L}_{jet} &= 
    (\partial_\mu \phi_i)^\dagger (\partial^\mu \phi_i)
    - i g A_\mathrm{ext}^{\mu a} \:
    \Big(
        (\partial_\mu \phi_i)^\dagger \, (t^a_\mathrm{fund})_{i j} \, \phi_j -
        \phi_i^\dagger \, (t^a_\mathrm{fund})_{i j} \, (\partial_\mu \phi_j)
    \Big) \: ,
\end{align}
\end{subequations}
respectively.\footnote{
We do not explicitly include the ``seagull'' vertex of scalar QCD because it does not contribute to the cross section at the order we consider here.  Any corrections they introduce are independent of the velocity, scaling at most as $\ord{\frac{\bot}{E}}$.  Moreover, the seagull corrections can enter only for the double-Born diagrams and have no unitary counterpart in the single-Born diagrams.  As such, at any order in $\frac{\bot}{E}$ the seagull diagrams form a unitary subset among themselves and do not affect the rest of the calculation.
}
The Lagrangians \eqref{e:Ljet} are sufficient to compute the velocity corrections to jet momentum broadening, which we detail in Sec.~\ref{sec:Broadening}.  For the calculation of medium-induced branching, we can further exploit the interchangeability of light-front wave functions by examining the $\phi \rightarrow \phi \, \phi$ branching process with a $\lambda \phi^3$ vertex. The appearance of this branching vertex will be different depending on the color representation of the jet partons, giving for the case of scalar gluon/gluon splitting and for the mixed case of scalar quark/gluon splitting
\begin{subequations}    \label{e:Ljet2}
\begin{align}   \label{e:Ljet2a}
    \mathcal{L}_{jet} &= \half (\partial_\mu \phi^a)(\partial^\mu \phi^a) 
    - i g A_\mathrm{ext}^{\mu a} \: (\partial^\mu \phi^b) \, (t^a_\mathrm{adj})_{b c} \, \phi^c
    - \frac{1}{3!} \lambda \, d^{a b c} \, \phi^a \phi^b \phi^c
    \\      \label{e:Ljet2b}
    \mathcal{L}_{jet} &= \half (\partial_\mu \phi^a)(\partial^\mu \phi^a) +
    (\partial_\mu \phi_i)^\dagger (\partial^\mu \phi_i)
    - i g A_\mathrm{ext}^{\mu a} \: (\partial_\mu \phi^b) \, (t^a_\mathrm{adj})_{b c} \, \phi^c \:,
    \notag \\ &\hspace{1cm}
    - i g A_\mathrm{ext}^{\mu a} \:
    \Big(
        (\partial_\mu \phi_i)^\dagger \, (t^a_\mathrm{fund})_{i j} \, \phi_j -
        \phi_i^\dagger \, (t^a_\mathrm{fund})_{i j} \, (\partial_\mu \phi_j)
    \Big) 
    - \lambda \phi^a \: \phi_i^\dagger (t^a_\mathrm{fund})_{i j} \phi_j \: ,
\end{align}
\end{subequations}
respectively.  

Our choice to study medium-induced $\phi \rightarrow \phi \, \phi$ branching and map the LFWF and color factors back to QCD afterward allows us to study the kinematic impact of the velocity corrections on the LPM interference pattern in its simplest form, without the additional complications from the numerator algebra of real QCD.  The exception to this simple mapping between our scalar theory and real QCD is the $\phi \rightarrow \phi \, \phi$ branching of the adjoint scalars in \eqref{e:Ljet2a}.  While this adjoint branching channel mimics in a sense the $g \rightarrow g g$ splitting process in QCD, it is interesting to note that the scalar $\phi^3$ vertex couples to a different color structure than in real QCD.  In QCD, the triple-gluon vertex couples to the antisymmetric structure constants $i f^{a b c}$ via a derivative coupling,  whereas the $\phi^3$ vertex in \eqref{e:Ljet2a} couples instead to the \textit{symmetric} structure constants $d^{a b c}$.  While this essential difference makes the $\phi \rightarrow \phi \, \phi$ adjoint scalar branching process not directly comparable to the $g \rightarrow g g$ process in QCD, the different color structures it produces mediated by $d^{a b c}$ are interesting to study in their own right.  We leave a detailed revisiting of the $g \rightarrow g g$ channel in real QCD for future work.

While the replacement of the jet and medium partons with scalars significantly simplifies the calculation of the velocity corrections, it is important that we do not oversimplify the problem and fundamentally change the gluon-mediated interaction between the jet and the medium.  The $t$-channel exchange of non-Abelian vector bosons is essential to obtaining an energy-independent eikonal limit of the elastic cross section $\frac{d\sigma}{d|t|} \sim \frac{1}{s^2} \langle | \mathcal{M} |^2 \rangle \sim const$ discussed in Appendix~\ref{app:Regge}.  If we had replaced even the exchanged gluons with scalars, the resulting cross section would have been doubly suppressed, scaling as $\frac{d\sigma}{d|t|} \sim E^{-2}$ \cite{Regge:1959mz}.  For this reason, we choose to work with scalar QCD in the medium \eqref{e:Lmed} in either the fundamental or adjoint representations, dressing those gluons with an effective Debye mass $\mu \sim g T$ proportional to the temperature of the thermal medium as in the Gyulassy-Wang potential~\cite{Gyulassy:1993hr}.

Finally, we note that there is one important class of sub-eikonal corrections which can be sensitive to the medium velocity which is not captured by our treatment here using a scalar theory: the exchange of quarks in the $t$-channel rather than gluons.  The same power counting \cite{Regge:1959mz} which indicates that the elastic cross section scales like $\sigma \sim E^{-2}$ for $t$-channel scalars reveals that the exchange of quarks goes as $\sigma \sim E^{-1}$ and is the same order as the sub-eikonal corrections we wish to keep.  Indeed $t$-channel quark exchange, corresponding to the QCD Reggeon \cite{Kirschner:1983di, Griffiths:1999dj, Itakura:2003jp}, is known to enter at the same order as the subeikonal component of gluon exchange and can carry information about the flavor, spin, and velocity of the medium particles \cite{Kovchegov:2015pbl, Kovchegov:2017lsr, Kovchegov:2018znm}.  These quark-exchange contributions do not appear in the scalar theory we consider here and make a separate class of contributions to the dependence on the medium velocity.  We defer an analysis of these interesting contributions for future work.

%
\subsection{External Potential of a Moving Medium}
\label{sec:potl}
%

%
\begin{figure}[t]
    \centering
	\includegraphics[width=0.4\textwidth]{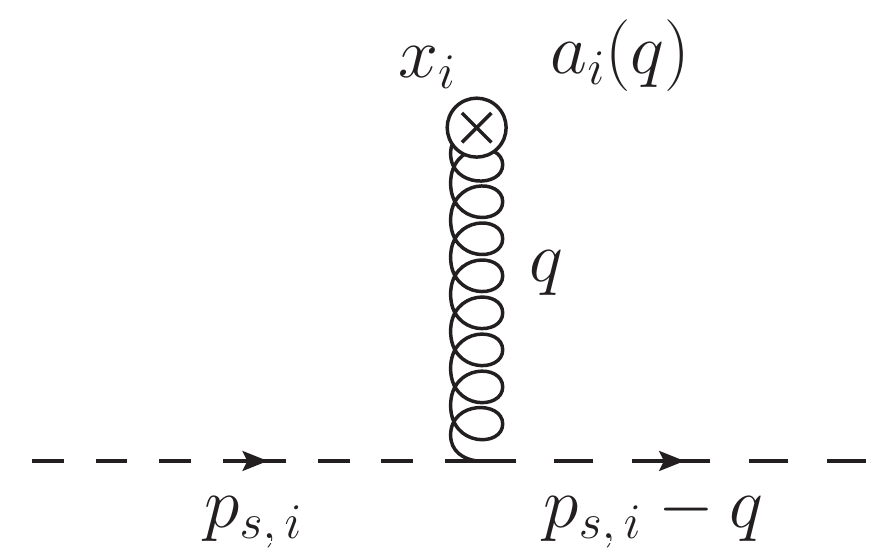}
	\caption{Gluonic potential $a_i^{\mu a} (q)$ produced by medium particle $i$.} 
	\label{f:Potential}
\end{figure}
%

Consider the gluon field-mediated potential of a QCD medium made up of particles in motion. We permit individual particles which compose that medium to move with a space-time-dependent four velocity field $U^\mu (x)$.  The total potential $A_{ext}$ of the medium is composed of a superposition of the individual potentials $a_i$ of the various particles at space-time positions $x_i^\mu = (0 , \vec{x}_i)$:\footnote{Before we continue, it is instructive to comment on the nature of the phase shifts in (\ref{e:potl1}), which are essential for our discussion. Considering a classical color charge current $j^{a\,\m}(x)=\sum_i (igt^a_i) \int\,dt\,\frac{dx_i^\m}{dt}\delta^{(4)}(x^\a-x_i^\a(t))$ of point-like sources moving with constant velocities $\vec{u}_i=\frac{d\vec{x}_i}{dt}$, one may find that its Fourier transform reads $j^{a\,\m}(q)= (2\p)\sum_i (igt^a_i) u_i^\m e^{-i\vec{q}\cdot\vec{x}_i^{(0)}}\d\left(q^0-\vec{u}_i\cdot \vec{q}\right)$, where $\vec{x}_i^{(0)}$ is the position of the source at $t=0$, and $u^\mu=(1, \vec{u})$. It is this phase shift in the current, expressed through the positions of the sources at a given moment, which appears in the color field later on. Notice that the phase factor is invariant under a shift in the initial time due to the relation between the momentum components, and $x^{(0)}_i$ can be taken to be purely spatial. For brevity we omit the superscript $(0)$ in all the other formulas throughout the paper.}
\begin{align}
\label{e:potl1}
    A_\mathrm{ext}^{\mu a} (q) = \sum_i e^{i q \cdot x_i} a_i^{\mu a} (q)\, .
\end{align}
The individual potentials $a_i$ can be computed from the diagram drawn in Fig.~\ref{f:Potential}:
\begin{align}
\label{e:potl2}
    a_i^{\mu a} (q) = (i g \, t_i^a) \, (2 p_{s \, i} - q)_\nu
    \left( \frac{-i g^{\mu \nu}}{q^2 - \mu^2 + i \epsilon} \right)
    \: (2\pi) \, \delta\Big( (p_{s \, i} - q)^2 - M^2 \Big) \,  .
\end{align}
Here, $t_i^a$ is the generator of $SU(N_c)$ in the appropriate representation in the color space of medium particle $i$, $g$ is the strong coupling constant, and $\mu$ is the Debye mass of the $t$-channel gluon. The medium particles (or ``scattering centers'') have invariant mass $M$, which defines their relativistic velocity $U_i^\mu = U^\mu (x_i)$  in terms of their momenta $p_{s \, i}^\mu$ by
\begin{align}
\label{e:flow1}
    p_{s \, i}^\mu = M \Big( \sqrt{1 + \vec{U}_i^2} \, , \, 
    \vec{U}_{i \, \bot} \, , \, U_{i \, z} \Big)
    \equiv M U_i^\mu \, . 
\end{align}
The on-shell condition $p_{s \, i}^2 = M^2$ implies that the velocity is time-like: $U_i^2 = 1$, which is satisfied by requiring $U_i^0 = \sqrt{1 + \vec{U}_i^2}$.

Using the kinematics \eqref{e:flow1} in the potential \eqref{e:potl2} gives
\begin{align}
\label{e:potl3}
    a_i^{\mu a} (q) &= g \, t_i^a \, 
    \left( \frac{2 M U_i^\mu - q^\mu}{q^2 - \mu^2 + i \epsilon} \right)
    \: (2\pi) \, \delta\Big( 2 M U_i \cdot q - q^2 \Big)
    \notag \\ &=
    g \, t_i^a \, \left( \frac{ U_i^\mu / U_i^0 }
    {q^2 - \mu^2 + i \epsilon} \right)
    \: (2\pi) \, \delta\left( q^0 - \frac{\vec{U}_{i}}{U_i^0} \cdot \vec{q} \right) ,
\end{align}
where we have neglected $q^\mu$ compared to $2 M U_i^\mu$ and $q^2$ compared to $2 M U_i \cdot q$.  In the original construction \cite{Gyulassy:1993hr,Gyulassy:2000er}, this is justified by considering the mass of the particles to be very large $M \rightarrow \infty$.  While we make the same approximation here, we note that in a more symmetric frame, the relevant suppression parameter for medium recoil will be the jet energy $E$ \cite{Sievert:2018imd}.  

From \eqref{e:potl3} we see that the velocity enters through the combination $U_i^\mu / U_i^0$, corresponding to the \textit{nonrelativistic velocity field} (i.e., with the relativistic $\gamma$-factor removed)
\begin{align}
\label{e:flow2}
    u_i^\mu \equiv U_i^\mu / U_i^0 = \left( 1 \, , \, \frac{1}{\sqrt{1 + \vec{U}_i^2}} \, \vec{U}_{i \, \bot} \, , \, \frac{1}{\sqrt{1 + \vec{U}_i^2}} \, U_{i \, z} \right) \, ,
\end{align}
with $U_i^0 = \sqrt{1 + \vec{U}_i^2} = (1 - \vec{u}_i^2)^{-1/2}$ the relativistic $\gamma$-factor.  In terms of the nonrelativistic velocity $u_i^\mu$, the potential takes on the simple form 
\begin{align}
\label{e:potl4}
    g \, a_i^{\mu a} (q) &= 
    t_i^a \, u_i^\mu \: v_i (q) \: 
    (2\pi) \, \delta\left( q^0 - \vec{u}_{i} \cdot \vec{q} \right) \, , 
\end{align}
with
\begin{align}
\label{e:potl5}
    v_i (q) \equiv v_i\left(\vec{q}^2 - (\vec{u}_i \cdot \vec{q})^2 \right) \equiv \frac{- g^2}{
    \vec{q}^2 + \mu_i^2 - ( \vec{u}_i \cdot \vec{q} )^2 - i \epsilon }\, .
\end{align}
The extra factor of $g$ in \eqref{e:potl4} is associated with the vertex coupling the potential to the active parton.
Note that we use the compact notation $v_i (q)$ to denote the general dependence on the position/scattering center $i$ and the four-momentum transfer $q$, as well as the explicit notation $v_i\left(\vec{q}^2 - (\vec{u}_i \cdot \vec{q})^2 \right)$ indicating the functional form of that dependence on the components of $q$.

%

\subsection{Underlying Assumptions and Calculation Accuracy}

While all the assumptions used in this work are introduced around the corresponding calculations, it is convenient to summarize them here. As has been already mentioned, we rely on the GLV approach \cite{Gyulassy:1999zd,Gyulassy:2000er,Gyulassy:2000fs}, and focus on the first order in the opacity expansion for the scalar QCD. Following \cite{Gyulassy:1993hr}, we assume that the in-medium sources are heavy $M\to\infty$, and neglect all the contributions suppressed by this mass. We also use the eikonal approximation, but include the sub-leading corrections, which are sensitive to the medium flow, omitting only the terms of the form $\mathcal{O}\left(\frac{\bot^2}{E^2}\right)$. However, we allow the medium to be large along the $z$-direction, and we generally keep the terms scaling as $\mathcal{O}\left(\left(\frac{\bot^2}{E}z\right)^n\right)$. As in \cite{Baier:1996sk} we assume that the mean free path is sufficiently large such that $\m (z_i-z_0)\gg 1$. This approximation allows one to neglect the poles of the in-medium potentials entering the amplitudes, which have residues exponentially suppressed due to the Debye mass $\m$. We also assume that the interactions of the energetic parton with the sources are local, and color correlations between different sources can be neglected -- a local color neutrality condition. 

Apart from the approximations listed above, in our derivations we have additional new freedom to take into account the change in the medium properties in the transverse directions. In this work we explicitly consider two types of situations -- the medium is either assumed to be translationally invariant, or treated within the hydrodynamic gradient expansion up to the linear order. In the latter case, the novel contributions can be of two types scaling as $\mathcal{O}\left(\frac{\bot}{E}\,z\pa_\bot\right)$ or $\mathcal{O}\left(\frac{\pa_\bot}{E}\right)$, and we neglect the terms not enhanced by the medium length. However, one should notice that in the limit of an arbitrarily narrow initial jet distribution such contributions could be important, and we leave the corresponding discussion for future studies. 

While the methods developed in this work are rather general and can be applied both to the jet broadening and gluon emission problems, the radiation amplitudes are more involved. Thus, we restrict the consideration of the gluon emission to the case of translationally invariant medium, omitting all the gradient effects. Moreover, we additionally neglect the terms scaling as $\mathcal{O}\left(\frac{\bot}{E}\left(\frac{\bot^2}{E}z\right)^n\right)$, leaving some of the new interference structures for the future work.

%

%
\section{Jet Momentum Broadening  Results}
\label{sec:Broadening}

We consider an intrinsic distribution of ultra-relativistic jets
\begin{align} \label{e:N0}
    E \, \frac{dN^{(0)}}{d^3 p} \equiv \frac{1}{2(2\pi)^3} \left| J(p) \right|^2 \, , 
\end{align}
produced by a hard-scattering event in the medium at position $x_0=0$, with most of the jet momentum directed along the $+z$ axis: $p_z \approx E$.  This intrinsic distribution $E \frac{dN^{(0)}}{d^3 p} = E \frac{dN^{(0)}}{d^2 p_\perp\,dE}$ may measure the transverse momentum $\vec{p}_\bot$ relative to an external axis (such as the direction of a measured photon in the final state for $\gamma + \mathrm{jet}$ correlations or an initial-state virtual photon in deep inelastic scattering), or such an external axis may be integrated out to give the initial energy distribution of jets $E \frac{dN^{(0)}}{dE}$.  For maximum generality, we will keep the initial jet distribution $E \frac{dN^{(0)}}{d^3 p}$ differential in both energy and transverse momentum.  For processes such as deep inelastic scattering or $\gamma + \mathrm{jet}$ correlations, where the final-state jet $\vec{p}_\bot$ can be measured relative to an external axis, this transverse momentum distribution is broadened due to scattering in the total external potential \eqref{e:potl1}.

To derive the final-state distribution of jets we have to calculate the diagrams shown in the top row of Fig.~\ref{f:usualdiags}.  For our equivalent scalar theory, the distinct final-state cuts correspond to the  single- and double-Born diagrams shown in detail in Fig. \ref{f:Broad1}. The calculation is technical and presented in Appendix~\ref{app:snglBorn} and Appendix~\ref{app:dblBorn}, respectively.

%
\begin{figure}
    \centering
	\includegraphics[width=0.4\textwidth]{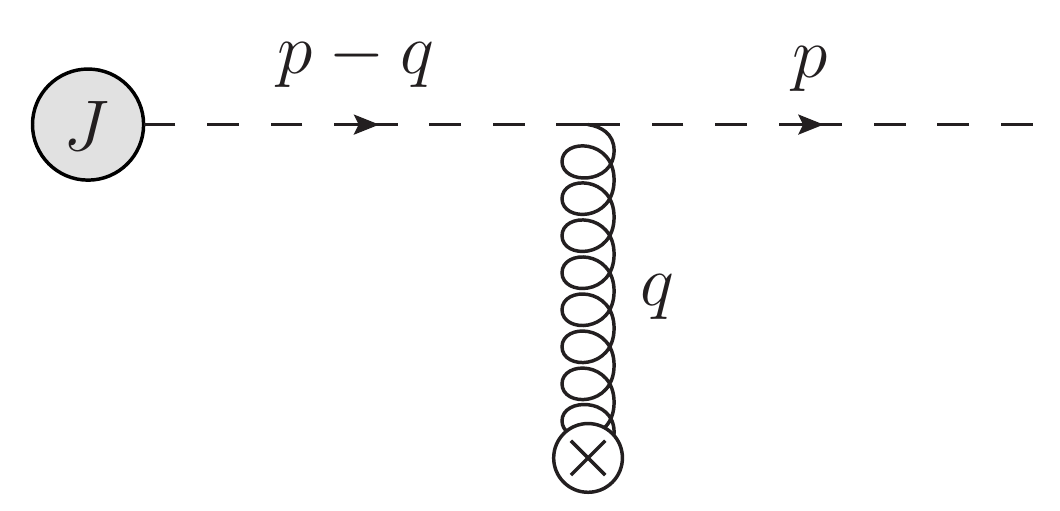}
	\includegraphics[width=0.4\textwidth]{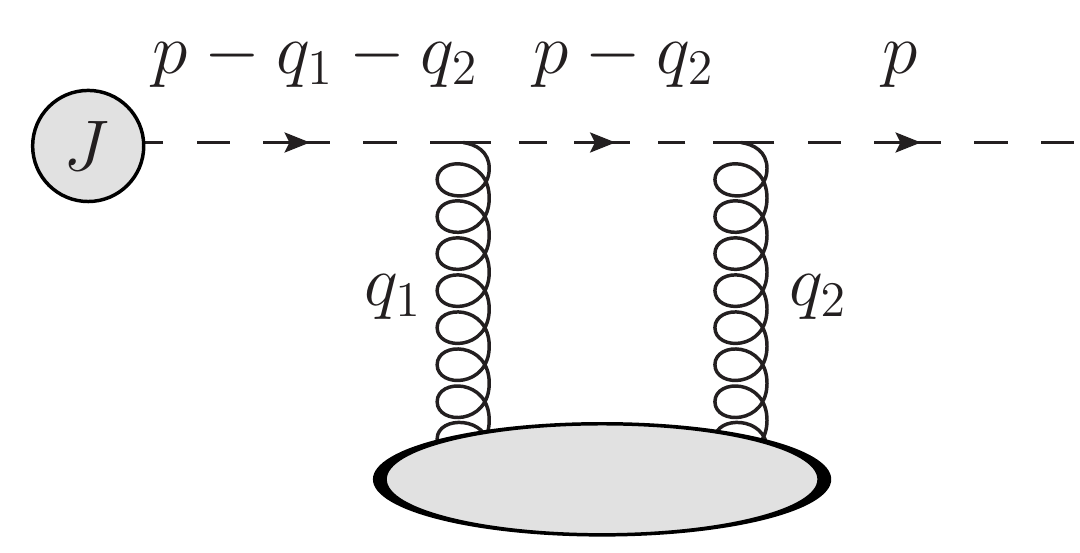}
	\caption{The Born-level amplitude $M_1$ (left) and double-Born amplitude $M_2$ (right) for transverse-momentum broadening in an external potential (left).}
	\label{f:Broad1}
	\label{f:DblBorn}
\end{figure}
%

%
\subsection{Final-State Distribution and Unitarity}

With the results from Appendix~\ref{app:snglBorn} and   Appendix~\ref{app:dblBorn}, one can write the full jet distribution at the first order in opacity. At the 0th order in the gradient expansion from (\ref{e:BroadBorn12}) and (\ref{e:DblBorn10}) 
we obtain
\begin{align}
\label{e:untrt1}
E \frac{dN^{(1)}}{d^3 p} 
    &= \int dz \, d^2q_\bot \,\r(z)\: {\bar\s}(q^2_\perp)\Bigg[\left( E \frac{dN^{(0)}}{d^2 (p-q)_\perp \, dE} \right) \bigg( 1
    + \vec{u}_{\bot} (z) \cdot  \vec{\Gamma} (\vec{q}_\bot) \bigg)\notag\\
    & \hspace*{4cm} - \left( E \frac{dN^{(0)}}{d^2 p_\perp \, dE} \right) \bigg( 1 + \vec{u}_{\bot} (z) \cdot  \vec{\Gamma}_{DB} (\vec{q}_\bot) \bigg)\Bigg]+\mathcal{O}\left(\pa_\bot\right) \, .
\end{align}
Here and in what follows $\mathcal{O}\left(\pa_\perp\right)$ indicates that all transverse gradients are neglected. The velocity corrections $\vec{\G}$ and $\vec{\G}_{DB}$ correspond to the sub-eikonal contributions
\begin{subequations}
\begin{align}
    \label{e:GamNew}
    \vec{\Gamma} (\vec{q}_\bot) &=
    - 2 \frac{\vec{p}_\bot - \vec{q}_\bot}{(1-u_{z})E}
    + \frac{\vec{q}_\bot}{(1-u_{z}) E} \:
    \left( \frac{(p-q)_\bot^2 - p_\bot^2}{\bar\sigma(q_\bot^2)} \right) \: 
    \frac{\pa\bar\sigma}{\pa q_\bot^2}
    \notag \\ & \hspace{1cm}
    - \frac{\vec{q}_\bot}{1-u_z} \left( \frac{1}{\bar{N}_0 (E, \vec{p}_\bot - \vec{q}_\bot)} \frac{\partial \bar{N}_0}{\partial E} \right) \, , \\
    \label{e:GamDB}
    \vec{\Gamma}_{DB}(\vec{q}_\perp)&=
    -2\frac{\vec{p}_\bot}{(1-u_{z})E} - \frac{\vec{p}_\bot}{(1-u_{z}) E} \: \frac{q_\bot^2}
    {\bar{\sigma}(q_\bot^2)} \:\frac{\pa\bar{\sigma}}{\pa q_\bot^2} \,.
\end{align}
\end{subequations}
We have also introduced a shorthand notation $\bar{\sigma}(\vec{q}_\bot) \equiv \frac{d\sigma}{d^2q_\bot} = \frac{1}{(2\pi)^2} \mathcal{C} |v(q_\bot^2)|^2$, where $\mathcal{C}$ is the overall color factor defined in (\ref{mathcalC}). Finally, we denote for brevity $\bar{N}_0(E,\vec{p}_\bot) \equiv E\frac{dN}{d^3 p} = \frac{1}{2(2\pi)^2} |J(p)|^2$.  The important point here is that while we write intermediate results in terms of the current $J(p)$ and the potential $v(q)$, the final corrections can be expressed as derivatives of observable quantities such as the scattering cross section and the distribution of jets.

The unitarity of this expression, guaranteeing that scattering only affects the distribution of jets and not the total number, can be explicitly checked now. In the static limit $\tvec{u} = 0$ considered in the original GLV calculation, the double-Born terms have the same form as the single-Born terms, differing only in that there is no net shift in transverse momentum $\tvec{q}$ of the initial momentum distribution.  Because the double-Born terms are $\tvec{q}$ independent, they couple to the total elastic scattering cross-section $\sigma_0 = \int d^2 q_\bot \bar{\sigma} (\tvec{q})$.  This allowed both single- and double-Born terms to be simply combined through an effective shift of the elastic scattering cross section $\bar\s(\vec{q}_\perp) \rightarrow \bar\s(\vec{q}_\perp) - \sigma_0 \delta^2 (\vec{q}_\bot)$, leaving a broadened jet distribution which is manifestly unitary: $\int d^3 p \frac{dN^{(1)}}{d^3 p} = 0$.  Thus, one sees explicitly that the first-order rescattering in the medium merely broadens the momentum distribution of the produced jets while conserving their total number.

After including the velocity-dependent corrections $\tvec{u} \cdot \vec{\Gamma} (\tvec{q})$ and $\tvec{u} \cdot \vec{\Gamma}_{DB}$, however, the situation is more involved.  While the double-Born diagrams still provide no net shift in the jet transverse momentum, they are no longer $\tvec{q}$-independent because of the $\tvec{q}$-dependent shift in the scattering potential contained in \eqref{e:GamDB}.  Because the unitary structure of the single- and double-Born terms has changed, it is important to verify the unitarity of the velocity corrections explicitly as a test of internal consistency.

To show the unitarity of \eqref{e:untrt1} explicitly, we have to integrate $\left(E \frac{dN^{(1)}}{d^3 p} \right)$ over $\vec{p}$ checking that the number of jets is unmodified and the integral is zero. 
In the single-Born term $\left( E \frac{dN^{(0)}}{d^2 (p-q)_\perp \, dE} \right)$ we shift the momentum $\tvec{p} \rightarrow \tvec{p} + \tvec{q}$ and combine it with the double-Born term $\left( E \frac{dN^{(0)}}{d^2 p_\perp \, dE} \right)$. The velocity-independent term and the first term of the velocity corrections from \eqref{e:GamNew} and \eqref{e:GamDB} cancel exactly, and the remaining terms
\begin{align}
    \int\,d^2p_\perp\,d^2q_\perp\, \bar\sigma(q_\bot^2)\left(\frac{dN^{(0)}}{d^2 p_\perp \, dE} \right)\: \Bigg[\frac{\vec{q}_\perp \left(q^2_\bot+2(\vec{p}_\perp\cdot\vec{q}_\perp)\right)-\vec{p_\perp}q_\bot^2}{\bar{\s}(q_\bot^2)}\frac{d\bar{\s}(q_\bot^2)}{dq^2_\bot}+\vec{q}_\perp\frac{E}{\bar{N}_0} \frac{\partial \bar{N}_0}{\partial E}\Bigg]=0 \, , \notag
\end{align}
vanish after angular averaging over the directions of $\vec{q}_\perp$. This explicitly demonstrates the unitarity of the jet distribution at the first order in opacity and including the velocity effects, in the absence of transverse gradients.

While the maximally general expressions contained in Appendixes~\ref{app:snglBorn} and~\ref{app:dblBorn},  can be directly applied to space-time profiles of arbitrary complexity, we can express the results in closed form only under certain simplifying approximations.  One such approximation is the gradient expansion, similar to what is used traditionally to justify the effective theory of hydrodynamics.  For comparison to such a gradient expansion and to gain some physical insight into the results, we consider here the first nontrivial corrections from medium gradients only in the case when the medium velocity is zero.  Combining \eqref{e:BroadBorn15} and \eqref{e:DblBorn12} in this way yields
\begin{align}
\label{e:untrt2}
    &\left( E \frac{dN^{(1)}}{d^3 p} \right)^\mathrm{(linear)}= 
    \int dz\, \int d^2 q_\bot\,\bar\s(q_\bot^2)\left(\pa^j\r+\r\,\frac{1}{\bar\s(q_\bot^2)}\frac{\pa\bar\s}{\pa\m^2}\,\pa^j\m^2\right)\notag\\
    &\hspace{3.5cm} \times \left\{\left( E \frac{dN^{(0)}}{d^2 (p-q)_\perp\, dE} \right)\left[\frac{(p-q)_\perp^j}{E}z\right]-\left( E \frac{dN^{(0)}}{d^2p_\perp\, dE} \right)\left[\frac{p_\perp^j}{E}z\right]\right\} \, .
\end{align}
The unitarity of this expression readily follows from its form since the two terms in brackets are equal under a shift of the $p_\perp$ integration. It is clear that in the case of uniform density and temperature  the whole gradient correction term vanishes. While the variation in the medium parameters along the direction of propagation is captured by the longitudinal $z-$integral exactly, without the need for a gradient expansion, the linear corrections shown above take into account the variations of the density in the transverse directions in the vicinity of the jet.

\subsection{Moments of the Jet Distribution}

\begin{figure}[t]
    \centering
	\includegraphics[width=0.795\textwidth]{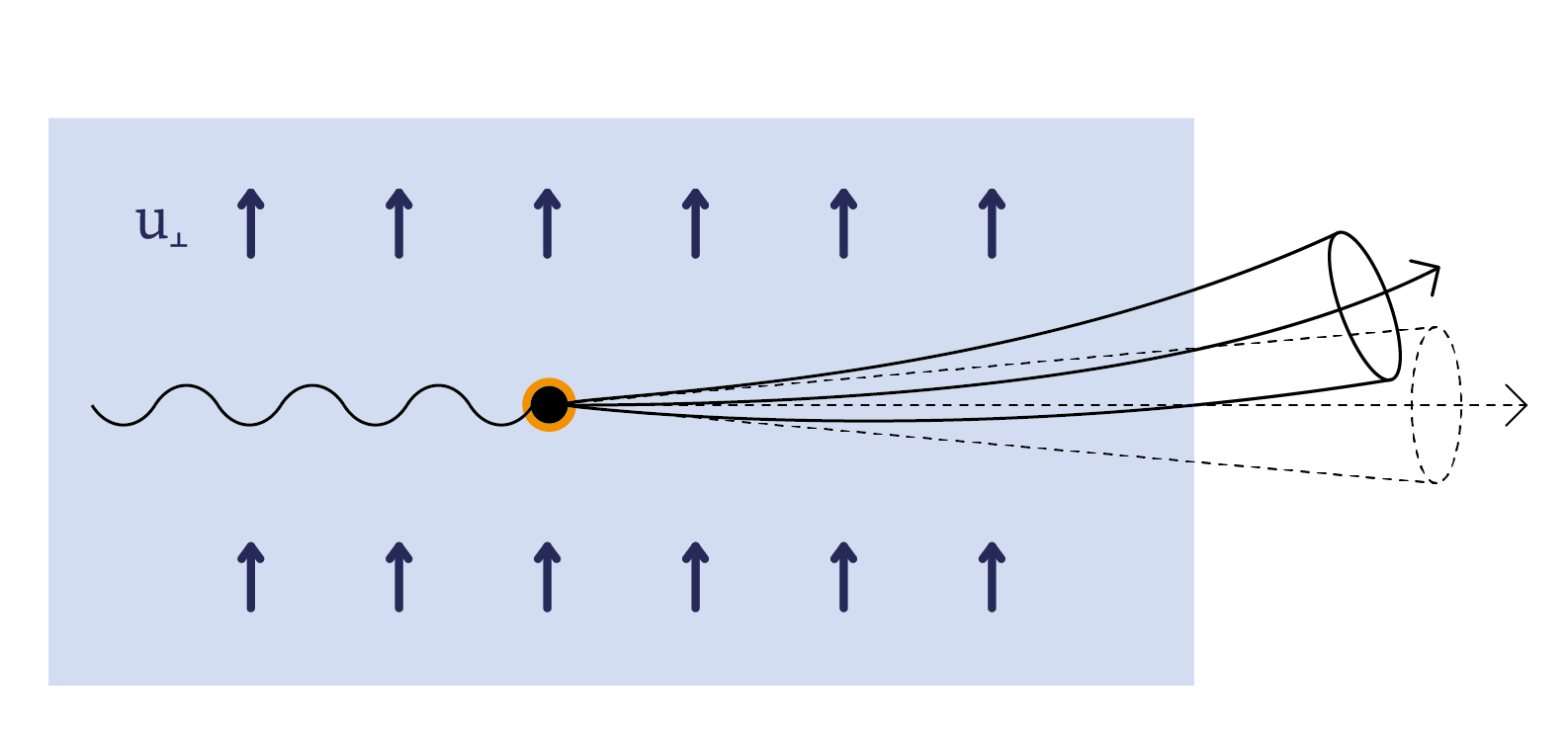}
	\caption{The coupling of jets  to the medium velocity presented in this work.  The transverse momentum of the jet is deflected in the direction of the medium velocity. } 
	\label{f:JETSbroad}
\end{figure}

In the case when the medium is not flowing~\cite{Gyulassy:2002yv,Qiu:2003pm} it is easy to see that there is no preferred direction and the transverse-momentum broadening is isotropic: $\langle \vec{p}_\perp \rangle = 0$.  One of the main results of our work is the explicit breaking of this isotropy due to the directional coupling to the medium gradients and medium velocity, as illustrated in Fig.~\ref{f:JETSbroad}
and by the following example. Suppose the initial transverse-momentum profile of jets produced by the hard scattering is highly collimated, with an energy dependence we leave arbitrary:
\begin{align}   \label{e:deltfn}
    E \frac{dN^{(0)}}{d^3 p} = \frac{1}{2(2\pi)^3} | J(p) |^2 = f(E) \, \delta^{(2)} (\tvec{p}) \, .
\end{align}
Then we can compute averages of various functions of the transverse momentum $\tvec{p}$ weighted by the broadened jet distribution \eqref{e:untrt1}
\begin{align}
    \langle \cdots \rangle \equiv 
    \frac{
        \int d^2 p_\bot \, (\cdots) \, E \frac{dN^{(1)}}{d^2 p_\bot \, dE}
    }{
        \int d^2 p_\bot \, E \frac{dN^{(0)}}{d^2 p_\bot \, dE}
    }
    = \frac{1}{f(E)} \:
    \int d^2 p_\bot \, (\cdots) \, E \frac{dN^{(1)}}{d^2 p_\bot \, dE} \,,
\end{align}
where $(\cdots)$ denotes the function of $\tvec{p}$ being averaged, which we assume for simplicity is zero at $\tvec{p}=0$.  This allows us to evaluate the leading contribution to the numerator at first order in opacity, while the denominator is fully given by the $0^\mathrm{th}$ order in opacity due to unitarity.

Unlike the isotropic broadening which occurs for a static medium, for any odd moment $\langle \tvec{p} F(p_\bot^2) \rangle$, where $F$ is an arbitrary function of $p_\bot^2$, we have
\begin{align}   \label{e:moments1}
    \langle \vec{p}_\perp F\left(p_\perp^2\right) \rangle&\simeq
    \int\,dz\,\rho(z)\int\,d^2 p_\bot\,d^2q_\bot\:\vec{p}_\perp F\left(p_\perp^2\right)\, \bar{\s}(q_\perp^2)\notag\\
    &\hspace*{-1cm}\times \Bigg[ \delta^{(2)}(\vec{p}_\perp-\vec{q}_\perp)\bigg( 1 + \vec{u}_{\bot} (z) \cdot  \vec{\Gamma} (\vec{q}_\bot) \bigg)  
    \vspace{1cm}- \delta^{(2)}(\vec{p}_\perp) \bigg( 1 + \vec{u}_{\bot} (z) \cdot  \vec{\Gamma}_{DB} (\vec{q}_\bot) \bigg)\Bigg] \; .
\end{align}
By construction, the double-Born diagrams do not broaden the jet profile, which comes only from the single-Born contributions in the first term.  Using the explicit form of $\vec{\Gamma}$ from \eqref{e:GamNew} with $\tvec{p} = \tvec{q}$, we have
\begin{align}
\label{e:untrt3}
    \hspace{-0.5cm}\langle \vec{p}_\perp F\left(p_\perp^2\right) \rangle&\simeq
    \int\,dz\,\rho(z)\int\,d^2q_\bot\:\vec{q}_\perp F\left(q_\perp^2\right)\, \bar{\s}(q_\perp^2)\bigg( 1 + \vec{u}_{\bot} (z) \cdot  \vec{\Gamma}\big|_{\vec{p}_\perp=\vec{q}_\perp} \bigg)
    \notag\\
    &= - \frac{1}{2E} \int\,d^2 q_\bot\:  q_\bot^4 \, F(q_\bot^2) \frac{\pa}{\pa q_\bot^2}
    \,
    \left[ 
    \int dz\,  \frac{\tvec{u}(z)}{(1-u_z (z))} \, \bar{\s} (z, q_\bot^2) \, \rho(z)
    \right]
    \notag \\ &\hspace{1cm}
    - \frac{1}{2 f(E)} \frac{\pa f}{\pa E} \int\,d^2 q_\bot\: q_\bot^2 \, F(q_\bot^2) 
    \,
    \left[ 
    \int dz \, \frac{\tvec{u} (z) }{(1-u_z (z))}
    \, \bar{\s}(z, q_\perp^2) \, \rho(z) \right]
    \notag \\ &=
    - \half \int\,d^2 q_\bot\: q_\bot^2 \, F(q_\bot^2) 
    \left[
    \langle \tvec{u} (q_\bot^2) \rangle \, 
    \frac{1}{f(E)} \frac{\pa f}{\pa E} +
    \frac{1}{E} \: 
    q_\bot^2 \frac{\pa \langle \tvec{u} (q_\bot^2) \rangle }{\pa q_\bot^2}
    \right] \, , 
\end{align}
where the preferred direction which breaks the isotropy of the broadened jet profile comes from the weighted average of the velocity distribution
\begin{align}
    \langle \tvec{u} (q_\bot^2) \rangle = 
    \int dz \, \frac{\tvec{u} (z) }{(1-u_z (z))}
    \, \bar{\s}(z, q_\perp^2) \, \rho(z) \, ,
\end{align}
which is an interesting combination of the density, the differential cross section as a function of $q_\bot^2$, and both the longitudinal and transverse components of the velocity.

To illustrate the features of the broadening anisotropy more clearly, let us assume that the medium properties are $z$-independent. Then we have
\begin{align}
    \langle \vec{p}_\perp F\left(p_\perp^2\right) \rangle &=
    - \frac{\tvec{u}}{(1-u_z)} \frac{L}{2 \lambda}  
    \left(
    \frac{1}{E} \int\,d^2 q_\bot\:  q_\bot^4 \, F(q_\bot^2) \, 
    \frac{1}{\sigma_0} \frac{\pa \bar{\s}}{\pa q_\bot^2}
    +
    \frac{1}{f(E)} \frac{\pa f}{\pa E} \int\,d^2 q_\bot\: q_\bot^2 \, F(q_\bot^2) 
    \frac{\bar{\s} (q_\bot^2)}{\sigma_0} 
    \right) \, .
\end{align}
If we choose a specific model for the elastic scattering cross section $\bar{\s}$, then we can evaluate the integrals directly.  For the Gyulassy-Wang potential \eqref{e:potl5},
\begin{align}   \label{e:GWrat}
    \frac{\bar{\s}(q_\bot^2)}{\sigma_0} = \frac{1}{\pi} \frac{\mu^2}{(q_\bot^2 + \mu^2)^2} \, ,
\end{align}
we have
\begin{align}
    \langle \vec{p}_\perp F\left(p_\perp^2\right) \rangle &=
    - \frac{\tvec{u}}{(1-u_z)} \frac{L}{2 \lambda}  \mu^2
    \bigg(
    -2 \frac{1}{E}
    \int_0^\infty d\xi \, F(\xi) \frac{\xi^2}{(1+\xi)^3}
    +
    \frac{1}{f(E)} \frac{\pa f}{\pa E} 
    \int_0^\infty \,d\xi \, F(\xi) \frac{\xi}{(1+\xi)^2}
    \bigg) \, , 
\end{align}
where we introduced $\xi = q_\bot^2 / \mu^2$.  If $F(q_\bot^2) = (q_\bot^2)^k = (\mu^2)^k \, \xi^k$, then
\begin{align}   \label{e:moments2}
    \langle \vec{p}_\perp (p_\perp^2)^k \rangle &=
    - \frac{\tvec{u}}{(1-u_z)} \frac{L}{2 \lambda}  (\mu^2)^{k+1}
    \bigg(
    -\frac{2}{E}
    \int_0^\infty d\xi \frac{\xi^{k+2}}{(1+\xi)^3}
    +
    \frac{1}{f(E)}\frac{\pa f}{\pa E} 
    \int_0^\infty \,d\xi \frac{\xi^{k+1}}{(1+\xi)^2}
    \bigg) \, .
\end{align}
For $-2 < k < 0$ both integrals are convergent, becoming logarithmically divergent in the UV as $k \rightarrow 0$ and in the IR as $k \rightarrow -3$ for the first integral and as $k \rightarrow -2$ for the second. Taking for illustration purposes the cases $k=-1$ and $k=0$, and assuming that the hard scattering follows the perturbative tail $f(E) \propto E^{-4}$ at large $E$, we have
\begin{subequations}
\begin{align}
    \left\langle \frac{\vec{p}_\perp}{p_\bot^2} \right\rangle &=
    - \frac{\tvec{u}}{(1-u_z)} \frac{L}{2 \lambda}
    \bigg( \frac{1}{f(E)} \frac{\pa f}{\pa E} - \frac{1}{E} \bigg)
    = 
    + \frac{5}{2} \frac{\tvec{u}}{(1-u_z)} \frac{L}{\lambda}
    \: \frac{1}{E} \; , 
    \\
    \frac{\left\langle \vec{p}_\perp \right\rangle}{\mu^2} &=
    - \frac{\tvec{u}}{(1-u_z)} \frac{L}{2 \lambda} 
    \bigg( \frac{1}{f(E)} \frac{\pa f}{\pa E}  - 2 \frac{1}{E} \bigg) \ln\frac{E}{\mu}
    =
    + 3 \frac{\tvec{u}}{(1-u_z)} \frac{L}{\lambda} 
    \: \frac{1}{E} \ln\frac{E}{\mu}  \, ,
\end{align}
\end{subequations}
where for $k=0$ we regulate the UV divergence with the cutoff $q_{\bot , max} \sim \sqrt{E \mu}$ corresponding to $\xi_{max} \sim \frac{E}{\mu}$.  Because $\ord{1}$ uncertainty in this logarithmic cutoff leads to $\ord{1}$ corrections to the finite terms, we keep only the terms proportional to $\ln\frac{E}{\mu}$ in the leading-log approximation.  By studying moments of the form \eqref{e:moments2} as a function of the continuous parameter $-2 \leq k \leq 0$, we can extract information about the shape of the scattering potential from the UV to the IR regimes.

While the velocity corrections lead to the generation of nonzero anisotropic momentum broadening \eqref{e:untrt3}, at this order they do not affect the width of the broadened distribution, as evidenced by the vanishing of the corrections to the even moments:
\begin{align}
\label{e:untrt4}
    &\langle F(p_\perp^{2}) \rangle-\langle F(p_\perp^{2}) \rangle|_{\vec{u}_\perp=0}\simeq 
    \int\,dz\,\rho(z)\int\,d^2 p_\bot\,d^2q_\bot\: F\left(p_\perp^2\right)\, \bar{\s}(q_\perp^2)\notag\\
    &\hspace{1.5cm}\times\Bigg[ \delta^{(2)}(\vec{p}_\perp-\vec{q}_\perp)\bigg(\vec{u}_{\bot} (z) \cdot  \vec{\Gamma} (\vec{q}_\bot) \bigg)- \delta^{(2)}(\vec{p}_\perp) \bigg(\vec{u}_{\bot} (z) \cdot  \vec{\Gamma}_{DB} (\vec{q}_\bot) \bigg)\Bigg]=0\,,
\end{align}
with the zero following directly from the angular integration over $d^2 p_\bot$ and $d^2 q_\bot$.  While at this order the correction to the even moments vanishes, it is clear that such a correction may be possible at $\ord{\frac{\bot^2}{E^2}}$.

Finally, let us consider the leading gradient corrections at zero velocity to the broadening anisotropy (\ref{e:moments1}) and symmetric measures of the width (\ref{e:untrt4}). If we take the initial distribution to be a highly-collimated delta function as in \eqref{e:deltfn}, then the moments arising from gradients are trivially zero.  Instead we take the initial distribution to be a Gaussian of finite width $w$:
\begin{align}   \label{e:Gaussprofile}
    E \frac{dN^{(0)}}{d^3 p} = \frac{1}{2(2\pi)^3} | J(p) |^2 = \frac{f(E)}{2\pi w^2} \, e^{-\frac{p_\perp^2}{2w^2}} \, .
\end{align}
Using the expression \eqref{e:untrt2} for the corrections linear in the gradients at zero velocity (that is, corrections of $\ord{\partial_\bot}$ but at eikonal accuracy $\ord{\left(\frac{\bot}{E}\right)^0}$ in the jet energy), we have
\begin{align}
    \langle \vec{p}_\perp\,F(p_\perp^2)\rangle^{(linear)}&\simeq 
    \int\,dz\int\,d^2 p_\bot\,d^2q_\bot\: \vec{p}_\perp F\left(p_\perp^2\right)
    \left( \pa^j\r \, \bar{\s} (q_\perp^2) +
    \r \, \frac{\pa \bar{\s}}{\pa\m^2}\,\pa^j\m^2\right)\notag\\
    &\hspace{1cm}\times
    \left\{
    \frac{e^{-\frac{(p-q)_\perp^2}{2w^2}}}{2\p w^2}  
    \left[\frac{(p-q)_\perp^j}{E}z\right] - \frac{e^{-\frac{p_\perp^2}{2w^2}}}{2\p w^2} 
    \left[\frac{p_\perp^j}{E}z\right]\right\} \, ,
    \label{gradmoments1}
\end{align}
where we have set the velocity corrections $\vec{\Gamma} = 0$ to eikonal accuracy.  Quantities such as $\rho(z)$ are evaluated at the initial transverse position $\tvec{x} = 0$, with the linear corrections from the transverse position dependence written explicitly.

Under the assumption that the $d^2 p_\bot$ integral is convergent, we shift the momentum $\tvec{p} \rightarrow \tvec{p} + \tvec{q}$ in the first term in braces (the single-Born term), giving
\begin{align}   \label{e:gradmom1}
    \langle \vec{p}_\perp\,F(p_\perp^2)\rangle^{(linear)}&\simeq 
    \int\,dz\int\,d^2 p_\bot\,d^2q_\bot\: 
    \frac{e^{-\frac{p_\perp^2}{2w^2}}}{2\p w^2}  
    \left[\frac{p_\perp^j}{E}z\right]
    \left( \pa^j\r \, \bar{\s} (q_\perp^2) +
    \r \, \frac{\pa \bar{\s}}{\pa\m^2}\,\pa^j\m^2\right)\notag\\
    &\hspace{1cm}\times
    \Bigg\{
    (\tvec{p} + \tvec{q}) \, F\left( (p + q)_\perp^2\right)
    - 
    \vec{p}_\perp F\left(p_\perp^2\right)
    \Bigg\} \, .
\end{align}
After integration over $d^2 p_\bot$ and $d^2 q_\bot$, the integral \eqref{e:gradmom1} remains a vector-valued function depending only on the gradients $\tvec{\nabla}\rho$ and $\tvec{\nabla}\mu^2$.  By rotational symmetry, after integration the result must be proportional to $\tvec{\nabla}\rho$ or $\tvec{\nabla}\mu^2$ in the corresponding term.  Schematically
\begin{align}
    \tvec{\mathcal{I}} (\tvec{\nabla}h) = \tvec{\nabla}h \: \mathcal{I}\left((\nabla h)^2\right)
    \qquad \longleftrightarrow \qquad
    \mathcal{I}\left((\nabla h)^2\right) = \frac{1}{(\nabla h)^2} \tvec{\nabla}h \cdot \tvec{\mathcal{I}} (\tvec{\nabla}h) 
\end{align}
for some function $h$.  Since the scalar part $\mathcal{I}$ of the integral is independent of the direction of $\tvec{\nabla} h $, it is invariant under averaging over the directions of the gradient.  Following this logic, we pull out the overall direction $\tvec{\nabla}\rho \, \bar{\s}(q_\bot^2) + \rho \frac{\pa \bar{\s}}{\pa \mu^2} \, \tvec{\nabla}\mu^2$ which survives after integration and angular-average the remaining scalar:
\begin{align}   \label{e:gradmom2}
    \langle \vec{p}_\perp\,F(p_\perp^2)\rangle^{(linear)}&\simeq 
    \int\,dz\int\,d^2 p_\bot\,d^2q_\bot\: 
    \frac{e^{-\frac{p_\perp^2}{2w^2}}}{2\p w^2} \:\: \half
    \left( 
    \tvec{\nabla}\rho \, \bar{\s}(q_\bot^2) + \rho \frac{\pa \bar{\s}}{\pa \mu^2} \, \tvec{\nabla}\mu^2
    \right) \frac{z}{E}
    \notag\\
    &\hspace{1cm}\times
    \Bigg\{
    (p_\bot^2 + \tvec{p} \cdot \tvec{q}) \, 
    F\left( (p + q)_\perp^2\right)
    - p_\bot^2 F\left(p_\perp^2\right)
    \Bigg\} \, .
\end{align}
Eq.~\eqref{e:gradmom2} shows that the gradients of the density and of the temperature (and hence of the Debye mass $\mu^2$) produce a preferred direction for the momentum broadening, breaking the isotropy and leading to the possibility of nonzero vector moments.

To further illustrate the dependence of the result, let us consider the case where the medium is homogeneous in the direction of the jet, so that the $z$ dependence becomes trivial.  Then we immediately perform the integral $\int_0^L dz \, z = \half L^2$ to obtain
\begin{align}   
    \langle \vec{p}_\perp\,F(p_\perp^2)\rangle^{(linear)}&\simeq 
    \frac{L}{\lambda} \, \frac{L}{4E} \,
    \int\,d^2 p_\bot\,d^2q_\bot\: 
    \frac{e^{-\frac{p_\perp^2}{2w^2}}}{2\p w^2} \:
    \left( 
    \frac{\tvec{\nabla}\rho}{\rho} \, \frac{\bar{\s}(q_\bot^2)}{\sigma_0} 
    + \frac{1}{\sigma_0} \frac{\pa \bar{\s}}{\pa \mu^2} \, \tvec{\nabla}\mu^2
    \right) 
    \notag\\
    &\hspace{1cm}\times
    \Bigg\{
    (p_\bot^2 + \tvec{p} \cdot \tvec{q}) \, 
    F\left( (p + q)_\perp^2\right)
    - p_\bot^2 F\left(p_\perp^2\right)
    \Bigg\} \, .
\end{align}
Further specifying to the Gyulassy-Wang potential and using the ratio \eqref{e:GWrat} we obtain
\begin{align}   
    \langle \vec{p}_\perp \, (p_\perp^2)^k \rangle^{(linear)}&\simeq 
    \frac{L}{\lambda} \, \frac{L}{4E} \,
    \int\,d^2 p_\bot\,d^2q_\bot\: 
    \frac{e^{-\frac{p_\perp^2}{2w^2}}}{2\p w^2} \:
    \left( 
    \frac{\tvec{\nabla}\rho}{\rho} \, 
    \frac{1}{\pi} \frac{\mu^2}{(q_\bot^2 + \mu^2)^2}
    - \tvec{\nabla}\mu^2
    \frac{2}{\pi} \frac{\mu^2}{(q_\bot^2 + \mu^2)^3}
    \right) 
    \notag\\
    &\hspace{1cm}\times
    \Bigg\{
    (p_\bot^2 + \tvec{p} \cdot \tvec{q}) \, 
    \left( (p + q)_\perp^2 \right)^k
    - (p_\bot^2)^{k+1}
    \Bigg\} \, ,
\end{align}
where we have taken $F(p_\bot^2) = (p_\bot^2)^k$ for some power $k$.  The angular integrals over the directions of $\tvec{p}$ and $\tvec{q}$ are significantly more complicated than in the velocity case, but they can be done analytically in terms of hypergeometric functions.  For $-1 < k < 1$ the integrals over the magnitudes $d p_\bot^2$ and $d q_\bot^2$ are absolutely convergent, with terms logarithmic in either the UV or IR beginning to appear at the endpoints $k = \pm 1$.  As with the velocity-dependent case, dialing the weight parameter $k$ used in the moments allows one to select on the UV properties of the scattering potential, the IR properties, or anything in between.

Now let us consider a few illustrative values of $k$.  For $k=0$ we have just the average transverse momentum $\langle \tvec{p} \rangle$, but the angular average vanishes
\begin{align*}
    \int \frac{d \phi}{2\pi} \, \Bigg\{ p_\bot^2 - \tvec{p} \cdot \tvec{q} - p_\bot^2 \Bigg\} = 0\,.
\end{align*}
Interestingly, we find that unlike the medium velocity, gradients do not produce a net $\langle \tvec{p} \rangle$ at this order.  For the case $k=1$, we have the cubic moment $\langle \vec{p}_\perp \, p_\perp^2 \rangle$, and the angular integral,
\begin{align*}
    \int \frac{d \phi}{2\pi} \, \Bigg\{ (p_\bot^2 + \tvec{p} \cdot \tvec{q})(p+q)_\bot^2 - p_\bot^4 \Bigg\} = 2 p_\bot^2 q_\bot^2 \, , 
\end{align*}
does not vanish, giving
\begin{align}   
    \langle \vec{p}_\perp \, p_\perp^2 \rangle^{(linear)}&\simeq 
    \frac{L}{\lambda} \, \frac{L}{4E} \,
    (2\pi)
    \int \, d p_\bot^2 \, d q_\bot^2 \: 
    p_\bot^2 \, \frac{e^{-\frac{p_\perp^2}{2w^2}}}{2\p w^2} \:
    \left( 
    \frac{\tvec{\nabla}\rho}{\rho} \, 
    \frac{\mu^2 q_\bot^2}{(q_\bot^2 + \mu^2)^2}
    - 2 \: \tvec{\nabla}\mu^2
    \frac{\mu^2 q_\bot^2}{(q_\bot^2 + \mu^2)^3}
    \right)  \, .
\end{align}
Evaluating the Gaussian integral, introducing the dimensionless variable $\xi \equiv q_\bot^2 / \mu^2$, and regulating the UV behavior with a cutoff $\xi < E/\mu$, we have
\begin{align}   \label{e:gradmom3}
    \langle \vec{p}_\perp \, p_\perp^2 \rangle^{(linear)}&\simeq 
    \frac{L}{\lambda} \, \frac{L}{4E} \,
    (4 w^2 \mu^2)
    \left( 
    \frac{\tvec{\nabla}\rho}{\rho} \, 
    \int\limits_0^\frac{E}{\mu} d\xi
    \frac{\xi}{(1+\xi)^2}
    - 2 \: \frac{\tvec{\nabla}\mu^2}{\mu^2}
    \int\limits_0^\frac{E}{\mu} d\xi
    \frac{\xi}{(1 + \xi)^3}
    \right) \, .
\end{align}
Observing that only the integral multiplying $\tvec{\nabla}\rho$ is logarithmic in the UV, we work to leading-logarithmic accuracy and obtain
\begin{align}   
    \langle \vec{p}_\perp \, p_\perp^2 \rangle^{(linear)}&\simeq 
    \frac{L}{\lambda} \, \frac{L}{E} \,
    w^2 \mu^2
    \frac{\tvec{\nabla}\rho}{\rho} \, 
    \ln\frac{E}{\mu} \, .
\end{align}
The dependence on the $\tvec{\nabla}\rho$ gradient only is a consequence of the cubic moment selecting on the UV behavior of the scattering cross section.

Similarly, we find that the gradient corrections to the average of an even power of the jet transverse momentum is zero -- $\langle p_\perp^{2k} \rangle^{(linear)}=0$. Indeed, the double angular averaging in \eqref{gradmoments1} cancels terms involving odd powers of $\vec{p}_\perp$ or $\vec{q}_\perp$ but for even moments of the jet momentum all terms are odd at least in one of the two momenta after the shift $(\vec{p_\perp}-\vec{q_\perp})\to\vec{p_\perp}$ in the first term under the integral.  Thus,  we conclude that transverse gradients (at zero velocity) generate transverse momentum anisotropies without modifying the width of the transverse momentum distribution, just as in the case of the transverse velocity.

We end this section with a few physics comments. The equations that give non-zero values for functions of moments of $\vec{p}_\perp$ are best applied for low moments where the integrals are convergent. While the integrals with higher $p_\bot$ weights can be performed because the maximum momentum transfer between the jet and the medium is limited to $q_\bot \lesssim \sqrt{E\mu}$, high moments will exhibit power law sensitivity to the jet energy $E$.  It is instructive to note that neither the transverse velocity $\tvec{u}$ nor the transverse gradients $\tvec{\nabla}\rho$ or $\tvec{\nabla}\mu^2$ contribute to the root-mean square (RMS) jet broadening at the considered accuracy.\footnote{We note that if \textit{two} directions are present simultaneously, e.g. velocity and one of the gradients, the even moments could be generated from the interference of both effects.} Instead, the odd vector moments get a contribution proportional to the velocity $\bf u_\perp$ or the gradients $\tvec{\nabla}\rho , \tvec{\nabla}\mu^2$, deflecting the jet in the corresponding direction and producing an anisotropy of the jet transverse momentum (see Fig.~\ref{f:JETSbroad}). This deflection depends on how fast the medium flows or how large the gradients are and grows with the average number of scatterings, but decreases with the jet energy.  For a medium in thermal equilibrium, both types of gradients $\tvec{\nabla}\rho$ and $\tvec{\nabla}\mu^2$ are proportional to gradients of the temperature, $\tvec{\nabla}T$.  This is the case, for instance, in the model used in Ref.~\cite{Sievert:2019cwq} where $\rho \propto T^3$ and $\mu \propto T$.  On the other hand, far from equilibrium, $\tvec{\nabla}\rho$ and $\tvec{\nabla}\mu^2$ can encode independent information about the medium, as exemplified in the case of cold nuclear matter at the EIC.

\section{Medium-Induced Branching Results}
\label{sec:Rad}

Next we wish to investigate the role of velocity corrections to the pattern of medium-induced radiation and subsequent radiative energy loss.  The formalism derived here can also be used to derive the anisotropic corrections to the radiation pattern arising from pure gradients; however, due to the length of the corresponding derivations, we reserve a dedicated analysis of these for future study.

%
\begin{figure}[!t]
    \centering
	\includegraphics[width=0.4\textwidth]{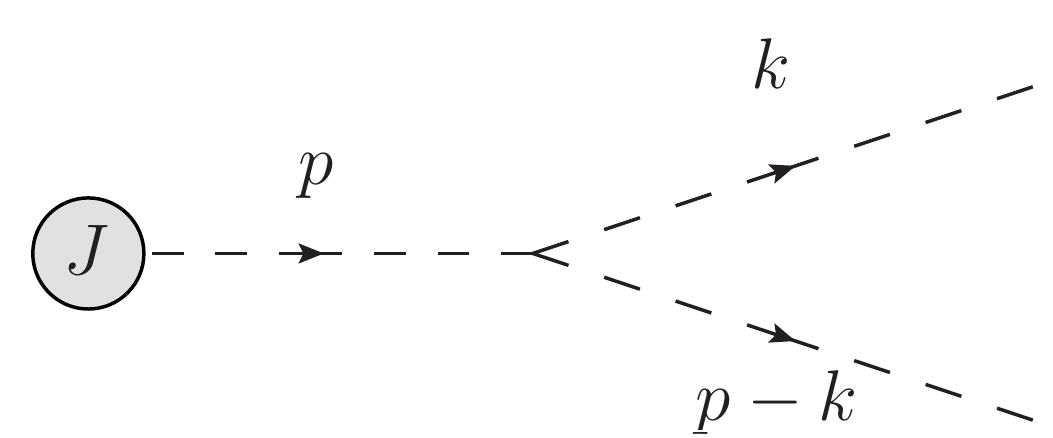}	
	\caption{Elementary splitting vertex for the emission of scalar radiation from a scalar jet.  We do not distinguish diagrammatically the color representations of the scalars, which may correspond to either choice in \eqref{e:Ljet2}.} 
	\label{f:Scalar_Vertex}
\end{figure}
%

With either choice of representations \eqref{e:Ljet2} for the jet partons, we consider the production of a jet of total momentum $p^\mu$, composed of a radiated parton with momentum $k^\mu$ and a counterpart with momentum $(p-k)^\mu$.  Then, the vacuum-like radiation amplitude shown in Fig.~\ref{f:Scalar_Vertex}, corresponding to the $0^\mathrm{th}$ order in opacity, is given by
\begin{align}
    R_0 = \mathcal{C}_0 \frac{-i \lambda}{p^2 + i \epsilon} J(p) \, ,
\end{align}
where $\mathcal{C}_0$ is a color matrix depending on the color representation  
chosen in \eqref{e:Ljet2}.  For the real adjoint scalars in \eqref{e:Ljet2a} we have $\mathcal{C}_0 \rightarrow d^{a b c}$, while for real adjoint + complex fundamental scalars in \eqref{e:Ljet2b} we have $\mathcal{C}_0 \rightarrow (t^a_\mathrm{fund})_{i j}$, with the color indices appropriately identified with the initial and final-state partons.  For the calculation of the velocity corrections to the LPM interference pattern, we will keep the color factor $\mathcal{C}_0$ generic, and we tabulate the appropriate color factors in Table~\ref{e:colortable} at the end of the calculation.

The final-state jet constituents $k^\mu$ and $(p-k)^\mu$ are on shell, while the parent parton $p^\mu$ is off shell, leading to the the kinematics
\begin{subequations} \label{e:kin1}
\begin{align}
    k^\mu &= \left(x E \, , \, \vec{k}_\bot \, , \, x E - \frac{k_\bot^2}{2 x E}\right), \\
    (p-k)^\mu &= \left( (1-x) E \, , \, \vec{p}_\bot - \vec{k}_\bot \, , \, (1-x) E - \frac{(p-k)_\bot^2}{2 (1-x) E}\right), \\
    p^\mu &= \left(E \, , \, \vec{p}_\bot \, , \, E - \frac{k_\bot^2}{2 x E} - \frac{(p-k)_\bot^2}{2 (1-x) E} \right) .
\end{align}
\end{subequations}
Here $x = k^0 / p^0$ is the fractional energy of the radiated parton, and we consider arbitrary $x$, excluding a small region near the endpoints $x \rightarrow 0, 1$ where the eikonal approximation for either $k^\mu$ or $(p-k)^\mu$ fails.

From the kinematics \eqref{e:kin1} we obtain
\begin{align} \label{e:scalvac1}
    R_0 = -i \lambda \mathcal{C}_0 \frac{x (1-x)}{( k - x p)_\bot^2} \, J(p) =
    - \mathcal{C}_0 \psi (x , \vec{k}_\bot - x \vec{p}_\bot ) \, J(p) \, , 
\end{align}
where we have identified the scalar splitting light-front wave function $\psi$, which is manifestly boost invariant.  Squaring Eq.~\eqref{e:scalvac1} and including the final-state phase space factors, we obtain the differential distribution of produced partons within the jet:
\begin{align} \label{e:scalvac2}
    E \frac{dN^{(0)}}{d^2 k \, dx \: d^2 p \, dE} &=
    \frac{1}{[ 2 (2\pi)^3 ]^2} \frac{1}{x (1-x)} 
    \left\langle \left| R_0 \right|^2 \right\rangle
    \notag \\ &=
    \frac{1}{[ 2 (2\pi)^3 ]^2} \frac{
    \langle \mathcal{C}_0 \mathcal{C}_0^\dagger \rangle
    }{x (1-x)} 
    \left| \psi(x, \vec{k}_\bot - x \vec{p}_\bot) \right|^2
    \left| J(p) \right|^2
     \notag \\ &\equiv
    \frac{\mathcal{C}_{(0,0)}}{2 (2\pi)^3\,x (1-x)} 
    \left| \psi(x, \vec{k}_\bot - x \vec{p}_\bot) \right|^2
    \left( E \frac{dN^{(0)}}{d^2 p_\perp \, dE} \right) .
\end{align}
Here, 
\begin{align}
\mathcal{C}_{(0,0)} \equiv
\langle \mathcal{C}_0 \mathcal{C}_0^\dagger \rangle \equiv
\frac{1}{d_\mathrm{proj}} \tr \left( \mathcal{C}_0 \mathcal{C}_0^\dagger \right)
\end{align}
denotes the averaging over the jet ("projectile") color states, with $d_\mathrm{proj}$ the dimension of its color representation.  
Eq.~\eqref{e:scalvac2} expresses the distribution of partons after the branching in terms of the intrinsic distribution $E \frac{dN^{(0)}}{d^3 p}$ of jets produced by the scattering event and the branching probability encoded through the light-front wave function.

Next we will consider the corrections to the vacuum-like radiation pattern described by \eqref{e:scalvac1} due to scattering in the moving medium. The single-Born and double-Born  amplitudes that we have to evaluate are the scalar analogs of the diagrams shown in Fig.~\ref{f:usualdiags}.  Just like for jet broadening, their calculation is technical and can be found in the appendixes. The single-Born diagrams and double-Born diagrams are presented in Appendix~\ref{sec:BRsnglBorn} and Appendix~\ref{sec:BRdblBorn}, respectively, and include the effects of medium 
velocity.

\subsection{Medium-Induced Branching: Final-State Distribution}

Let us combine all the results of the velocity corrections to the in-medium branching process throughout    Appendix~\ref{sec:BRsnglBorn} and Appendix~\ref{sec:BRdblBorn}.  To make the final result more compact let us introduce a number of shorthand notations.  For the wave functions we associate the kinematics of the branching with the single-Born diagrams $A,B,C$ of Figs.~\ref{f:Born_Rad_A} and \ref{f:Born_Rad_BC}: $\psi_A \equiv \psi(x, \vec{k}_\perp-x\vec{p}_\perp)$, $\psi_B \equiv \psi(x, \tvec{k} - x \tvec{p} + x \tvec{q})$, and $\psi_C \equiv \psi(x, \tvec{k} - x \tvec{p} - (1-x) \tvec{q})$.  For the double-Born diagrams, only diagram G of Fig.~\ref{f:DblBorn_G} introduces a wave function with a different argument $\psi_G \equiv \psi(x, \tvec{k} - x \tvec{p} + \tvec{q})$.  For the LPM phases we denote the pole values in the static limit as 
$$ q^-_{p} \equiv \frac{(k - x p)_\bot^2}{2 x (1-x)E(1-u_z)}, \;  q^-_{p-q} \equiv \frac{(k - x p)_\bot^2}{2 x (1-x)E(1-u_z)} - \frac{(p-q)^2_\bot-p_\bot^2}{2E(1-u_z)}, $$ 
$$ q^-_{p-k-q} \equiv -\frac{(p-k-q)^2_\bot - (p-k)_\bot^2}{2(1-x)E(1-u_z)}, \; q^-_{k-q} \equiv -\frac{(k-q)^2_\bot-k_\bot^2}{2xE(1-u_z)}, \; q^-_{k+q} \equiv -\frac{(k+q)^2_\bot-k_\bot^2}{2xE(1-u_z)} \; .$$ When we square the amplitude and average over the colors, the in-medium source averaging produces a color factor $\frac{1}{2 C_{\bar{R}}}$ as in \eqref{e:Avg}, with the averaging in the projectile being denoted by a combined color factor as in $\mathcal{C}_{(A,A)} \equiv  \frac{1}{2 C_{\bar{R}}} \langle \mathcal{C}_A^a \, \mathcal{C}_A^{a \, \dagger} \rangle$, where the single color factors $\mathcal{C}_A^a$ are introduced for the single-Born diagrams in the Appendixes~\ref{sec:BRsnglBorn} and \ref{sec:BRdblBorn}.  The explicit color factors for the Lagrangian \eqref{e:Ljet2a} are tabulated in Table~\ref{e:colortable} and may be compared, for example,  with the explicit color factors from  Eq.~(55) of Ref.~\cite{Sievert:2018imd}.  These color factors are also generally accompanied by the color factor $\mathcal{C}$ from the conversion between $|v(q_\bot^2)|^2$ and the elastic cross section $\bar{\sigma}(q_\bot^2)$.

With this notation we proceed to square the sum of the single-Born amplitudes \eqref{e:R1Afinal}, \eqref{e:RBfinal}, and \eqref{e:RCfinal}.  We convert the discrete sum over medium particles into a continuous integral over densities, and neglect transverse gradients. For interference terms like $\langle R_{1,A} R_{1,B}^* \rangle$, we note that the product of wave functions is real and symmetric: $\psi_A \psi_B^* = \psi_B \psi_A^*$; this is also true of the wave functions of particles other than scalars, after the external quantum numbers have been summed over (see Eq.~(31) of Ref.~\cite{Sievert:2018imd}).

The resulting distribution $E \frac{dN^{(1)}}{d^2 k_\bot \, dx \: d^2 p_\bot \, dE}$, which for the moment we keep in separate terms, is given as follows: 
\begin{subequations} \label{e:ans1}
\begin{align}
    \left( E \frac{dN^{(1)}}{d^2 k_\bot \, dx \: d^2 p_\bot \, dE} \right)_A &=
    \frac{1}{2(2\pi)^3 x (1-x)} \frac{\mathcal{C}_{(A,A)}}{\mathcal{C}} 
    \int_0^L dz \, \rho \,  \int d^2 q_\bot \bar{\s}(q_\bot^2) |\psi_A|^2
    \notag \\ &\hspace{-3cm} \times    \left( E \frac{dN^{(0)}}{d^2 (p-q)_\bot \, dE} \right)
    \bigg[ 1 + 2 \tvec{u} \cdot \vec{\Omega}_A \bigg] \, , 
    \\
    \left( E \frac{dN^{(1)}}{d^2 k_\bot \, dx \: d^2 p_\bot \, dE} \right)_B &=
    \frac{1}{2(2\pi)^3 x (1-x)} \frac{2 \mathcal{C}_{(B,B)}}{\mathcal{C}} 
    \int_0^L dz \, \rho \,  \int d^2 q_\bot \bar{\s}(q_\bot^2) |\psi_B|^2
    \notag \\ &\hspace{-3cm} \times    \left( E \frac{dN^{(0)}}{d^2 (p-q)_\bot \, dE} \right)
    \bigg[ 1 + \tvec{u} \cdot (\vec{\Omega}_{I B} + \vec{\Omega}_{II B}) \bigg]
    \bigg[ 1 - \cos\left( (q_{p-k-q}^- - q_{p-q}^-) z \right) \bigg] \, , 
    \\ 
    \left( E \frac{dN^{(1)}}{d^2 k_\bot \, dx \: d^2 p_\bot \, dE} \right)_C &=
    \frac{1}{2(2\pi)^3 x (1-x)} \frac{2 \mathcal{C}_{(C,C)}}{\mathcal{C}} 
    \int_0^L dz \, \rho \,  \int d^2 q_\bot \bar{\s}(q_\bot^2) |\psi_C|^2\notag \\ &\hspace{-3cm} \times    \left( E \frac{dN^{(0)}}{d^2 (p-q)_\bot \, dE} \right)
    \bigg[ 1 + \tvec{u} \cdot (\vec{\Omega}_{I C} + \vec{\Omega}_{II C}) \bigg]
    \bigg[ 1 - \cos\left( (q_{k-q}^- - q_{p-q}^-) z \right) \bigg] \, , 
    \\ 
    \left( E \frac{dN^{(1)}}{d^2 k_\bot \, dx \: d^2 p_\bot \, dE} \right)_{AB} &=
    \frac{1}{2(2\pi)^3 x (1-x)} \frac{2 \mathcal{C}_{(A,B)}}{\mathcal{C}} 
    \int_0^L dz \, \rho \,  \int d^2 q_\bot \bar{\s}(q_\bot^2) (\psi_A \psi_B^*)
    \notag \\ &\hspace{-3cm} \times   \left( E \frac{dN^{(0)}}{d^2 (p-q)_\bot \, dE} \right)
    \left\{
    \bigg[ 1 + \tvec{u} \cdot (\vec{\Omega}_{A} + \vec{\Omega}_{I B}) \bigg]
    \cos\left( (q_{p-k-q}^- - q_{p-q}^-) z \right) \right.
    \notag \\ &\hspace{-3cm}  \left. -
    \bigg[ 1 + \tvec{u} \cdot (\vec{\Omega}_{A} + \vec{\Omega}_{II B}) \bigg] \, 
    \right\} \, , 
    \\ 
    \left( E \frac{dN^{(1)}}{d^2 k_\bot \, dx \: d^2 p_\bot \, dE} \right)_{AC} &=
    \frac{1}{2(2\pi)^3 x (1-x)} \frac{2 \mathcal{C}_{(A,C)}}{\mathcal{C}} 
    \int_0^L dz \, \rho \,  \int d^2 q_\bot \bar{\s}(q_\bot^2) (\psi_A \psi_C^*)
    \notag \\ &\hspace{-3cm} \times \left( E \frac{dN^{(0)}}{d^2 (p-q)_\bot \, dE} \right)
    \left\{
    \bigg[ 1 + \tvec{u} \cdot (\vec{\Omega}_{A} + \vec{\Omega}_{I C}) \bigg]
    \cos\left( (q_{k-q}^- - q_{p-q}^-) z \right)\right.
        \notag \\ &\hspace{-3cm}  \left. -
    \bigg[ 1 + \tvec{u} \cdot (\vec{\Omega}_{A} + \vec{\Omega}_{II C}) \bigg]
    \right\} \, ,
    \\ 
    \left( E \frac{dN^{(1)}}{d^2 k_\bot \, dx \: d^2 p_\bot \, dE} \right)_{BC} &=
    \frac{1}{2(2\pi)^3 x (1-x)} \frac{2 \mathcal{C}_{(B,C)}}{\mathcal{C}} 
    \int_0^L dz \, \rho \,  \int d^2 q_\bot \bar{\s}(q_\bot^2) (\psi_B \psi_C^*)
    \notag \\ &\hspace{-3cm} \times    \left( E \frac{dN^{(0)}}{d^2 (p-q)_\bot \, dE} \right)
    \bigg\{
    \bigg[ 1 + \tvec{u} \cdot (\vec{\Omega}_{I B} + \vec{\Omega}_{I C}) \bigg]
    \cos\left( (q_{p-k-q}^- - q_{k-q}^-) z \right)
     \notag \\ &\hspace{-3cm}   -
    \bigg[ 1 + \tvec{u} \cdot (\vec{\Omega}_{I B} + \vec{\Omega}_{II C}) \bigg]
    \cos\left( (q_{p-k-q}^- - q_{p-q}^-) z \right)
    \notag \\ &\hspace{-3cm} 
    -
    \bigg[ 1 + \tvec{u} \cdot (\vec{\Omega}_{II B} + \vec{\Omega}_{I C}) \bigg]
    \cos\left( (q_{k-q}^- - q_{p-q}^-) z \right)
    +
    \bigg[ 1 + \tvec{u} \cdot (\vec{\Omega}_{II B} + \vec{\Omega}_{II C}) \bigg]
    \bigg\} \, , 
    \\ 
    \left(E \frac{dN^{(1)}}{d^2 k \, dx \: d^2 p_\perp \, dE} \right)_D  &=
    - \frac{1}{2(2\pi)^3\,x (1-x)}\frac{\mathcal{C}_{(D, 0)}}{\mathcal{C}}
    \int_0^L dz \, \rho \: \int d^2q_\bot \: \bar\sigma(q_\bot^2) |\psi_A|^2
    \notag \\ &\hspace{-2cm} \times    \left( E \frac{dN^{(0)}}{d^2p_\perp \, dE} \right)
    \cos\left( q_p^- z \right)
    \left[ 1 + \vec{u}_\bot \cdot \vec{\Gamma}_{DB} (\vec{q}_\bot)
    \right] \, , 
\\
    \left( E \frac{dN^{(1)}}{d^2 k \, dx \: d^2 p \, dE} \right)_E &=
    - \frac{1}{2 (2\pi)^3\,x(1-x)}\frac{\mathcal{C}_{(E,0)}}{\mathcal{C}}
    \int_0^L dz \, \rho \, \int d^2q_\bot \, \bar\sigma(q_\bot^2) |\psi_A|^2
    \notag \\ &\hspace{-2cm} \times  \left( E \frac{dN^{(0)}}{d^2 p \, dE} \right)
    \left[ 1 - \cos\left(q_p^- z \right) \right]
    \left[ 1 + \vec{u}_\bot \cdot \vec{\Gamma}_{DB}^{(p-k)} \right] \, , 
\\
    \left( E \frac{dN^{(1)}}{d^2 k \, dx \: d^2 p \, dE} \right)_F &=
    - \frac{1}{2 (2\pi)^3\,x(1-x)}\frac{\mathcal{C}_{(F,0)}}{\mathcal{C}}
    \int_0^L dz \, \rho \, \int d^2q_\bot \, \bar\sigma(q_\bot^2) |\psi_A|^2
    \notag \\ &\hspace{-2cm} \times \left( E \frac{dN^{(0)}}{d^2 p \, dE} \right)
    \left[ 1 - \cos\left(q_p^- z \right) \right]
    \left[ 1 + \vec{u}_\bot \cdot \vec{\Gamma}_{DB}^{(k)} \right] \, , 
\\
    \left( E \frac{dN^{(1)}}{d^2k \, dx \: d^2p_\perp \, dE} \right)_G &=
    \frac{1}{2(2\pi)^3\,x(1-x)}\frac{2 \mathcal{C}_{(G,0)}}{\mathcal{C}}
    \int_0^L dz \, \rho \: \int d^2 q_{\bot} \, \bar\sigma(q_{\bot}^2) (\psi_G \psi_A^*)
    \notag \\ &\hspace{-2cm} \times  \left( E \frac{dN^{(0)}}{d^2 p \, dE} \right) 
    \bigg[ 1 + \vec{u}_\bot \cdot \vec{\Gamma}_G \bigg]    
    \bigg[
    \cos\left(q_p^- z \right)
    -
    \cos\left( (q^-_{k+q} + q^-_{p-k-q})z \right)
    \bigg] \, . 
\end{align}
\end{subequations}
Here $\vec{\O}_A$ is given in (\ref{e:OA}), $\vec{\O}_{I B}$ and $\vec{\O}_{II B}$ are given in (\ref{e:OB}), $\vec{\O}_{I C}$ and $\vec{\O}_{II C}$ can be obtained from $\vec{\O}_{I B}$ and $\vec{\O}_{II B}$ under the substitution $k\leftrightarrow (p-k)$ and $x\leftrightarrow 1-x$, $\vec{\Gamma}_{DB}^{(p-k)}$ and $\vec{\Gamma}_{DB}^{(k)}$ defined in (\ref{e:GamDB2}) and (\ref{e:GamDB3}) correspond to $\vec{\Gamma}_{DB}$ under appropriate momentum substitution, and $\vec{\Gamma}_{G}$ is given by (\ref{e:GamDBG}).

Combining all the contributions together, the total distribution at first order in opacity is
\begin{align}   \label{e:fullanswer}
    &E \frac{dN^{(1)}}{d^2 k_\bot \, dx \: d^2 p_\bot \, dE} =
    \frac{1}{2(2\pi)^3 x (1-x)} 
    \int_0^L dz \, \rho \,  \int d^2 q_\bot \bar{\s}(q_\bot^2)
    \Bigg\{
    \left( E \frac{dN^{(0)}}{d^2 (p-q)_\bot \, dE} \right)
    \notag \\ &\hspace{0cm}\times
    \Bigg[
    \frac{\mathcal{C}_{(A,A)}}{\mathcal{C}} 
    |\psi_A|^2
    \bigg( 1 + 2 \tvec{u} \cdot \vec{\Omega}_A \bigg)
     \notag \\ &    +
    \frac{2 \mathcal{C}_{(B,B)}}{\mathcal{C}} 
    |\psi_B|^2
    \bigg( 1 + \tvec{u} \cdot (\vec{\Omega}_{I B} + \vec{\Omega}_{II B}) \bigg)
    \bigg( 1 - \cos\left( (q_{p-k-q}^- - q_{p-q}^-) z \right) \bigg)
    \notag \\ &
    +
    \frac{2 \mathcal{C}_{(C,C)}}{\mathcal{C}} 
    |\psi_C|^2
    \bigg( 1 + \tvec{u} \cdot (\vec{\Omega}_{I C} + \vec{\Omega}_{II C}) \bigg)
    \bigg( 1 - \cos\left( (q_{k-q}^- - q_{p-q}^-) z \right) \bigg)
    \notag \\ &
    +
    \frac{2 \mathcal{C}_{(A,B)}}{\mathcal{C}} 
     (\psi_A \psi_B^*)
    \left[
    \bigg( 1 + \tvec{u} \cdot (\vec{\Omega}_{A} + \vec{\Omega}_{I B}) \bigg)
    \cos\left( (q_{p-k-q}^- - q_{p-q}^-) z \right)
    -
    \bigg( 1 + \tvec{u} \cdot (\vec{\Omega}_{A} + \vec{\Omega}_{II B}) \bigg)
    \right]
    \notag \\ &
    +
    \frac{2 \mathcal{C}_{(A,C)}}{\mathcal{C}} 
    (\psi_A \psi_C^*)
    \left[
    \bigg( 1 + \tvec{u} \cdot (\vec{\Omega}_{A} + \vec{\Omega}_{I C}) \bigg)
    \cos\left( (q_{k-q}^- - q_{p-q}^-) z \right)
    -
    \bigg( 1 + \tvec{u} \cdot (\vec{\Omega}_{A} + \vec{\Omega}_{II C}) \bigg)
    \right]
    \notag \\ &
    +
    \frac{2 \mathcal{C}_{(B,C)}}{\mathcal{C}} 
    (\psi_B \psi_C^*)
    \bigg[
    \bigg( 1 + \tvec{u} \cdot (\vec{\Omega}_{I B} + \vec{\Omega}_{I C}) \bigg)
    \cos\left( (q_{p-k-q}^- - q_{k-q}^-) z \right)
    +
    \bigg( 1 + \tvec{u} \cdot (\vec{\Omega}_{II B} + \vec{\Omega}_{II C}) \bigg)
    \notag \\ & \hspace{3.5cm}-
    \bigg( 1 + \tvec{u} \cdot (\vec{\Omega}_{I B} + \vec{\Omega}_{II C}) \bigg)
    \cos\left( (q_{p-k-q}^- - q_{p-q}^-) z \right)
      \notag \\ & \hspace{3.5cm}-
    \bigg( 1 + \tvec{u} \cdot (\vec{\Omega}_{II B} + \vec{\Omega}_{I C}) \bigg)
    \cos\left( (q_{k-q}^- - q_{p-q}^-) z \right)
    \bigg] \Bigg]
    \notag \\ &+
    \left( E \frac{dN^{(0)}}{d^2 p_\bot \, dE} \right)
    \Bigg[
    - \frac{\mathcal{C}_{(D, 0)}}{\mathcal{C}}
    |\psi_A|^2
    \cos\left( q_p^- z \right)
    \bigg( 1 + \vec{u}_\bot \cdot \vec{\Gamma}_{DB} 
    \bigg)
    \notag \\ &
    - \frac{\mathcal{C}_{(E,0)}}{\mathcal{C}}
    |\psi_A|^2
    \bigg( 1 - \cos\left(q_p^- z \right) \bigg)
    \left( 1 + \vec{u}_\bot \cdot \vec{\Gamma}_{DB}^{(p-k)} \right)
    - \frac{\mathcal{C}_{(F,0)}}{\mathcal{C}}
    |\psi_A|^2
    \left( 1 - \cos\left(q_p^- z \right) \right)
    \left( 1 + \vec{u}_\bot \cdot \vec{\Gamma}_{DB}^{(k)} \right)
    \notag \\ & +
    \frac{2 \mathcal{C}_{(G,0)}}{\mathcal{C}}
    (\psi_G \psi_A^*)
    \bigg( 1 + \vec{u}_\bot \cdot \vec{\Gamma}_G \bigg)    
    \bigg(
    \cos\left(q_p^- z \right)
    -
    \cos\left( (q^-_{k+q} + q^-_{p-k-q})z \right)
    \bigg)
    \Bigg]
    \Bigg\} \, .
\end{align}

While \eqref{e:fullanswer} contains the full answer for the velocity corrections to the medium-induced branching, it is instructive to consider various simplifying limits.  To begin, we note that as discussed in \eq{e:regime} and the paragraph thereafter, in the approximation which allows us to neglect the velocity corrections times sines of the LPM phases, for consistency we can replace the cosine with unity.  While it would be preferable to keep the cosine structures derived in \eqref{e:fullanswer} and augment them with the additional sine terms, we can instead use this approximation to simplify the interference pattern by dropping the velocity corrections multiplying phase structures of the form $(1 - \cos\phi)$ and $(\cos\phi_1 - \cos\phi_2)$:
\begin{align} 
    &E \frac{dN^{(1)}}{d^2 k_\bot \, dx \: d^2 p_\bot \, dE} =
    \frac{1}{2(2\pi)^3 x (1-x)} 
    \int_0^L dz \, \rho \,  \int d^2 q_\bot \bar{\s}(q_\bot^2)
    \Bigg\{
    \left( E \frac{dN^{(0)}}{d^2 (p-q)_\bot \, dE} \right)
    \notag \\ &\times
    \Bigg[
    \frac{\mathcal{C}_{(A,A)}}{\mathcal{C}} 
    |\psi_A|^2
    \bigg( 1 + 2 \tvec{u} \cdot \vec{\Omega}_A \bigg)
    +
    \frac{2 \mathcal{C}_{(B,B)}}{\mathcal{C}} 
    |\psi_B|^2
    \bigg( 1 - \cos\left( (q_{p-k-q}^- - q_{p-q}^-) z \right) \bigg)
    \notag \\ &
    +
    \frac{2 \mathcal{C}_{(C,C)}}{\mathcal{C}} 
    |\psi_C|^2
    \bigg( 1 - \cos\left( (q_{k-q}^- - q_{p-q}^-) z \right) \bigg)
    \notag \\ &
    +
    \frac{2 \mathcal{C}_{(A,B)}}{\mathcal{C}} 
     (\psi_A \psi_B^*)
    \bigg[
    \cos\left( (q_{p-k-q}^- - q_{p-q}^-) z \right) - 1
    + \tvec{u} \cdot (\vec{\Omega}_{I B} - \vec{\Omega}_{II B})    
    \bigg]    
    \notag \\ &
    +
    \frac{2 \mathcal{C}_{(A,C)}}{\mathcal{C}} 
    (\psi_A \psi_C^*)
    \bigg[
    \cos\left( (q_{k-q}^- - q_{p-q}^-) z \right) - 1 
    + \tvec{u} \cdot (\vec{\Omega}_{I C} - \vec{\Omega}_{II C})
    \bigg]
    \notag \\ &
    +
    \frac{2 \mathcal{C}_{(B,C)}}{\mathcal{C}} 
    (\psi_B \psi_C^*)
    \bigg[
    1 +
    \cos\left( (q_{p-k-q}^- - q_{k-q}^-) z \right)
    -
    \cos\left( (q_{p-k-q}^- - q_{p-q}^-) z \right)
    -
    \cos\left( (q_{k-q}^- - q_{p-q}^-) z \right)
    \bigg] \Bigg]
    \notag \\ &+
    \left( E \frac{dN^{(0)}}{d^2 p_\bot \, dE} \right)
    \Bigg[
    - \frac{\mathcal{C}_{(D, 0)}}{\mathcal{C}}
    |\psi_A|^2
    \bigg( 
    \cos\left( q_p^- z \right)
    + \vec{u}_\bot \cdot \vec{\Gamma}_{DB} 
    \bigg)
    - \frac{\mathcal{C}_{(E,0)}}{\mathcal{C}}
    |\psi_A|^2
    \left( 1 - \cos\left(q_p^- z \right) \right)
    \notag \\ &
    - \frac{\mathcal{C}_{(F,0)}}{\mathcal{C}}
    |\psi_A|^2
    \left( 1 - \cos\left(q_p^- z \right) \right)
    +
    \frac{2 \mathcal{C}_{(G,0)}}{\mathcal{C}}
    (\psi_G \psi_A^*)
    \bigg(
    \cos\left(q_p^- z \right)
    -
    \cos\left( (q^-_{k+q} + q^-_{p-k-q})z \right)
    \bigg)
    \Bigg]
    \Bigg\}.
\end{align}
In this limit, many of the velocity corrections drop out, but a few remain.  Of particular note are the combinations
\begin{subequations}
\begin{align}
    \vec{\Omega}_{I B} - \vec{\Omega}_{II B} &= 
    \frac{2 \tvec{q}}{1-u_z} \frac{(k - x p + x q)_\bot^2}{2 x (1-x) E} \frac{1}{v} \frac{\partial v}{\partial q_\bot^2} \, ,
    \\
    \vec{\Omega}_{I C} - \vec{\Omega}_{II C} &= 
    \frac{2 \tvec{q}}{1-u_z} \frac{(k - x p - (1-x) q)_\bot^2}{2 x (1-x) E} \frac{1}{v} \frac{\partial v}{\partial q_\bot^2} \, ,
\end{align}
\end{subequations}
for which all corrections except the virtuality shift of the potential have canceled out.

Alternatively we may consider the small-$x$ limit $x \ll 1$ in which one of the radiated partons is very soft.  In this limit $\psi_A = \psi_B = \psi(x, \tvec{k})$, $\psi_C = \psi(x, \tvec{k} - \tvec{q})$, and $\psi_G = \psi(x, \tvec{k} + \tvec{q})$.  We also have 
\begin{subequations}
\begin{align}
    \vec{\Omega}_{II B} &= \vec{\Omega_A} = 
    - \frac{\tvec{q}}{1-u_z} \frac{k_\bot^2}{x E} \frac{1}{v}\frac{\pa v}{\pa q_\bot^2}\, , 
    \\
    \vec{\Gamma}_G (-\tvec{q}) &= \vec{\Omega}_{I C} =
    - \frac{\tvec{k}}{(1-u_z) x E}+\frac{\tvec{q}}{1-u_z} \frac{(k-q)_\bot^2 - k_\bot^2}{x E} \frac{1}{v} \frac{\partial v}{\partial q_\bot^2}\, , 
    \\
    \vec{\Omega}_{II C} &=
    - \frac{\tvec{k}}{(1-u_z) x E} 
    - \frac{\tvec{q}}{1-u_z} \frac{k_\bot^2}{x E}
    \frac{1}{v} \frac{\partial v}{\partial q_\bot^2}\, , 
\end{align}
\end{subequations}
and $\vec{\Omega}_{I B} = \vec{\Gamma}_{DB}^{(p-k)} = \vec{\Gamma}_{DB}= 0$, and the phases simplify to 
$$ q^-_{p} = q^-_{p-q} = \frac{k_\bot^2}{2 x E(1-u_z)}, \, q^-_{k-q} = - \frac{(k-q)^2_\bot-k_\bot^2}{2 x E(1-u_z)}, \; q^-_{k+q} = -\frac{(k+q)^2_\bot-k_\bot^2}{2xE(1-u_z)}, \; q^-_{p-k-q} = 0 \; .$$
If we impose both the small-$x$ limit and set $\cos\phi = 1$ for the velocity corrections, we obtain
\begin{align}   
    &E \frac{dN^{(1)}}{d^2 k_\bot \, dx \: d^2 p_\bot \, dE} =
    \frac{1}{2(2\pi)^3 x (1-x)} 
    \int_0^L dz \, \rho \,  \int d^2 q_\bot \bar{\s}(q_\bot^2)
    \Bigg\{
    \left( E \frac{dN^{(0)}}{d^2 (p-q)_\bot \, dE} \right)
    \notag \\ &\times
    \Bigg[
    \frac{\mathcal{C}_{(A,A)}}{\mathcal{C}} 
    |\psi_A|^2
    \bigg( 1 + 2 \tvec{u} \cdot \vec{\Omega}_A \bigg)
    +
    \frac{2 \mathcal{C}_{(B,B)}}{\mathcal{C}} 
    |\psi_A|^2
    \bigg( 1 - \cos q_p^- z  \bigg)
    \notag \\ &
    +
    \frac{2 \mathcal{C}_{(C,C)}}{\mathcal{C}} 
    |\psi_C|^2
    \bigg( 1 - \cos\left( (q_{k-q}^- - q_p^-) z \right) \bigg)
    +
    \frac{2 \mathcal{C}_{(A,B)}}{\mathcal{C}} 
     |\psi_A|^2
    \bigg(
    \cos q_p^- z - 1 - \tvec{u} \cdot \vec{\Omega}_{A}     
    \bigg)
    \notag \\ &
    +
    \frac{2 \mathcal{C}_{(A,C)}}{\mathcal{C}} 
    (\psi_A \psi_C^*)
    \bigg(
    \cos\left( (q_{k-q}^- - q_p^-) z \right) - 1
    + \tvec{u} \cdot (\vec{\Omega}_{I C} - \vec{\Omega}_{II C}) 
    \bigg)
    \notag \\ &
    +
    \frac{2 \mathcal{C}_{(B,C)}}{\mathcal{C}} 
    (\psi_A \psi_C^*)
    \bigg(
    1 + \cos q_{k-q}^- z - \cos q_p^- z 
    - \cos\left( (q_{k-q}^- - q_p^-) z \right)
    \bigg) 
    \Bigg]
    \notag \\ &+
    \left( E \frac{dN^{(0)}}{d^2 p_\bot \, dE} \right)
    \Bigg[
    - \frac{\mathcal{C}_{(D, 0)}}{\mathcal{C}}
    |\psi_A|^2
    \cos q_p^- z 
    - \frac{\mathcal{C}_{(E,0)}}{\mathcal{C}}
    |\psi_A|^2
    \bigg( 1 - \cos q_p^- z \bigg)
    \notag \\ &
    - \frac{\mathcal{C}_{(F,0)}}{\mathcal{C}}
    |\psi_A|^2
    \left( 1 - \cos q_p^- z \right)    
    + \frac{2 \mathcal{C}_{(G,0)}}{\mathcal{C}}
    (\psi_A \psi_C^*)
    \bigg(
    \cos q_p^- z 
    -
    \cos q^-_{k-q} z
    \bigg)
    \Bigg]
    \Bigg\} \, .
\end{align}

\vspace{1cm}

\begin{table}[h]
\hspace*{-0.8cm} \begin{tabular}{|c||c|c|c||c|c|c|c|}    
    \multicolumn{3}{l}{Generic color factors:} \\
    \hline
    $\mathcal{C}_0$ &
    $\mathcal{C}_A$ &
    $\mathcal{C}_B$ &
    $\mathcal{C}_C$ &
    $\mathcal{C}_D^{d d'}$ &
    $\mathcal{C}_E^{d d'}$ &
    $\mathcal{C}_F^{d d'}$ &
    $\mathcal{C}_G^{d d'}$ \\
    $\mathcal{C}_{(0,0)}$ &
    $\mathcal{C}_{(A,A)}$ &
    $\mathcal{C}_{(B,B)}$ &
    $\mathcal{C}_{(C,C)}$ &
    $\mathcal{C}_{(D,0)}$ &
    $\mathcal{C}_{(E,0)}$ &
    $\mathcal{C}_{(F,0)}$ &
    $\mathcal{C}_{(G,0)}$ \\
    $\mathcal{C}$ &
    $\mathcal{C}_{(A,B)}$ &
    $\mathcal{C}_{(B,C)}$ &
    $\mathcal{C}_{(A,C)}$ &
    &
    &
    &
    \\ \hline \multicolumn{1}{c}{\,} \\
    \multicolumn{3}{l}{Adjoint scalars \eqref{e:Ljet2a}:} \\
    \hline
    $d^{a b e}$ &
    $d^{a b c} \, i f^{e c d}$ &
    $d^{a c e} \, i f^{c b d}$ &
    $d^{e c b} \, i f^{c a d}$ &
    $d^{a b c} \, i f^{c' c d} \, i f^{e c' d}$ &
    $d^{e a c'} \, i f^{c' c d} \, i f^{c b d}$ &
    $d^{e c' b} \, i f^{c' c d} \, i f^{c a d'}$ &
    $d^{e c c'} \, i f^{c a d'} \, i f^{c' b d}$ \\
    $\frac{N_c^2 - 4}{N_c}$ &
    $\frac{N_c}{2 C_{\bar{R}}} \, \mathcal{C}_{0,0}$ &
    $\frac{N_c}{2 C_{\bar{R}}} \, \mathcal{C}_{0,0}$ &
    $\frac{N_c}{2 C_{\bar{R}}} \, \mathcal{C}_{0,0}$ &
    $\frac{N_c}{2 C_{\bar{R}}} \, \mathcal{C}_{0,0}$ &
    $\frac{N_c}{2 C_{\bar{R}}} \, \mathcal{C}_{0,0}$ &
    $\frac{N_c}{2 C_{\bar{R}}} \, \mathcal{C}_{0,0}$ &
    $\frac{N_c}{4 C_{\bar{R}}} \, \mathcal{C}_{0,0}$ \\
    $\frac{N_c}{2 C_{\bar{R}}}$ &
    $\frac{N_c}{4 C_{\bar{R}}} \, \mathcal{C}_{0,0}$ &
    $\frac{N_c}{4 C_{\bar{R}}} \, \mathcal{C}_{0,0}$ &
    $\frac{N_c}{4 C_{\bar{R}}} \, \mathcal{C}_{0,0}$ &
    &
    &
    &
    \\ \hline \multicolumn{1}{c}{\,} \\
    \multicolumn{6}{l}{
    Fundamental + adjoint scalars \eqref{e:Ljet2b} / 
    QCD $q \rightarrow qg$:
    } \\    
    \hline
    $- t^a$ &
    $- t^a t^d$ &
    $- t^d t^a$ &
    $- i t^c f^{a d c}$ &
    $- t^a t^{d'} t^d$ &
    $- t^{d'} t^d t^a$ &
    $t^c f^{d c c'} f^{a  d' c'}$ &
    $- i t^{d'} t^c f^{a d c}$ \\
    $C_F$ &
    $\frac{C_F}{2 C_{\bar{R}}} \, \mathcal{C}_{0,0}$ &
    $\frac{C_F}{2 C_{\bar{R}}} \, \mathcal{C}_{0,0}$ &
    $\frac{N_c}{2 C_{\bar{R}}} \, \mathcal{C}_{0,0}$ &
    $\frac{C_F}{2 C_{\bar{R}}} \, \mathcal{C}_{0,0}$ &
    $\frac{C_F}{2 C_{\bar{R}}} \, \mathcal{C}_{0,0}$ &
    $\frac{N_c}{2 C_{\bar{R}}} \, \mathcal{C}_{0,0}$ &
    $-\frac{N_c}{4 C_{\bar{R}}} \, \mathcal{C}_{0,0}$ \\
    $\frac{C_F}{2 C_{\bar{R}}}$ &
    $-\frac{1}{4 N_c C_{\bar{R}}} \, \mathcal{C}_{0,0}$ &
    $-\frac{N_c}{4 C_{\bar{R}}} \, \mathcal{C}_{0,0}$ &
    $\frac{N_c}{4 C_{\bar{R}}} \, \mathcal{C}_{0,0}$ &
    &
    &
    &
    \\ \hline
\end{tabular}
\caption{
\label{e:colortable}
Color matrices and factors for the Lagrangian \eqref{e:Ljet2a} with all real adjoint scalars and for the Lagrangian \eqref{e:Ljet2b} with mixed real/adjoint scalars which is identical to the QCD branching channel $q \rightarrow q g$.  For the former case, the final-state scalars $k, p-k$ have adjoint color indices $a, b$, respectively, and the initial-state scalar $p$ has color $e$.  For the latter case, the final-state gluon $k$ has adjoint color index $a$.  In both cases,  color indices $d , d'$ correspond to the insertion of the external potentials $A_\mathrm{ext}^{\mu d}$.}
\end{table}
\vspace{1cm}

\begin{figure}[t]
    \centering
	\includegraphics[width=0.795\textwidth]{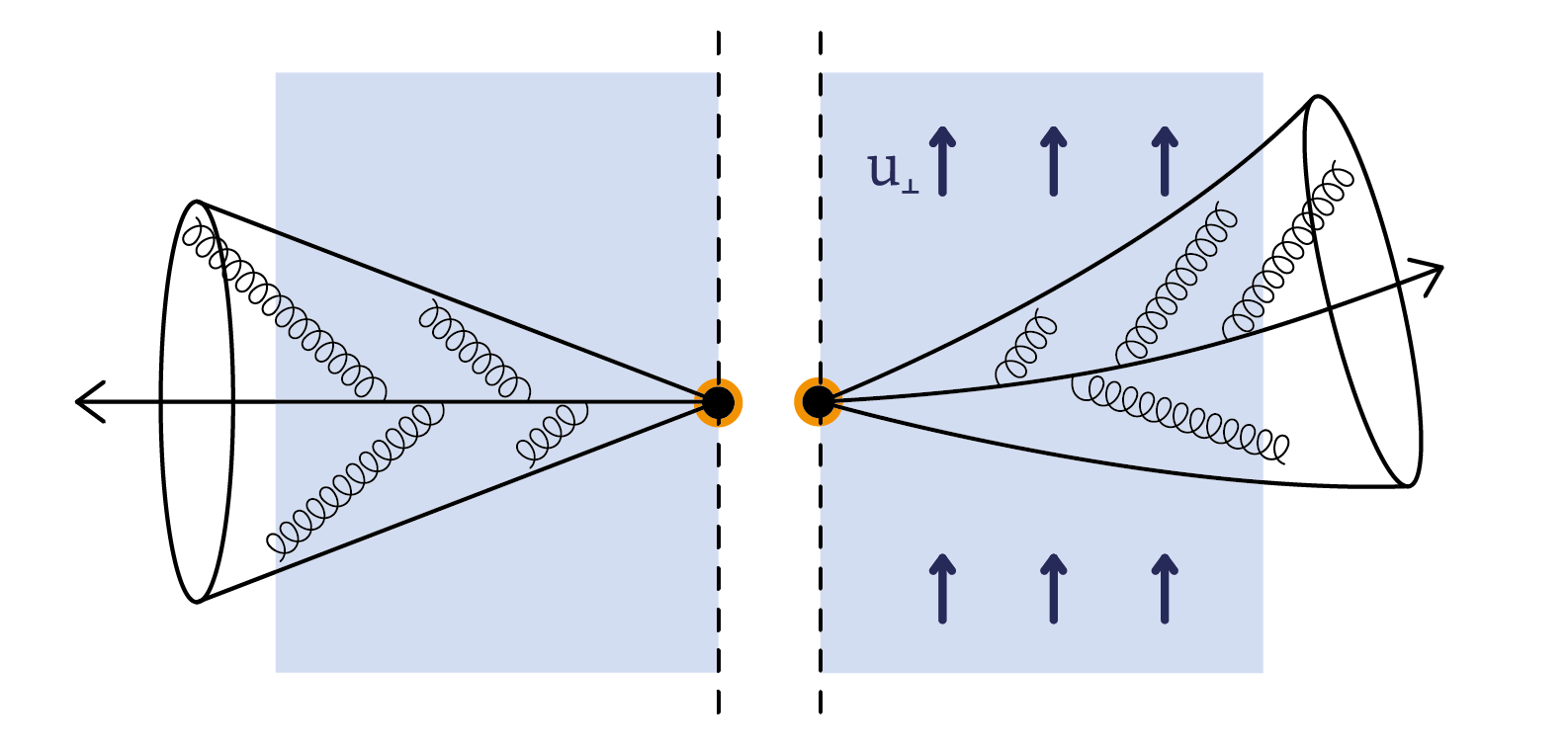}
	\caption{The coupling of  soft radiation to the medium velocity presented in this work.  The medium-induced radiation is emitted preferentially in the direction of the medium velocity as schematically shown by the number of gluons along the direction of the flow in comparison to the ``standard'' symmetric unbent jet.} 
	\label{f:JETSrad}
\end{figure}

Finally, it is instructive to substitute color factors for \eqref{e:Ljet2b} which are motivated by the QCD branching channel $q \rightarrow q g$ as enumerated in Table~\ref{e:colortable} and to employ the so-called ``broad source approximation'' $\frac{{dN}^{(0)}}{d^2 (p-q)_\bot dE} \approx \frac{{dN}^{(0)}}{d^2 p_\bot dE}$ used in the original GLV calculation \cite{Gyulassy:2000er}.  With these substitutions, we have

\begin{align}   \label{e:smallxbroadsrc}
    E \frac{dN^{(1)}}{d^2 k_\bot \, dx \: d^2 p_\bot \, dE} &=
    \frac{C_F}{2(2\pi)^3 x (1-x)} \left(\frac{2 N_c}{C_F}\right)
    \int_0^L dz \, \rho \,  \int d^2 q_\bot \bar{\s}(q_\bot^2) \,  \left( E \frac{dN^{(0)}}{d^2 p_\bot \, dE} \right)
    \notag \\ & \times
    \Bigg\{
    \bigg( |\psi_C|^2 - (\psi_A \psi_C^*) \bigg)    
    \bigg( 1 - \cos\left( (q_{k-q}^- - q_p^-) z \right) \bigg)
    \notag \\ &
    + \half |\psi_A|^2 \left( \tvec{u} \cdot \vec{\Omega}_{A} \right)
    + \half (\psi_A \psi_C^*)
    \bigg( \tvec{u} \cdot (\vec{\Omega}_{I C} - \vec{\Omega}_{II C}) \bigg)
    \Bigg\} \,  .
\end{align}
Since $\vec{\Omega}_A \propto \tvec{q}$ with all other factors independent of the direction of $\tvec{q}$, this term vanishes after angular averaging.  Similarly, in the small-$x$ limit the combination
\begin{align}
    \half (\psi_A \psi_C^*)  \, \tvec{u} \cdot (\vec{\Omega}_{I C} - \vec{\Omega}_{II C}) &= 
    \psi(x, \tvec{k}) \psi^*(x, \tvec{k} - \tvec{q}) \:\:
    \frac{\tvec{u} \cdot \tvec{q}}{1-u_z} \frac{(k - q)_\bot^2}{2 x E} \frac{1}{v} \frac{\partial v}{\partial q_\bot^2}
\end{align}
simplifies.  Interestingly, for the strictly scalar theory, this quantity also averages to zero, since the scalar wave function is isotropic:
\begin{align}
    \half (\psi_A \psi_C^*)  \, \tvec{u} \cdot (\vec{\Omega}_{I C} - \vec{\Omega}_{II C}) &= 
    \left( \frac{\lambda^2 x^2}{k_\bot^2 (k-q)_\bot^2} \right) \:\:
    \frac{\tvec{u} \cdot \tvec{q}}{1-u_z} \frac{(k - q)_\bot^2}{2 x E} \frac{1}{v} \frac{\partial v}{\partial q_\bot^2} .
\end{align}
However, for the $q \rightarrow q g$ splitting wave function, the splitting is not isotropic (see e.g. Eq.~(31) of Ref.~\cite{Sievert:2018imd}) and leads to a nonzero term which survives angular averaging:
\begin{align}
    \half (\psi_A \psi_C^*)  \, \tvec{u} \cdot (\vec{\Omega}_{I C} - \vec{\Omega}_{II C}) &= 
    16 \pi \alpha_s \left( \frac{\tvec{k} \cdot (\tvec{k} - \tvec{q})}{k_\bot^2 (k-q)_\bot^2} \right) \:\:
    \frac{\tvec{u} \cdot \tvec{q}}{1-u_z} \frac{(k - q)_\bot^2}{2 x E} \frac{1}{v} \frac{\partial v}{\partial q_\bot^2}
    \notag \\ &\rightarrow
    - 8 \pi \alpha_s \left( \frac{q_\bot^2}{k_\bot^2} \right) \:\:
    \frac{\tvec{u} \cdot \tvec{k}}{1-u_z} \frac{1}{2 x E} \frac{1}{v} \frac{\partial v}{\partial q_\bot^2} .
\end{align}
As a result, we find that in this simplest case (small $x$ and broad source) the velocity correction to the purely scalar theory is zero, while the velocity correction to real $q \rightarrow q g$ branching leads to anisotropic emission of gluons based on the preferred direction provided by the medium velocity:
\begin{align}   \label{e:QCDfinalanswer}
    &E \frac{dN^{(1)}}{d^2 k_\bot \, dx \: d^2 p_\bot \, dE} =
    \frac{\alpha_s \: N_c}{\pi^2 x} \left( E \frac{dN^{(0)}}{d^2 p_\bot \, dE} \right)
    \int_0^L dz \, \rho \,  \int d^2 q_\bot \bar{\s}(q_\bot^2)
    \notag \\ & \hspace{2cm}\times
    \Bigg\{
    \frac{2 \tvec{k} \cdot \tvec{q}}{k_\bot^2 (k-q)_\bot^2}    
    \bigg( 1 - \cos\left( \frac{(k-q)_\bot^2}{2 x E(1-u_z)} z \right) \bigg)
    + \frac{q_\bot^2}{k_\bot^2 (q_\bot^2 + \mu^2)} 
    \frac{\tvec{u} \cdot \tvec{k}}{2 (1-u_z) x E}
    \Bigg\} \, ,
\end{align}
where we have evaluated the derivative of the potential for the specific choice of the Gyulassy-Wang potential \eqref{e:potl5}.  \eq{e:QCDfinalanswer} explicitly reproduces the original GLV result of Ref.~\cite{Gyulassy:2000er} in the same limit, with a single surviving velocity correction which leads to the preferential emission of the radiated gluon in the direction of the medium velocity as schematically shown in Fig. \ref{f:JETSrad}.  This correction is suppressed by one power of the energy, but enhanced in the small-$x$ limit.

It is instructive to note that $u_z$ dependance of the first term in (\ref{e:QCDfinalanswer}) can be readily understood as a longitudinal boost of the GLV result from the frame, where the matter has zero $u_z$. This longitudinal boost transforms the parameters of the problem in the matter rest frame resulting in $p_z$, $E$, and $L$ of the lab frame. After the $z-$integration is rescaled the first term of (\ref{e:QCDfinalanswer}) is reproduced. However, the transverse velocity correction cannot be obtained by merely boosting the static result since the eikonal expansion and transverse boosts do not commute -- the transverse momentum gains a large shift $\g\tvec{u}E$ under such a boost. Thus, to obtain the full (\ref{e:QCDfinalanswer}) from its static limit one has to start with an expression including the second sub-leading order in the eikonal expansion in the matter rest frame.

\subsection{Moments of the Radiation Spectrum in the Soft Gluon Emission Limit}

It is instructive to consider vector-valued moments of the transverse momentum distribution of radiated gluons as well, just as we did for jet broadening.  Defining the average as
\begin{align}\label{average2}
    \left\langle \cdots \right\rangle \equiv
    \frac{
        \int d^2 k_\bot \, (\cdots) \, E \frac{dN^{(1)}}{d^2 k dx \, d^2 p dE}
        }{
        \int d^2 k_\bot \, E \frac{dN^{(0)}}{d^2 k dx \, d^2 p dE}
        } \, ,
\end{align}
where we have evaluated both the numerator and denominator to the first nonvanishing order, let us consider the vector-valued moments $\left\langle \tvec{k} \, (k_\bot^2)^n \right\rangle$ for some power $n$.  For the numerator, integrating \eqref{e:QCDfinalanswer} with the vector $\tvec{k}$ leads to no contribution from the first term which corresponding to the static, eikonal limit.  For the anisotropic term coupling to the medium velocity, we have
\begin{align}
    \int d^2 k_\bot \, &\tvec{k} (k_\bot^2)^n \, E \frac{dN^{(1)}}{d^2 k dx \, d^2 p dE} =
    \notag \\ &=
    \frac{\alpha_s \: N_c}{\pi^2 x} \left( E \frac{dN^{(0)}}{d^2 p_\bot \, dE} \right)  \frac{L}{\lambda} 
    \frac{\tvec{u}}{4 (1-u_z) x E}
    \int d^2 q_\bot \frac{\bar{\s}(q_\bot^2)}{\s_0}
    \frac{q_\bot^2}{q_\bot^2 + \mu^2} 
    \int d^2 k_\bot \, (k_\bot^2)^n 
    \notag \\ &=
    \frac{\alpha_s \: N_c}{\pi x} \left( E \frac{dN^{(0)}}{d^2 p_\bot \, dE} \right)  \frac{L}{\lambda} 
    \frac{\tvec{u}}{8 (1-u_z) x E}
    \int\limits_{k_{min}^2}^{k_{max}^2} d k_\bot^2 \, (k_\bot^2)^n \, ,
\end{align}
where we have utilized angular averaging and taken the medium to be homogeneous for simplicity.  For the cross section ratio \eqref{e:GWrat} corresponding to the Gyulassy-Wang potential the $d^2 q_\bot$ integral trivially gives $1/2$.  For the denominator of \eqref{average2} which normalizes the average, we use the vacuum distribution \eqref{e:scalvac2} in the small-$x$ limit with the $q \rightarrow q g$ light-front wave functions (see for instance Eq.~(31) of Ref.~\cite{Sievert:2018imd}), obtaining
\begin{align}
    \int d^2 k_\bot \, E \frac{dN^{(0)}}{d^2 k dx \, d^2 p dE} &=
    \frac{C_F}{2 (2\pi)^3 x} 
    \left( E \frac{dN^{(0)}}{d^2 p_\bot dE} \right)
    \int d^2 k_\bot \, \left| \psi(x, \tvec{k}) \right|^2 \,
    \notag \\ &=
    \frac{\alpha_s C_F}{\pi x} 
    \left( E \frac{dN^{(0)}}{d^2 p_\bot dE} \right)
    \ln\frac{k_{max}^2}{k_{min}^2} \, .
\end{align}
The vector-valued moment of the radiated momentum is then given by the ratio
\begin{align}
    \left\langle \tvec{k} (k_\bot^2)^n \right\rangle &=
    \frac{N_c}{C_F} \frac{L}{\lambda} 
    \frac{\tvec{u}}{8 (1-u_z) x E}
    \frac{
        \int_{k_{min}^2}^{k_{max}^2} d k_\bot^2 \, (k_\bot^2)^n
    }{
        \ln k_{max}^2 / k_{min}^2
    } \, .
\end{align}
For moments of simple powers $\tvec{k} \, (k_\bot^2)^n$, a good illustrative choice is $n=-1$, for which the dependence on the integration limits $k_{min} , k_{max}$ cancels exactly with the vacuum:
\begin{align}   \label{e:radmoment}
    \left\langle \frac{\tvec{k}}{k_\bot^2} \right\rangle &=
    \frac{N_c}{C_F} \frac{L}{\lambda} 
    \frac{\tvec{u}}{8 (1-u_z) x E} \, .
\end{align}
Any other choice of $n$ will lead to power-law sensitivity to the endpoints $k_{min} , k_{max}$, although more complicated functions can also be chosen which are insensitive to the endpoints.  The final result \eqref{e:radmoment} demonstrates that the soft-gluon radiation is emitted preferentially in the direction of the medium velocity, with the softest gluons (smallest energy $\omega = x E$) being deflected the most (see Fig.~\ref{f:JETSrad}).  This demonstrates that the soft radiation produced by the jet which dominates the radiative energy loss actually flows along with the medium itself. By replacing the standard eikonal expression for the medium-induced spectrum \cite{Gyulassy:2000er} with our extended result \eqref{e:QCDfinalanswer}, one can immediately generalize existing phenomenology to include the velocity coupling, such as in the Boltzmann transport framework of Ref.~\cite{Luo:2018pto}.

\subsection{Comments on Model Dependence}

With the derivation of velocity and gradient effects on jet broadening and in-medium branching complete, a few comments about the model dependence (and independence) of our results are in order.  First, we note that our most general expressions for the single-Born diagrams \eqref{e:BroadBorn10} for broadening and \eqref{e:R1Afinal}, \eqref{e:RBfinal}, and \eqref{e:RCfinal} for radiative branching do not make use of the specific form of the scattering potential at all.  They are therefore valid for any model of the medium potential and are not limited to the particular form of the Gyulassy-Wang potential \eqref{e:potl5}.  Accordingly, \eqref{e:BroadBorn12} and the single-Born terms in \eqref{e:fullanswer} are valid for any model of the elastic scattering cross section.  For the double-Born diagrams, however, contact integrals such as \eqref{e:DblBorn5} were sensitive to the poles and details of the potential itself. These results have therefore incorporated features which are specific to the Gyulassy-Wang potential \eqref{e:potl5}, both in the broadening case and in the radiation case.  

Despite this, the model dependence which may be present in the double-Born velocity corrections~\eqref{e:GamDB} to jet broadening cannot be too great, or it would spoil the unitarity which we showed explicitly in~\eqref{e:untrt1}.  This delicate cancellation between the (model-independent) single-Born diagrams and the (possibly model-dependent) double-Born diagrams suggests that deviations from \eqref{e:GamDB} for $\vec{\Gamma}_{DB}$ in other models of the potential are, at worst, highly constrained.  Moreover, in the calculation of the moments \eqref{e:moments1} of the jet transverse momentum, the double-Born diagrams do not enter at all since by construction they do not contribute to the broadening of the momentum.  As a result, the general expressions for the transverse momentum moments \eqref{e:untrt3} and \eqref{e:untrt4} do not depend on the choice of the Gyulassy-Wang potential and are fully independent of the model chosen for the medium.  Thus, we find that our conclusions for the jet broadening calculation are quite general and independent of the assumption of the Gyulassy-Wang potential in particular.  For the case of radiative branching, however, the sensitivity to the details of the pole structure associated with the medium potential may be greater, and it will be interesting to investigate in future work the corresponding expressions with other choices of the medium potential.

\section{Discussion and Conclusions}
\label{sec:concl}

In this paper, we performed a first-principles calculation of the effects of medium motion on parton propagation and branching in strongly interacting matter. We also calculated the effects of gradients in density and velocity on the in-medium propagation.  The investigation of these effects in perturbative calculations of jet broadening and energy loss is well motivated. In the case of heavy ion collisions, where a rapidly expanding QGP is created, the overwhelming majority of produced hard partons will be moving at an angle relative to the direction of medium's transverse expansion. Even for the rare cases when the partons propagate radially outward from the center of the collisions region, we might expect that the fluid velocity fields and the jet direction of motion might not be always collinear.  Furthermore, in non-central collisions the medium will have a non-zero angular momentum, resulting in directed flow $v_1$. This is yet another source of motion that will affect the propagation of jets, especially away from midrapidity.   

While the application of our results is most transparent for heavy ion collisions, the  approach is general and applicable to both the QGP and cold nuclear matter. For the  upcoming EIC, as theory and phenomenology in e+A collisions progress~\cite{Arratia:2019vju,Li:2020zbk,Li:2020rqj,Zhang:2021tcc}, we expect this study to be relevant for the inclusion of nuclear matter density and color charge fluctuations and gradients. Furthermore, there is always the orbital motion of nucleons and in addition it might be interesting to consider the motion of the nucleon remnants set by the nuclear break-up in large $Q^2$ deep inelastic scattering processes.  

To address these questions, we developed a new approach to incorporate velocity and gradient corrections into pQCD calculations that differentially take into account the interactions with the scattering centers in the medium, such as the opacity expansion. This allowed us to address the jet/flow coupling for the QGP and set the stage for future studies of color charge fluctuations and orbital motion at the EIC. While collisional energy losses have been studied extensively for leading partons and parton 
showers~\cite{Gossiaux:2006yr,Adil:2006ei,Neufeld:2011yh},  in this work we 
derived the effect of collisional energy transfer to/from the jet by the medium and an angle-dependent shift in the exchanged gluon virtuality on the broadening pattern and the in-medium splitting kernels. We found that these lead to a deflection of the jet in the direction of medium motion. 
While medium velocity and gradients \com{(separately)} do not affect the root mean square broadening of final-state jets, $\sqrt{\langle p_\bot^2 \rangle}$, they do lead to anisotropies, such as $\langle \vec{p}_\bot \rangle \neq 0$. We showed that  the deflection of jets is proportional to the transverse velocity ${\bf u}_\perp$, is enhanced by the medium opacity $L/
\lambda$, and is suppressed by the jet energy.  Not only is 
the distribution of jets affected, but higher odd moments also receive non-trivial contributions.

We further investigated the formation of the parton shower in matter and derived the velocity corrections on in-medium branching.  We found that these corrections directly couple to the interference phases that reflect the formation times of parton splitting to the typical length scales in the medium and are at the heart of the LPM effect. While in this study we used scalar QCD as the underlying field theory, our results are general and full QCD is recovered by substituting the relevant wavefunctions and color factors. To gain physics insight into the formal results, we considered simplifying limits such as soft gluon emission and linear velocity corrections. A peculiar feature of scalar QCD is that branching is isotropic and in this case the velocity correction would vanish in the simplest kinematic limit. Going to full QCD, however, in the soft branching limit we found the leading correction that results in gluon emission preferentially in the direction of  ${\bf u}_\perp$. Interestingly, the corrections are larger for soft gluons emitted at a large angle relative to the jet axis. 

Using the techniques developed here, it may be possible to deconvolute information about the medium motion and spatial distribution (gradients) from the observed distributions of jets and jet substructure.  As such, we believe that this work constitutes a significant step toward full-fledged, first-principles velocity and gradient tomography using jets.

\section*{Acknowledgments}

A.V.S. would like to thank C. Andres, N. Armesto, J. Barata, F. Dominguez, H.T. Li, Y. Makris, J. Reiten, and C. Salgado for discussions and comments on this work. M.D.S. wishes to thank W. Horowitz, J. Noronha, J. Noronha-Hostler, and D. Wertepny for useful discussions. I.V. acknowledges illuminating discussion with P. Jacobs and X.N. Wang.  The work of A.V.S. on Secs. I, II, and III (and related Appendixes A, B, and C) is supported by the Russian Science Foundation Grant RSF 21-12-00237 and on Secs. IV and V (and related Appendixes D and E) by the European Research Council project ERC-2018-ADG-835105. A.V.S. is also grateful for support from LANL LDRD office, from Xunta de Galicia (Centro singular de investigaci\'on de Galicia accreditation 2019-2022),  from the European Union ERDF,  from the Spanish Research State Agency by “Mar{\'{i}}a de Maeztu” Units of Excellence program MDM-2016-0692,  project FPA2017-83814-P,  and from European Union’s Horizon 2020 research and innovation program under the Grant Agreement No.~82409. M.D.S. gratefully acknowledges support from the US-DOE Nuclear Science Grant No. DE-SC0020633. I.V. is supported by the U.S. Department of Energy under Contract No. 89233218CNA000001 and by the LDRD program at LANL.


\newpage

\appendix

%
\section{Universality of Regge Scattering Beyond Eikonal Order}
\label{app:Regge}
%

Consider the $2 \rightarrow 2$ scattering of particles with momenta $p_1 + p_2 \rightarrow p_1^\prime + p_2^\prime$.  We will compare three cases, in which the particles are scalars, spin-$1/2$ fermions, and vector bosons (gluons).  In all three cases we consider the scattering by the exchange of $t$-channel gluons, either in real QCD or in scalar QCD, depending on the theory.  We wish to compare the kinematic dependence of the $2 \rightarrow 2$ cross section on the ratio $t/s$ in the forward-scattering Regge limit $|t/s| \ll 1$, after averaging over any spin quantum numbers.

The usual Regge limit corresponds to $s \rightarrow \infty$, with all other scales including $t$ and the masses held fixed and small.  This differs, however, from the assumptions which underlie the Gyulassy-Wang potential we employ.  The Gyulassy-Wang potential assumes that the scattering centers of the medium are very heavy, possessing a mass $M \rightarrow \infty$ which suppresses the recoil of the medium particles. 
These assumptions can be relaxed, see for example~\cite{Ovanesyan:2011xy}.

The $2 \rightarrow 2$ scattering kinematics are compactly expressed in terms of the Mandelstam invariants
\begin{subequations}
\begin{align}
    s   &\equiv (p_1 + p_2)^2 = (p_1^\prime + p_2^\prime) 
    && = && 
    2 p_1 \cdot p_2 + M^2 = 2 p_1^\prime \cdot p_2^\prime + M^2  \,  ,
    \\
    t &\equiv q^2 \equiv (p_1^\prime - p_1)^2 = (p_2 - p_2^\prime)^2 
    && = && 
    - 2 p_1 \cdot p_1^\prime = - 2 p_2 \cdot p_2^\prime + 2 M^2 \, , 
    \\
    u &\equiv (p_1 - p_2^\prime) = (p_1^\prime - p_2)
    && = && 
    - 2 p_1 \cdot p_2^\prime + M^2 = - 2 p_1^\prime \cdot p_2 + M^2 \, , 
\end{align}
\end{subequations}
where we have taken the jet particle to be massless $p_1^2 = p_1^{\prime \, 2} = 0$ and the medium particle $p_2^2 = p_2^{\prime \, 2} = M^2$ to be heavy.  The sum of the Mandelstam invariants is $s + t + u = 2 M^2$, and for concreteness we will work in Feynman gauge and treat the $t$-channel gluon as having a mass $\mu$ for consistency with the Gyulassy-Wang potential.  For direct comparison with the heavy-mass assumption of the Gyulassy-Wang potential, we can explicitly write the initial momenta $p_1$ of the jet and $p_2$ of the medium particle as
\begin{subequations}
\begin{align}
    p_1^\mu &= (E, \vec{p}_{1\bot}, p_{1z}) \, , \\
    p_2^\mu &= \gamma M (1, \tvec{u}, u_z) \, ,
\end{align}
\end{subequations}
such that 
\begin{align}   \label{e:vtxcorrect}
    s - M^2 = 2 \gamma \, (1-u_z) \, M E
    \left[ 1 - \frac{\tvec{u} \cdot \vec{p}_{1\bot}}{(1 - u_z) E} + \ord{\frac{p_{1\bot}^2}{E^2} } \right] \, .
\end{align}
After computing the square of the scattering amplitude, we will first take the $M \rightarrow \infty$ limit, dropping all terms which are subleading in powers of $M$ for consistency with the Gyulassy-Wang potential.  Then we will expand the kinematics in the Regge limit in powers of the small quantity $p_{1\bot} / E$, comparing the results up to first sub-eikonal order $\ord{\frac{\bot}{E}}$.

For quarks scattering by gluon exchange, the amplitude is
\begin{align}
    i \mathcal{M}_\mathrm{quarks} = \frac{i g^2}{t - \mu^2} \: (t^a \otimes t^a) \: 
    \left[ \bar{U}(p_1^\prime) \gamma_\mu U(p_1) \right] \:
    \left[ \bar{U}(p_2^\prime) \gamma^\mu U(p_2) \right],
\end{align}
leading to the spin- and color-averaged amplitude squared
\begin{align}   \label{e:spinor22}
    \left\langle \left| \mathcal{M} \right|^2 \right\rangle_\mathrm{quarks} &=
    \frac{1}{4} \frac{g^4}{(t-\mu^2)^2} \: \frac{C_F}{2 N_c} \:
    \tr\left[ \slashed{p_1} \gamma_\nu \slashed{p_1}^\prime \gamma_\mu \right] \:
    \tr\left[ (\slashed{p_2} + M) \gamma^\nu (\slashed{p_2}^\prime + M) \gamma^\mu \right]
    \notag \\ &=
    \frac{g^4}{(t-\mu^2)^2} \: \frac{C_F}{2 N_c} \: 
    \Big[ 4 (s - M^2)^2 + 4 (s - M^2) t + 4 M^2 t + 2 t^2\Big] 
    \notag \\ &=
    \frac{g^4}{(t-\mu^2)^2} \: \frac{C_F}{2 N_c} \: 
    \Bigg[ 
        16 \gamma^2 (1-u_z)^2 \, M^2 E^2 \left(1 - \frac{\tvec{u} \cdot \vec{p}_{1\bot}}{(1 - u_z) E}\right)^2
        \notag \\ &\hspace{1cm}
       + 8 \gamma (1-u_z) \, M E t \, \left(1 - \frac{\tvec{u} \cdot \vec{p}_{1\bot}}{(1 - u_z) E}\right)
       + 4 M^2 t + 2 t^2 \Bigg] 
    \notag \\ &\approx
    \frac{g^4}{(t-\mu^2)^2} \: \frac{C_F}{2 N_c} \: 
    \Bigg[ 16 \gamma^2 (1-u_z)^2 \, M^2 E^2 \left(1 - \frac{\tvec{u} \cdot \vec{p}_{1\bot}}{(1 - u_z) E}\right)^2
       + 4 M^2 t \Bigg] 
    \notag \\ &\approx
    \frac{g^4}{(t-\mu^2)^2} \: \frac{C_F}{2 N_c} \: 
    16 \gamma^2 (1-u_z)^2 \, M^2 E^2 \left(1 - 2 \frac{\tvec{u} \cdot \vec{p}_{1\bot}}{(1 - u_z) E}
    + \ord{\frac{p_{1\bot}^2}{E^2}}     \right) \, .
\end{align}
Similarly, for gluon scattering the amplitude is
\begin{align}   
    i \mathcal{M}_\mathrm{gluons} &= \frac{-i g^2}{t - \mu^2} \: (f^{a b c} \, f^{d c e}) \: 
    \notag \\ &\times
    \left[
    (p_1 - q) \cdot \epsilon_{\lambda_1^\prime}^* \: (\epsilon_{\lambda_1})_\mu
    - (p_1 + p_1^\prime)_\mu \: (\epsilon_{\lambda_1} \cdot \epsilon_{\lambda_1^\prime}^*)
    + (p_1^\prime + q) \cdot \epsilon_{\lambda_1} \: (\epsilon_{\lambda_1^\prime}^*)_\mu
    \right]
    \notag \\ &\times
    \left[
    (p_2 + p_2^\prime)^\mu \: (\epsilon_{\lambda_2} \cdot \epsilon_{\lambda_2^\prime}^*)
    - (p_2 + q) \cdot \epsilon_{\lambda_2^\prime}^* \: (\epsilon_{\lambda_2})^\mu    
    - (p_2^\prime - q) \cdot \epsilon_{\lambda_2} \: (\epsilon_{\lambda_2^\prime}^*)^\mu
    \right],
\end{align}
with $q = p_1^\prime - p_1 = p_2 - p_2^\prime$ and the various $\epsilon_\lambda$ factors denoting polarization vectors.  Squaring the amplitude and averaging over colors and spins gives
\begin{align}   \label{e:vector22}
    \left\langle \left| \mathcal{M} \right|^2 \right\rangle_\mathrm{gluons} &=
    \frac{g^4}{(t-\mu^2)^2} \: \frac{N_c}{2 C_F} \: 
    \Big[ 
    4 (s-M^2)^2 + 4(s-M^2) t + 2 t^2 - 8 M^2 t
    \Big]
    \notag \\ &=
    \frac{g^4}{(t-\mu^2)^2} \: \frac{N_c}{2 C_F} \: 
    \Bigg[ 
    16 \gamma^2 (1-u_z)^2 \, M^2 E^2 \left(1 - \frac{\tvec{u} \cdot \vec{p}_{1\bot}}{(1 - u_z) E}\right)^2 
        \notag \\ &\hspace{1cm}
        + 8 \gamma (1-u_z) \, M E t \left(1 - \frac{\tvec{u} \cdot \vec{p}_{1\bot}}{(1 - u_z) E}\right) 
        + 2 t^2 - 8 M^2 t
    \Bigg]    
    \notag \\ &\approx
    \frac{g^4}{(t-\mu^2)^2} \: \frac{N_c}{2 C_F} \: 
    \Bigg[ 
    16 \gamma^2 (1-u_z)^2 \, M^2 E^2 \left(1 - \frac{\tvec{u} \cdot \vec{p}_{1\bot}}{(1 - u_z) E}\right)^2 
    - 8 M^2 t    \Bigg]  
    \notag \\ &\approx
    \frac{g^4}{(t-\mu^2)^2} \: \frac{N_c}{2 C_F} \: 
    16 \gamma^2 (1-u_z)^2 \, M^2 E^2 \left(1 - 2 \frac{\tvec{u} \cdot \vec{p}_{1\bot}}{(1 - u_z) E}
    + \ord{\frac{p_{1\bot}^2}{E^2}}   \right)  \; , 
\end{align}
which agrees up to $\ord{\frac{\bot}{E}}$ with the expression for quark-quark scattering, up to the fundamental/adjoint color factor substitution.

For scalars scattering by gluon exchange in scalar QCD, we may choose the representation $R$ of the $SU(N_c)$ generators $t_R$ to be either fundamental or adjoint.  The resulting color factors will reproduce either the quark-scattering factor $\frac{C_F}{2 N_c}$ or the gluon-scattering factor $\frac{N_c}{2 C_F}$; we can write either case compactly in terms of the Casimir $C_R$ of the scalar representation $R$ and the complementary Casimir $C_{\bar{R}}$ of the opposite representation: $\frac{C_R}{2 C_{\bar{R}}}$.  The kinematic part of the scalar scattering amplitude is then straightforward:
\begin{align}
    i \mathcal{M}_\mathrm{scalars} &= \frac{i g^2}{t-\mu^2} \: (t_R^a \otimes t_R^a) \: 
    (p_1 + p_1^\prime) \cdot (p_2 + p_2^\prime) \, ,    
\end{align}
leading to
\begin{align}   \label{e:scalar22}
    \left\langle \left| \mathcal{M} \right|^2 \right\rangle_\mathrm{scalars} &=
    \frac{g^4}{(t-\mu^2)^2} \: \frac{C_R}{2 C_{\bar{R}}} \: 
    \Big[  (p_1 + p_1^\prime) \cdot (p_2 + p_2^\prime) \Big]^2
    \notag \\ &=
    \frac{g^4}{(t-\mu^2)^2} \: \frac{C_R}{2 C_{\bar{R}}} \: 
    \Big[ 4(s-M^2)^2 + 4 (s-M^2) t + t^2) \Big]
    \notag \\ &=
    \frac{g^4}{(t-\mu^2)^2} \: \frac{C_R}{2 C_{\bar{R}}} \: 
    \Bigg[ 16 \gamma^2 (1-u_z)^2 \, M^2 E^2 \left( 1 - \frac{\tvec{u} \cdot \vec{p}_{1\bot}}{(1 - u_z) E} \right)^2 
        \notag \\ &\hspace{1cm}
        + 8 \gamma (1-u_z) \, M E t \left( 1 - \frac{\tvec{u} \cdot \vec{p}_{1\bot}}{(1 - u_z) E} \right) + t^2) \Bigg]
    \notag \\ &\approx
    \frac{g^4}{(t-\mu^2)^2} \: \frac{C_R}{2 C_{\bar{R}}} \: 
    \Bigg[ 16 \gamma^2 (1-u_z)^2 \, M^2 E^2 \left( 1 - \frac{\tvec{u} \cdot \vec{p}_{1\bot}}{(1 - u_z) E} \right)^2 \Bigg]
    \notag \\ &\approx
    \frac{g^4}{(t-\mu^2)^2} \: \frac{C_R}{2 C_{\bar{R}}} \: 
    16 \gamma^2 (1-u_z)^2 \, M^2 E^2 \left( 1 - 2 \frac{\tvec{u} \cdot \vec{p}_{1\bot}}{(1 - u_z) E} 
    + \ord{\frac{p_{1\bot}^2}{E^2}}     \right) \; .
\end{align}

Comparing the amplitudes-squared for quarks \eqref{e:spinor22}, gluons \eqref{e:vector22}, and scalars \eqref{e:scalar22}, we see that not only do they agree in the strict eikonal limit, but the universality continues to hold at $\ord{\frac{\bot}{E}}$.  This assures us that in replacing the Regge scattering of quarks or gluons with scalars in the corresponding color representation, we do not modify the structure of the velocity corrections in any way.  Instead, all three cases receive the same correction to the eikonal vertex, arising entirely from the correction \eqref{e:vtxcorrect} to $(s - M^2)^2$.  One can also compare Eqs.~\eqref{e:spinor22}, \eqref{e:vector22}, and \eqref{e:scalar22} in the usual Regge limit by setting $M^2 = 0$ and expanding in powers of $|t/s|$.  Doing so again verifies that the universality of Regge scattering continues to hold at least through the first sub-eikonal order $\ord{|t/s|}$.  This analysis justifies our replacement of the quark and gluon degrees of freedom of QCD with a simpler scalar theory, combining the universal nature of Regge scattering and the interchangeability of the light-front wave functions.

\section{Jet Broadening: Single Born Diagram}
\label{app:snglBorn}

With either choice of Lagrangian \eqref{e:Ljet},
the lowest-order scattering amplitude for this process is shown in Fig.~\ref{f:Broad1} and gives
\begin{align}
\label{e:BroadBorn1}
    i M_1 (p) = \int \frac{d^4 q}{(2\pi)^4}
    \Big[ i g \, t^a_\mathrm{proj} A_\mathrm{ext}^{\mu a} (q) \: (2 p - q)_\mu \Big]
    \left[  \frac{i}{(p-q)^2 + i \epsilon}  \right] J(p-q) \, ,
\end{align}
where the jet with initial momentum $(p-q)^\m$ is scattered by the potential into the measured final-state distribution of momentum $p^\m$. The momentum of the jet is
\begin{align}
\label{e:jetkin1}
    p^\mu &= \Big( E \, , \, \vec{p}_\bot \, , \, \sqrt{E^2 - p_\bot^2} \Big)
    \approx
    \Big( E \, , \, \vec{p}_\bot \, , \, E \left( 1 - \tfrac{p_\bot^2}{2 E^2} \right) \Big) \, , 
\end{align}
where we have neglected higher-order terms $\ord{\tfrac{\bot^4}{E^4}}$ suppressed by the jet energy.  We want to ultimately keep all corrections of $\ord{\tfrac{\perp}{E}}$ and drop corrections of $\ord{\tfrac{\perp^2}{E^2}}$ and higher in our final answer, while making no assumptions about $u_i^\mu$ to allow for a highly relativistic medium.

Substituting \eqref{e:potl1} and \eqref{e:flow2} into \eqref{e:BroadBorn1} gives
\begin{align}
\label{e:BroadBorn2}
    M_1 (p) = i \sum_i t^a_\mathrm{proj} t_i^a \int \frac{d^4 q}{(2\pi)^4}
    e^{i q \cdot x_i} 
    \,\frac{u_i^\mu (2 p - q)_\mu}{(p-q)^2 + i \epsilon}\,
    v_i (q) \: J(p-q) \: (2\pi) \delta(q^0 - \vec{u}_i \cdot \vec{q})\, .
\end{align}
The delta function from the potential helps to perform the $dq^0$ integral immediately, and the numerator algebra gives
\begin{align}
\label{e:BroadBorn3}
    u_i^\mu (2 p - q)_\mu &= 2 u_i^\mu p_\mu = 
    2 (E - \vec{u}_{i \, \bot} \cdot \vec{p}_\bot - u_{i \, z} p_z)
    \notag \\ &\approx
    2 E (1-u_{i \, z}) \left[ 1 - \frac{\vec{u}_{i \, \bot} \cdot \vec{p}_\bot}{E (1-u_{i \, z})} +
    \ord{\frac{\perp^2}{E^2}} \right]\, .
\end{align}
After picking up the delta function in $q^0$, the next step is to perform the integration over $q_z$ by residues.  To this end, the propagator is
\begin{align}
\label{e:BroadBorn4}
    (p-q)^2 + i \epsilon &=
    (E - q^0)^2 - (p-q)_\bot^2 - (p_z - q_z)^2 + i \epsilon 
  \notag \\ &=
    (E - \vec{u}_{i \, \bot} \cdot \vec{q}_\bot - u_{i \, z} q_z)^2 
    - (p-q)_\bot^2 - \left( E - \tfrac{p_\bot^2}{2 E} - q_z \right)^2 + i \epsilon 
 \notag \\ &=
    -\Big( 1-u_{i \, z}^2 \Big) q_z^2 + 2 q_z \left( (1-u_{i \, z}) E - \tfrac{p_\bot^2}{2E} + (\vec{u}_{i \, \bot} \cdot \vec{q}_\bot) u_{i \, z} \right)
    \notag \\ & \hspace{1cm} +
    \Big( p_\bot^2 - (p-q)_\bot^2 - 2 E (\vec{u}_{i \, \bot} \cdot \vec{q}_\bot) + (\vec{u}_{i \, \bot} \cdot \vec{q}_\bot)^2 \Big) + i\epsilon
 \notag \\ &=
    -( 1-u_{i \, z}^2 )  \left[ q_z - Q_{p-q}^+ - i \epsilon \right]
    \left[ q_z - Q_{p-q}^- + i \epsilon \right]
\end{align}
with poles
\begin{subequations}
\label{e:BroadPoles1}
\begin{align}
    Q_{p-q}^+ &= \frac{2 E}{1 + u_{i \, z}} \left[ 1 -
    \frac{ \vec{u}_{i \, \bot} \cdot \vec{q}_\bot }{ 2 E } 
    + \ord{\frac{\bot^2}{E^2}}  \right] \, ,
    \\
    Q_{p-q}^- &= \frac{ \vec{u}_{i \, \bot} \cdot \vec{q}_\bot }{ 1 - u_{i \, z} } + \frac{(p-q)_\bot^2 - p_\bot^2}{2 E (1 - u_{i \, z})} + \ord{\frac{\bot^2}{E^2}}\, ,
\end{align}
\end{subequations}
where the subscript indicates that these poles are zeros of $(p-q)^2+i\e=0$. In principle, there can also be poles of $q_z$ associated with the potential \eqref{e:potl5}.  These poles occur at momenta $q$ with finite imaginary part, as there is no valid cut of the $t$-channel gluon carrying the potential.  As such, these poles of the potential lead to an exponentially-decaying amplitude in the position $x_i$ of the scattering center.  For a sufficiently large medium $\mu z_i \gg 1$, this decaying mode can be neglected \cite{Gyulassy:2000er, Kolbe:2019njo}, so here we only consider the explicit poles \eqref{e:BroadPoles1} arising from the cuttable propagator.  Such corrections can be safely neglected for these single-Born diagrams where the imaginary part of the pole leads to a decaying mode, but they will become important for the double-Born diagrams where no such exponential decay occurs.

Inserting \eqref{e:BroadBorn3} and \eqref{e:BroadBorn4} into \eqref{e:BroadBorn2} gives
\begin{align}
\label{e:BroadBorn5}
    M_1 (p) &= i \sum_i t^a_\mathrm{proj} t_i^a \int \frac{d^3 q}{(2\pi)^3}
    e^{i q \cdot x_i} 
    \Big[ 2 E (1-u_{i \, z}) \Big] \left[ 1 - \frac{\vec{u}_{i \, \bot} \cdot \vec{p}_\bot}{E (1-u_{i \, z})} \right]
    \notag \\ & \hspace{1cm} \times
    \left[ \frac{-1}{1-u_{i \, z}^2} \frac{1}{(q_z - Q_{p-q}^+ - i \epsilon) \, (q_z - Q_{p-q}^- + i \epsilon)} \right]
    v_i (q_i) \: J(p-q_i) \, .
\end{align}
Of the two poles $Q_{p-q}^\pm$, the large pole $Q_{p-q}^+$ leads to an amplitude which is highly suppressed by the jet energy $E$ and can be neglected.  The only unsuppressed contribution arises from collecting the residue of $Q_{p-q}^-$, which requires closing the contour below the real axis and hence $z_i > 0$.  Doing so yields
\begin{align}
\label{e:BroadBorn6}
    M_1 (p) &= \sum_i \frac{2 E \: \theta(z_i)}{1+u_{i \, z}} \: 
    t^a_\mathrm{proj} t_i^a \int \frac{d^2q_\bot} {(2\pi)^2} e^{i q_i \cdot x_i} 
    \left[ 
    1 - \frac{\vec{u}_{i \, \bot} \cdot \vec{p}_\bot}{E (1-u_{i \, z})}
    \right]
    \left[ \frac{1}{Q_{p-q}^+ - Q_{p-q}^-} \right]  v_i (q_i) \: J(p-q_i)
    \notag \\ &=
    \sum_i \frac{2 E \: \theta(z_i)}{1+u_{i \, z}} \: 
    t^a_\mathrm{proj} t_i^a \int \frac{d^2q_\bot} {(2\pi)^2} e^{i q_i \cdot x_i} 
    \left[ 
    1 - \frac{\vec{u}_{i \, \bot} \cdot \vec{p}_\bot}{E (1-u_{i \, z})}
    \right] 
    \notag \\ &\hspace{2cm} \times
    \left[ 
    \frac{2 E}{1 + u_{i \, z}} 
    - \frac{ \vec{u}_{i \, \bot} \cdot \vec{q}_\bot }{ 1 + u_{i \, z}}
    - \frac{ \vec{u}_{i \, \bot} \cdot \vec{q}_\bot }{ 1 - u_{i \, z} }
    \right]^{-1} v_i (q_i) \: J(p-q_i) \, , 
    \notag \\ M_1 (p) &=
    \sum_i \theta(z_i) \: 
    t^a_\mathrm{proj} t_i^a \int \frac{d^2q_\bot} {(2\pi)^2} e^{i q_i \cdot x_i} 
    \left[ 
    1 - \frac{\vec{u}_{i \, \bot} \cdot (\vec{p}_\bot - \vec{q}_\bot)}{E (1-u_{i \, z})}
    \right]  v_i (q_i) \: J(p-q_i) \, .
\end{align}
Note that in addition to the explicit correction factor we obtain to the eikonal vertex of the external potential, there are also other implicit corrections to the arguments of $v_i (q_i)$ and $J(p-q_i)$ after setting the momentum transfer $q$ equal to its pole value $q_i$:
\begin{align}
\label{e:qpoles1}
    q^\mu \rightarrow q_i^\mu &= (\vec{u}_i \cdot \vec{q}_i \, , \, \vec{q}_\bot \, , \, Q_{p-q}^-)
    =
    \Big( \vec{u}_{i \, \bot} \cdot \vec{q}_{\bot} + u_{i \, z} Q_{p-q}^- 
    \, , \, \vec{q}_\bot \, , \, Q_{p-q}^- \Big)    
\end{align}
from the delta function in the external potential and from the pole of the jet propagator.  We will address these implicit corrections after we have squared the amplitude.

We now proceed to square the amplitude and average over quantum numbers in the usual way.  First, performing the averaging over colors of the in-medium sources, we obtain
\begin{align} \label{e:Avg}
\left\langle t^a_i t^b_j \right\rangle &= 
\frac{1}{d_\mathrm{tgt}} \tr\left( t^a_i t^b_j \right) 
= \begin{cases}
    \frac{1}{2 N_c} \delta_{i j} \delta^{a b} &
    \mathrm{if \: target \: is \: fundamental} \\
    \frac{1}{2 C_F} \delta_{i j} \delta^{a b} &
    \mathrm{if \: target \: is \: adjoint}
\end{cases}
\notag \\ &\equiv
\frac{1}{2  C_{\bar{R}}} \delta_{i j} \delta^{a b} ,
\end{align}
where $d_\mathrm{tgt}$ is the dimension of the color representation of the in-medium sources (``target'') and $C_{\bar{R}}$ is the quadratic Casimir in the \textit{opposite} representation as defined above.  We note this color averaging enforces a color-neutrality condition $\delta_{i j}$ -- that is, that both gluon exchanges occur on the \textit{same} in-medium parton -- which is equivalent to performing a Gaussian averaging of the target fields $A_\mathrm{ext}^{\mu a}$ \cite{Sievert:2018imd, Blaizot:2012fh, CasalderreySolana:2007pr}.  Here, the particular color factors used assume that the scattering centers of the medium exist in the same representation of $SU(N_c)$.  Then, upon squaring the amplitude \eqref{e:BroadBorn6}, we generate two transverse integrals over $d^2 q$ and $d^2 q'_\bot$, but only a single summation $\sum_i$:
\begin{align}
\label{e:BroadBorn7}
    \left\langle \left| M_1 \right|^2 \right\rangle &= 
    \frac{C_\mathrm{proj}}{2 C_{\bar{R}}}
    \sum_i \theta(z_i) \: 
    \int \frac{d^2q_\bot} {(2\pi)^2} \frac{d^2 q'_\bot}{(2\pi)^2}
    e^{i (q_i - q_i^\prime) \cdot x_i} \:
    v_i (q_i) v_i^* (q_i^\prime) \: J(p-q_i) J^*(p-q_i^\prime)
    \notag \\ & \hspace{1cm} \times
    \left[ 
    1 - \frac{\vec{u}_{i \, \bot} \cdot (\vec{p}_\bot - \vec{q}_\bot)}{E (1-u_{i \, z})}
    - \frac{\vec{u}_{i \, \bot} \cdot (\vec{p}_\bot - \vec{q}_\bot^{\: \prime})}{E (1-u_{i \, z})}
    \right] \,,
\end{align}
where $C_\mathrm{proj} \mathbf{1} = t^a_\mathrm{proj} t^a_\mathrm{proj}$ is the quadratic Casimir in the representation of the projectile.  For compactness, we define the overall color factor of this process as
\begin{align}
\label{mathcalC}
    \mathcal{C} \equiv \frac{C_\mathrm{proj}}{2 C_{\bar{R}}}
    = \begin{cases}
        \frac{C_F}{2 N_c} &
    \mathrm{if \: proj \, = \, tgt \, = \, fundamental} \\
    \frac{N_C}{2 C_F} &
    \mathrm{if \: proj \, = \, tgt \, = \, adjoint} \\
    \:\: \frac{1}{2} &
    \mathrm{if \: proj \, \neq \, tgt} 
    \end{cases} .
\end{align}
Next, we convert in the usual way from the discrete summation over scattering centers to a continuous integral over their densities:
\begin{align}
    \sum_i f_i &= N \left(\frac{1}{N} \sum_i f_i \right)
    = N \, \left\langle f \right\rangle
 =
    N \, \int \frac{d^2 x_{i \, \bot}}{A_\bot} \int\frac{dz_i}{L} 
    \: f(\vec{x}_i)
=
    \int d^3 x \: \rho(\vec{x}) \: f(\vec{x}) \, ,
\end{align}
where $N$ is the number of scattering centers and $\rho$ is their number density over a transverse area $A_\bot$ and length $L$.  The last line is also often expressed in terms of the elastic scattering cross section $\sigma_0$ and mean free path $\lambda$ through $\rho = \frac{1}{\lambda \, \sigma_0}$.   With a continuous density profile \eqref{e:BroadBorn7} becomes
\begin{align}
\label{e:BroadBorn8}
    \left\langle \left| M_1 \right|^2 \right\rangle &= \mathcal{C}\,\int d^3 x \: \rho(\vec{x})    
    \int \frac{d^2q_\bot} {(2\pi)^2} \frac{d^2 q'_\bot}{(2\pi)^2} \:
    e^{i (q - q^\prime) \cdot x} \:
    v (q) v^* (q^\prime) \: J(p-q) J^*(p-q^\prime)
    \notag \\ & \hspace{1cm} \times
    \left[ 
    1 - \frac{\vec{u}_{\bot} \cdot (\vec{p}_\bot - \vec{q}_\bot)}{E (1-u_{z})}
    - \frac{\vec{u}_{\bot} \cdot (\vec{p}_\bot - \vec{q}_\bot^{\: \prime})}{E (1-u_{z})}
    \right] \, .
\end{align}

In the usual formulation, the dependence on the transverse coordinate $\vec{x}_{\bot}$ of quantities such as the density $\rho$ is often neglected, such that for example $\rho = \rho(z)$.  Then the integration over $d^2 x_\perp$ yields a delta function $\delta^2 (\vec{q}_\bot - \vec{q}_\bot^\prime)$ and collapses the momentum integration.  In our case, the velocity field $\vec{u}_{i} = \vec{u} ( \vec{x} )$ also contains a spatial dependence in both its transverse and longitudinal components, and neglecting those spatial variations would constitute a severe restriction on the kinds of velocity fields to which our formulas would apply.  This motivates us to retain the transverse spatial dependence of all such quantities: the density $\rho(\vec{x})$, the velocity field $\vec{u}(\vec{x})$, and the Debye mass $\mu (\vec{x})$.\footnote{The Debye mass $\mu \sim g T$ contains a spatial dependence implicitly through its dependence on the temperature profile $T(\vec{x})$.} To make the final expressions more tractable, we will expand them in powers of transverse gradients and analyze the various contributions.

It is noteworthy that, in the physical situations of interest to us, the applicability of a formal gradient expansion may be questionable.  In cold nuclear matter, the applicability of a perturbative calculation of the jet-medium interaction is limited to distances of $\ord{1/\Lambda_{QCD}} \sim 1 \, \mathrm{fm}$. In the QGP, for the exchange of Debye screened gluons, that distance is further shortened to be $\ord{1/\mu} \sim \ord{1/g T}$. If the medium density $\rho$ or velocity field $u^\mu$ is slowly varying over these scales, then a gradient expansion, starting at $0^{th}$ order, is appropriate.  In cold nuclear matter this can be justified based on a smooth optical Glauber picture of the density profile $\rho(\vec{x}_i)$, in which the density only varies over macroscopic scales $L \sim A^{1/3}$.  However nucleon-scale density fluctuations are known to play a crucial role in heavy-ion collisions (see, e.g., \cite{Luzum:2013yya}), making such an expansion questionable.  However, we emphasize that the general expressions derived at the level of \eq{e:BroadBorn8}, for instance, can be applied to medium properties which vary arbitrarily and are not limited to the gradient expansion.

The usual treatment of the transverse averaging then corresponds to the leading term in the gradient expansion:
\begin{align}
\label{e:BroadBorn9}
    \hspace{-0.3cm}\left\langle \left| M_1 \right|^2 \right\rangle &=
    \mathcal{C}\int dz\: \rho(z)\: 
    \int \frac{d^2q_\bot} {(2\pi)^2} \:
    | v (q) |^2  \: |J(p-q)|^2 
    \left[ 
    1  - 2 \frac{\vec{u}_\bot (z) \cdot (\vec{p}_\bot - \vec{q}_\bot)}{E (1-u_{z} (z))}
    \right] + \mathcal{O}\left(\pa_\perp\right) \, ,
\end{align}
where also the Debye mass $\mu = \mu(z)$ is independent of transverse position. Noting that neglecting transverse gradients is an especially poor approximation to the hydrodynamics of heavy ion collisions, we proceed by carefully isolating the spatial dependence implicit in $v(q)$ and $J(p - q)$ through the various poles \eqref{e:qpoles1}.  

First we note that the phase factor $e^{i (q - q^\prime) \cdot x}$ takes a simple form because of the lightlike kinematics of the jet:
\begin{align}
\label{e:BBphase1}
    \exp\left[i (q - q^\prime) \cdot x \right] &=
    \exp\left[
    - i (\vec{q}_\bot - \vec{q}_\bot^{\: \prime}) 
    \cdot \vec{x}_{\bot}
    - i (q_{z} - q_{z}^\prime) z
    \right]
    \notag \\ &=
    e^{- i (\vec{q}_\bot - \vec{q}_\bot^{\: \prime}) 
    \cdot \vec{x}_{\bot} }
    \exp\left[ - i \left(Q_{p-q}^- - Q_{p-q'}^-\right)z 
    \right]
    \notag \\ &=
    e^{- i (\vec{q}_\bot - \vec{q}_\bot^{\: \prime}) 
    \cdot \vec{x}_{\bot} }
    \exp\left[ -i\:\frac{\vec{u}_{\bot} \cdot (\vec{q}_{\bot} - \vec{q}_{\bot}^{\: \prime})}{1-u_z}z- i \: \frac{(p-q)_\bot^2 - (p-q')_\bot^2}{2 E (1-u_z)} z \right] \, ,
\end{align}
We also note that the last term in the exponential is suppressed by $1/E$ but enhanced by the position $z$ of the scattering, which may be large (on the order of the medium length $L$). As such we do not make any assumption at this stage about the smallness of that combined quantity.

For the scattering potential $v(q)$, the correction due to the $\vec{x}_{\bot}$-dependent shift in the argument is
\begin{align}
\label{e:BBpotl1}
    v (q) &= v\left(q_\bot^2 + (Q_{p-q}^-)^2 - q_{0}^2\right)
\approx
    v(q_\bot^2) \left[ 1 + \frac{(Q_{p-q}^-)^2 - q_{0}^2}{v(q_\bot^2)}  \frac{\pa v}{\pa q_\bot^2} \right]
    \notag \\ &\approx
    v(q_\bot^2) \left[ 1 + 
    \frac{\vec{u}_{\bot} \cdot \vec{q}_\bot}{(1-u_{z}) E} \:
    \frac{(p-q)_\bot^2 - p_\bot^2}{v(q_\bot^2)} \: \frac{\pa v}{\pa q_\bot^2} \right] \, ,
\end{align}
and similarly for $v (q^\prime)$.  Using both \eqref{e:BBphase1} and \eqref{e:BBpotl1} in \eqref{e:BroadBorn9} gives
\begin{align}
    \left\langle \left| M_1 \right|^2 \right\rangle &= 
    \mathcal{C}\,\int d^3 x \:
    \frac{d^2q_\bot} {(2\pi)^2} \frac{d^2 q'_\bot}{(2\pi)^2} \, \rho \: 
    e^{-i (\vec{q}_\bot - \vec{q}_\bot^{\: \prime}) \cdot \vec{x}_{\bot}} \:
     \notag \\ & \hspace{1cm} \times 
    \exp\left[ -i\:\frac{\vec{u}_{\bot} \cdot (\vec{q}_{\bot} - \vec{q}_{\bot}^{\: \prime})}{1-u_z}z- i \: \frac{(p-q)_\bot^2 - (p-q')_\bot^2}{2 E (1-u_z)} z \right]    \:   v (q_\bot^2) v^* (q_\bot^{\prime \, 2}) 
    \notag \\ & \hspace{1cm} \times
 \: J(p-q) J^*(p-q^\prime)
    \Bigg[ 1 
    - \frac{\vec{u}_{\bot} \cdot (\vec{p}_\bot - \vec{q}_\bot)}{E (1-u_{z})}
    %
    %
    - \frac{\vec{u}_{\bot} \cdot (\vec{p}_\bot - \vec{q}_\bot^{\: \prime})}{E (1-u_{z})}
    \notag \\ & \hspace{1cm}
    %
    + \frac{\vec{u}_{\bot} \cdot \vec{q}_\bot}{(1-u_{z}) E} \:
    \frac{(p-q)_\bot^2 - p_\bot^2}{v(q_\bot^2)} \: \frac{\pa v}{\pa q_\bot^2}  +  \frac{\vec{u}_{\bot} \cdot \vec{q}_\bot^{\: \prime}}{(1-u_{z}) E} \:
    \frac{(p-q')_\bot^2 - p_\bot^2}{v^*(q_\bot^{\prime \, 2})} \: \frac{\pa v^*}{\pa q_\bot^{\prime 2}}
    \Bigg] \, ,
\end{align}
where we have suppressed the explicit position dependence of $\rho, \vec{u}_\bot$, and $u_z$ for brevity.

Similarly, we can account for the shift of argument in the source terms $J(p-q)$, 
\begin{align}
    (p-q)^\mu &= \Big( E - \vec{u}_\bot \cdot \vec{q}_\bot - u_z Q_{p-q}^- \: , \: \vec{p}_\bot - \vec{q}_\bot \: , \: E - Q_{p-q}^- \Big)^\mu
    \notag \\ &\approx
    \left( E - \frac{\vec{u}_\bot \cdot \vec{q}_\bot}{1-u_z} \: , \: \vec{p}_\bot - \vec{q}_\bot \: , \: E - \frac{\vec{u}_\bot \cdot \vec{q}_\bot}{1-u_z} \right)^\mu \, ,
\end{align}
by recognizing that the medium motion enters through a shift in the energy.  We can extract the explicit energy shift by writing
\begin{align}
    J(p-q) &= J\left(E - \frac{\vec{u}_\bot \cdot \vec{q}_\bot}{1-u_z} , \vec{p}_\bot - \vec{q}_\bot \right)
    =J\left(E, \vec{p}_\bot - \vec{q}_\bot \right)
    - \left( \frac{\vec{u}_\bot \cdot \vec{q}_\bot}{1-u_z} \right)  
    \frac{\partial J}{\partial E} \; , 
\end{align}
which then gives
\begin{align}
\label{e:BroadBorn10}
    \left\langle \left| M_1 \right|^2 \right\rangle &= \mathcal{C}\,\int d^3 x \:
    \frac{d^2q_\bot} {(2\pi)^2} \frac{d^2 q'_\bot}{(2\pi)^2} \, \rho \:
    e^{-i (\vec{q}_\bot - \vec{q}_\bot^{\: \prime}) \cdot \vec{x}_{i \, \bot}} \:v (q_\bot^2) v^* (q_\bot^{\prime \, 2}) \: J(E, \vec{p}_\bot - \vec{q}_\bot) J^*(E, \vec{p}_\bot - \vec{q}_\bot^\prime) 
     \notag \\ & \hspace{1cm} \times
    \exp\left[ -i\:\frac{\vec{u}_{\bot} \cdot (\vec{q}_{\bot} - \vec{q}_{\bot}^{\: \prime})}{1-u_z}z- i \: \frac{(p-q)_\bot^2 - (p-q')_\bot^2}{2 E (1-u_z)} z\right] \:
    \notag \\ & \hspace{1cm} \times
    \Bigg[ 1 
    - \frac{\vec{u}_{\bot} \cdot (\vec{p}_\bot - \vec{q}_\bot)}{E (1-u_{z})}
    - \frac{\vec{u}_{\bot} \cdot (\vec{p}_\bot - \vec{q}_\bot^{\: \prime})}{E (1-u_{z})}
    + 
    \frac{\vec{u}_{\bot} \cdot \vec{q}_\bot}{(1-u_{z}) E} \:
    \frac{(p-q)_\bot^2 - p_\bot^2}{v(q_\bot^2)} \: \frac{\pa v}{\pa q_\bot^2}
    \notag \\ &\hspace{1cm} 
    + \frac{\vec{u}_{\bot} \cdot \vec{q}_\bot^{\: \prime}}{(1-u_{z}) E} \:
    \frac{(p-q')_\bot^2 - p_\bot^2}{v^*(q_\bot^{\prime \, 2})} \: \frac{\pa v^*}{\pa q_\bot^{\prime 2}}
    - \left( \frac{\vec{u}_{\bot} \cdot \vec{q}_\bot}{1-u_z} \right)  \frac{1}{J} \frac{\partial J}{\partial E}
     - \left( \frac{\vec{u}_{\bot} \cdot \vec{q}_\bot^\prime}{1-u_z} \right)  \frac{1}{J^*} \frac{\partial J^*}{\partial E}
    \Bigg] 
    \notag \\ \notag \\ &=
    \mathcal{C}\,\int d^3 x \:
    \frac{d^2q_\bot} {(2\pi)^2}  \frac{d^2 q'_\bot}{(2\pi)^2} \, \rho \:
    e^{-i (\vec{q}_\bot - \vec{q}_\bot^{\: \prime}) \cdot \vec{x}_{\bot}} \: v (q_\bot^2) v^* (q_\bot^{\prime \, 2}) \: J(E, \vec{p}_\bot - \vec{q}_\bot) J^*(E, \vec{p}_\bot - \vec{q}_\bot^\prime)
   \notag \\ & \hspace{1cm} \times   
    \exp\left[ -i\:\frac{\vec{u}_{\bot} \cdot (\vec{q}_{\bot} - \vec{q}_{\bot}^{\: \prime})}{1-u_z}z- i \: \frac{(p-q)_\bot^2 - (p-q')_\bot^2}{2 E (1-u_z)} z\right] \: \times
    \Bigg[ 1 
    + \vec{u}_{\bot} \cdot \vec{\Gamma}_\bot (\vec{q}_\bot, \vec{q}_\bot^\prime)
    \Bigg] \, , 
\end{align}
with the velocity profile $\vec{u}_\bot$ coupling to the correction factor
\begin{align} 
\label{e:Gamma}
    \hspace{-0.3cm}\vec{\Gamma} (\vec{q}_\bot , \vec{q}_\bot^{\: \prime}) &\equiv 
    - \frac{\vec{p}_\bot - \vec{q}_\bot}{(1-u_{z})E}
    - \frac{\vec{p}_\bot - \vec{q}_\bot^{\: \prime}}{(1-u_{z})E}
    + \frac{\vec{q}_\bot}{(1-u_{z}) E} \:
    \left( \frac{(p-q)_\bot^2 - p_\bot^2}{v(q_\bot^2)} \right) \: 
    \frac{\pa v}{\pa q_\bot^2}
    \notag \\ &
    \hspace{-0.5cm}+ \frac{\vec{q}_\bot^{\: \prime}}{(1-u_{z}) E} \:
    \left( \frac{(p-q')_\bot^2 - p_\bot^2}{v^*(q_\bot^{\prime \, 2})} \right) \: 
    \frac{\pa v^*}{\pa q_\bot^{\prime 2}}
    - \frac{\vec{q}_\bot}{1-u_z} \left( \frac{1}{J} \frac{\partial J}{\partial E} \right)
    - \frac{\vec{q}_\bot^{\: \prime}}{1-u_z} \left(   \frac{1}{J^*} \frac{\partial J^*}{\partial E} \right) .
\end{align}

We emphasize that the dependence on the spatial content of the medium is encoded explicitly in the density $\rho (\vec{x})$ and transverse and longitudinal velocity profiles $\vec{u}_\perp (\vec{x})$ and $u_z (\vec{x})$, as well as implicitly in the temperature dependence of the Debye mass $\mu (\vec{x}) \sim g T(\vec{x})$.  

The exchanged gluons which couple the jet to the medium have a characteristic transverse wavelength $1 / \mu$ over which they resolve the medium.  Because this calculation employs pQCD to describe the degrees of freedom, this must be a hard scale: $\mu \gtrsim \ord{1 \, \mathrm{GeV}}$ such that the length scale resolved by the interactions is $\Delta x \sim 1/\mu < 1 \, \mathrm{fm}$.  Depending on how rapidly the medium quantities vary over this perturbative distance, the gradients of these spatial densities may become increasingly important.

%
\subsection{No Gradients (Translational Invariance)}
%
At $0^{th}$ order in the gradient expansion, all medium quantities are taken to possess $2D$ translational invariance and are functions of $z$ only.  This gives
\begin{align}
        \int d^2 x_\perp \, e^{-i (\vec{q}_\bot - \vec{q}_\bot^{\: \prime}) \cdot \vec{x}_{\bot}} \rightarrow 
        (2\pi)^2 \delta^{(2)} (\vec{q}_\bot - \vec{q}_\bot^{\: \prime}) \,
\end{align}
and, thus, 
\begin{align}
\label{e:BroadBorn11}
    \left\langle \left| M_1 \right|^2 \right\rangle &=  
    \, \mathcal{C}\,\int dz \,\frac{d^2q_\bot} {(2\pi)^2} \, \r\:
    |v (q_\bot^2)|^2 \: |J(E, \vec{p}_\bot - \vec{q}_\bot)|^2 
    \bigg[1
    + \vec{u}_{\bot} \cdot  \vec{\Gamma} (\vec{q}_\bot) \Bigg] + \mathcal{O}\left(\pa_\bot\right) \, , 
\end{align}
with $\vec{\Gamma} (\vec{q}_\bot)=\vec{\Gamma} (\vec{q}_\bot , \vec{q}_\bot)$, which is explicitly given in (\ref{e:GamNew}). Using \eqref{e:N0} we can further write
\begin{align}
\label{e:BroadBorn12}
    \left(E \frac{dN^{(1)}}{d^3 p} \right)_\mathrm{SB} &= 
    \int dz \, d^2q_\bot \, \rho(z) \: \bar{\sigma}(q^2_\bot) \: \left( E \frac{dN^{(0)}}{d^2 (p-q)_\perp \, dE} \right) 
    \bigg[ 1
    +  \vec{u}_{\bot} (z) \cdot  \vec{\Gamma} (\vec{q}_\bot) \Bigg] + \mathcal{O}\left(\pa_\bot\right) \, , 
\end{align}
where SB stands for single Born. 

We will discuss in Appendix~\ref{app:dblBorn} the appearance of the double-Born diagrams which enforce unitarity.  Interestingly, we find that the manner in which unitarity is preserved in the more general case appears different; whereas in the static case, the double-Born diagrams effectively replace $\bar{\sigma}(\vec{q}_\bot) \rightarrow \bar{\sigma}(\vec{q}_\bot) - \s_0\delta^2 (\vec{q}_\bot)$ in \eqref{e:BroadBorn12} with $\s_0$ being the integral of $\bar{\s}(\vec{q}_\perp)$, after including medium flow and gradients we find a different, explicitly unitary form.

The essential physics of how the jet couples to the medium motion is already contained in the simple kinematic statement $q^0 = \vec{u} \cdot \vec{q}$ from the potential \eqref{e:potl4} of a moving source.  Unlike the static case where $q^0 \rightarrow 0$, now a finite energy $q^0$ is transferred between the source and the jet.  Depending on whether the momentum transfer $\vec{q}$ flows along with or against the flow $\vec{u}$, the medium can transfer energy to the jet or vice versa.  This exchange is best quantified by the light-cone momentum $q^+ \sim q^0 + q^z \sim \frac{ \vec{u}_\bot \cdot \vec{q}_\bot }{ 1 - u_z }$.  When $\vec{q}_\bot$ is parallel to the flow $\vec{u}_\bot$, the medium constructively transfers energy to the jet; conversely, when $\vec{q}_\bot$ is {\it{antiparallel}} to the transverse flow, the jet loses energy to the medium.

This collisional energy transfer to/from the jet has a few consequences.  Looking at the last two terms of \eqref{e:GamNew}, we can identify two of them.  One is that the energy transfer $q^+$ or $q^0$ leads to a small shift in the energy of the initial jet distribution $\bar{N}_0$ compared to the final-state jet distribution.  The other is that the longitudinal momentum $Q_z \approx \frac{\vec{u}_\bot \cdot \vec{q}_\bot}{1-u_z}$ at the on-shell pole leads to a shift in the transverse momentum spectrum of $\bar{\s} (\vec{q}_\bot)$.  

Finally, the first factor in \eqref{e:GamNew} comes from the explicit correction to the eikonal approximation, which itself comes in two pieces.  The first, as seen in \eqref{e:BroadBorn3}, is the penalty $- \frac{\vec{u}_\bot \cdot \vec{p}_\bot}{(1-u_z) E}$ for bending the high-energy jet.  The vector coupling of the jet to the medium is $2 u_\mu p^\mu$.  In the eikonal approximation, this is just $2E$, but the extra transverse momentum from the medium leads to a spacelike correction $- \frac{\vec{u}_\bot \cdot \vec{p}_\bot}{(1-u_z)E}$.  The other correction arises from the propagator $\frac{1}{(p-q)^2}$ and couples the recoil direction $\vec{q}_\bot$ to the flow.  When the momentum transfer $\vec{q}_\bot$ is parallel to the transverse flow $\vec{u}_\bot$, the scattering amplitude (and hence the probability) increases.

Together, these details comprise the effect of coupling the flow velocity $\vec{u}$ to the jet.  There will be additional subtleties when gradient corrections are considered and the averaging procedure is modified, but the basic physical mechanism is the finite energy transfer between the medium and the jet due to the flow.

%
\subsection{Linear Gradient Corrections}
%

It is instructive to illustrate the gradient effects in the simplest non-trivial limit. To first order in transverse gradients and at zero velocity, we expand the explicit spatial dependence of the density to linear order:
\begin{align}
    \rho(\vec{x}_\bot, z) \approx \rho_0(z) + \partial^j \rho(z) \: x_\bot^j \, ,
\end{align}
where we use roman superscripts for the transverse two dimensional subspace. To get a complete accounting of the gradient corrections, we also need to expand the implicit spatial dependence contained in the Debye mass $\mu \sim g T (\vec{x})$ embedded in the potential $v(q_\bot^2)$. Thus we make the comparable expansions
\begin{align}
    \mu^2 (\vec{x}_\bot, z) \approx \mu_0^2 (z) + \partial^j \mu^2(z) \: x_\bot^j  \, .   
\end{align}
The expansion of the implicit dependence in $v$ gives
\begin{subequations}
\begin{align}
    v\left(q_\bot^2, \mu^2 (\vec{x}_\bot)\right) &= v\left(q_\bot^2, \mu_0^2 + 
    \partial^j \mu^2(z)  x_\bot^j\right)
    \approx
    v\left(q_\bot^2, \mu_0^2\right) \left[ 1 + \frac{\partial^j \mu^2(z)\: x_\bot^j}{v} \frac{\partial v}{\partial \mu^2} \right] \,,
\end{align}
\end{subequations}
with
\begin{align}
    \left.\frac{\partial v}{\partial \mu^2}\right|_{\mu^2 = \mu_0^2} = 
    \frac{g^2}{(q_\bot^2 + \mu_0^2)^2} 
\end{align}
for the Gyulassy-Wang potential.

Then, the full spatial dependence of the medium variables is expanded to linear order, giving
\begin{align}
\label{e:IntByParts1}
       & \int d^2 x_\perp \,\rho(\vec{x})\,  e^{-i (\vec{q}_\bot - \vec{q}_\bot^{\: \prime}) \cdot \vec{x}_{\bot}} \,
        v\left(q_\bot^2; \mu^2(\vec{x}_\bot)\right)
        v^*\left(q_\bot^{\prime 2}; \mu^2(\vec{x}_\bot)\right)
        \notag \\ &\approx
        \int d^2 x_\perp \,\rho_0\, e^{-i (\vec{q}_\bot - \vec{q}_\bot^{\: \prime}) \cdot \vec{x}_{\bot}} \,
        v_0 (q_\bot^2)
        v_0^*(q_\bot^{\prime 2})
        \left[ 1 + \frac{1}{\rho_0}\pa^j\r\: x_\bot^j + \bigg(
        \frac{\pa^j\m^2}{v} \frac{\partial v}{\partial \mu^2} + \mathrm{c.c.} \bigg)x_\bot^j\right]
       \notag \\ &\approx
        (2\pi)^2 \, \delta^2 (\vec{q}_\bot - \vec{q}_\bot^{\: \prime} ) \:
        \Bigg\{  
        \rho_0\,v_0 (q_\bot^2) \, v_0^*(q_\bot^{\prime 2})
        \notag \\ & \hspace{2cm} 
        - i \,\rho_0\,\frac{\partial}{\partial (q - q')_\bot^j} 
        v_0 (q_\bot^2) \, v_0^*(q_\bot^{\prime 2})
        \Bigg[ \frac{1}{\rho_0}\pa^j\r
        + \bigg(
        \frac{\pa^j\m^2}{v} \frac{\partial v}{\partial \mu^2} + \mathrm{c.c.} \bigg)
        \Bigg]  \Bigg\}  \, , 
\end{align}
where the derivative in principle acts not only on the factors explicitly written here but also on the full $\vec{q}_\bot - \vec{q}_\bot^\prime$ dependence of \eqref{e:BroadBorn10}.

For these purposes it is useful to change variables in \eqref{e:BroadBorn10} from $\vec{q}_\bot , \vec{q}_\bot^\prime$ to the mean and relative coordinates
\begin{equation}
    \vec{Q}_\bot \equiv \thalf (\vec{q}_\bot + \vec{q}_\bot^\prime) \, \qquad 
    \vec{q}_{12\bot} \equiv \vec{q}_\bot - \vec{q}_\bot^\prime  \, , 
\end{equation}
giving
\begin{align}
\label{e:BroadBorn13}
    \left\langle \left| M_1 \right|^2 \right\rangle &=  
    \mathcal{C}\int d^3 x \frac{d^2q_\bot} {(2\pi)^2} \frac{d^2 q'_\bot}{(2\pi)^2} \,\rho(\vec{x})\:
    e^{-i (\vec{q}_\bot - \vec{q}_\bot^{\: \prime}) \cdot \vec{x}_{\bot}} \:
    \exp\left[ - i \: \frac{(p-q)_\bot^2 - (p-q')_\bot^2}{2 E} z \right] \:
    \notag \\ & \hspace{1cm} \times
    v (q_\bot^2) v^* (q_\bot^{\prime \, 2}) \: J(E, \vec{p}_\bot - \vec{q}_\bot) J^*(E, \vec{p}_\bot - \vec{q}_\bot^\prime)
 \notag \\ &=
    \mathcal{C} \int d^3 x \frac{d^2Q_\bot} {(2\pi)^2} \frac{d^2 q_{12\bot}}{(2\pi)^2} \,\rho(\vec{x})\:
    e^{-i \, \vec{q}_{12\bot} \cdot \vec{x}_{\bot}}
    \exp\left[ - i \: \frac{(p - Q - \thalf q_{12} )_\bot^2 - (p - Q + \thalf q_{12})_\bot^2}{2 E} z \right] \:
    \notag \\ & \hspace{1cm} \times
    v \Big( (Q + \thalf q_{12})^2 \Big)
    v \Big( (Q - \thalf q_{12})^2 \Big)
    J(E, \vec{p}_\bot - \vec{Q}_\bot - \thalf \vec{q}_{12\bot}) 
    J(E, \vec{p}_\bot - \vec{Q}_\bot + \thalf \vec{q}_{12\bot})\, ,
\end{align}
where in the last step we have used that $v$ and $J$ have at most constant imaginary phases. This is explicitly true for the Gyulassy-Wang potential, and it is also true for a tree-level 2-to-2 process for the source.

We note that the symmetry properties of many factors in the integrand mean that the derivative with respect to $q_{12}$ vanishes:
\begin{subequations}
\begin{align}
    \frac{\partial}{\partial q_{12}} \Big[ f(Q + q_{12}) + f(Q - q_{12}) \Big]_{\vec{q}_{12\perp} = 0} &= f'(Q) - f'(Q) = 0 \, , 
    \\
    \frac{\partial}{\partial q_{12}} \Big[ f(Q + q_{12}) \, f(Q - q_{12}) \Big]_{\vec{q}_{12\perp} = 0} &= 
    f'(Q) f(Q) - f(Q) f'(Q) = 0 \,  .
\end{align}
\end{subequations}
This means that the derivative does not act on the potential squared $v v^*$ or the source current squared $J J^*$; it only acts on the exponential
\begin{align}
    \left. \frac{\partial}{\partial q_{12\bot}^j} \:\left(\exp\left[ - i \: \frac{(p - Q - \thalf q_{12} )_\bot^2 - (p - Q + \thalf q_{12})_\bot^2}{2 E} z \right]\right)\right|_{\vec{q}_{12\perp} = 0} = i \frac{(p-Q)_\bot^j}{E} z \, .
\end{align}
Notice that while the leading gradient corrections are proportional to $\frac{(p-Q)^i_\perp}{E}z$, this factor is not additionally suppressed since $z$ may be large depending on the geometry of the nuclear matter.

The term which was independent of gradients was calculated previously; for the term linear in gradients, we obtain
\begin{align}
\label{e:BroadBorn14}
\left\langle \left| M_1 \right|^2 \right\rangle^{(\mathrm{linear})} &=
    \int dz\,d^2q_\bot \,\rho_0\:
    \bar{\s}_0( q_\bot^2 ) \: | J(E, \vec{p}_\bot - \vec{q}_\bot) |^2
    \left[  \frac{(p-q)_\bot^j}{E} z  \right]
    \notag \\ & \hspace{1cm}\times
    \Bigg[ \frac{1}{\rho_0}\pa^j\r
        + 2 \bigg(
        \frac{\pa^j\m^2}{v} \frac{\partial v}{\partial \mu^2} \bigg)
        \Bigg]\, ,
\end{align}
where one should notice the difference between $\bar{\s}_0$ and $\s_0$. Finally we convert to the jet multiplicity distribution
\begin{align}
\label{e:BroadBorn15}
    \left( E \frac{dN^{(1)}}{d^3 p} \right)_\mathrm{SB}^\mathrm{(linear)} &=
    \int dz \, d^2q_\bot \,\rho_0\: \bar{\sigma}_0 ( q_\bot^2 ) \: 
    \left( E \frac{dN^{(0)}}{d^2 (p-q)_\perp\, dE} \right) \:
    \left[  \frac{(p-q)_\bot^j}{E} z  \right]\notag\\
    &\hspace{1cm}\times\Bigg[ \frac{1}{\rho_0}\pa^j\r
    + \frac{\pa^j\m^2}{\bar{\sigma}_0} \frac{\partial \bar\sigma}{\partial \mu^2}
    \Bigg]\,,
\end{align}
where all of the quantities $\rho_0$,
$\mu_0^2$, $\pa_\bot^j \rho$, and $\pa_\bot^j \mu^2$ are considered functions of $z$. 

The corrections  due to an inhomogeneous but static medium,
\begin{align}
    \left[  \frac{(p-q)_\bot^j}{E} z  \right]
    \left[
        \pa^j \rho +
        \frac{\rho_0}{\bar{\sigma}_0} \frac{\partial \bar{\sigma}}{\partial \mu^2}
    \pa^j \mu^2
    \right] \, , 
\end{align}
account for either the increase in density $\rho$ or the increase in cross section due to changing Debye mass $\mu^2$ along the propagator $p-q$.  As drawn in Fig.~\ref{f:Broad1}, this describes the propagation from the source point (taken to be zero) to the point of interaction $\vec{x}$ with the medium.  Note that the ratio $(p-q)_\bot / E$ is related to an angle $\theta$ and that $\tan \theta = \Delta x_\bot / z \approx (p-q)_\bot / E$ provides the transverse displacement as a result of the nonzero angle $\theta$.  Therefore the factor $\left[ \frac{(p-q)_\bot^j}{E} z \right]$ is really the transverse displacement $\Delta x_\bot^j$ along the trajectory of $p-q$. 

If, for example,  the density increases by $d\rho = \vec{\nabla}_\bot \rho \cdot \Delta\vec{x}_\bot$ along the direction $p-q$, then this jet has an increased chance to scatter in the medium and acquire final momentum $p$.  Conversely, this means that a final state jet observed with momentum $p$ is more likely to have come from an initial jet that was moving along the direction of $\vec{\nabla}_\bot \rho$.  Although each scattering leads to isotropic broadening in this example, jets moving in this particular direction are more likely to suffer the broadening.

\section{Jet Broadening: Double Born Diagram}
\label{app:dblBorn}

Now we can turn to the double-scattering amplitude $i M_2$ which is needed to ensure unitarity of the problem. In the case of jet broadening unitarity is easy to understand -- it ensures that while the distribution of jets can change, their total number is conserved.   The corresponding diagram is sketched in Fig.~\ref{f:DblBorn} and involves two insertions of the external potential \eqref{e:potl4}:
\begin{align}
\label{e:DblBorn1}
    M_2 (p) &=
    - i \sum_{i \, j} (t^b_{proj} t^a_{proj}) t_j^b t_i^a  \: 
    \int \frac{d^4 q_1}{(2\pi)^4} \, \frac{d^4 q_2}{(2\pi)^4} \:
    e^{i q_1 \cdot x_i} \, e^{i q_2 \cdot x_j} \:
    v_i (q_1) \, v_j (q_2) \: J(p - q_1 - q_2) 
    \notag \\ &\hspace{0cm} \times
    \left[  \frac{2  u_j \cdot p}{(p-q_2)^2 + i \epsilon}  \right] 
    \left[  \frac{2 u_i \cdot (p - q_2)}{(p - q_1 - q_2)^2 + i \epsilon}  \right] 
    %
    %
    \left[ (2\pi) \, \delta\left( q_1^0 - \vec{u}_{i} \cdot \vec{q}_1 \right) \right]
    \left[  (2\pi) \, \delta\left( q_2^0 - \vec{u}_{j} \cdot \vec{q}_2 \right) \right]\,.
\end{align}
Note that, as discussed in \eqref{e:Ljet}, we only consider two sequential insertions of the external potential, neglecting the ``seagull'' diagrams particular to scalar QCD.  Like the contributions of the 4-gluon vertex in real QCD, these contributions would be the same order in the coupling but without the phase-space enhancement associated with having a propagator in between them.

Because both insertions of the potential occur at the amplitude level, it is trivial to multiply by the complex conjugate of the unmodified amplitude $M_0^* (p) = i\,\,J(p)$ and perform the averaging over quantum numbers.  The color averaging sets $i=j$ following \eqref{e:Avg}, giving
\begin{align}
\label{e:DblBorn2}
    \left\langle M_2 M_0^* \right\rangle &= 
    \mathcal{C} \sum_i  \: 
    \int \frac{d^4 q_1}{(2\pi)^4} \, \frac{d^4 q_2}{(2\pi)^4} \:
    e^{i (q_1 + q_2) \cdot x_i} \: v_i (q_1) \, v_i (q_2) \, J(p - q_1 - q_2) J^* (p)
    \notag \\ &\hspace{0cm} \times
    \left[  \frac{2  u_i \cdot p}{(p-q_2)^2 + i \epsilon}  \right] 
    \left[  \frac{2 u_i \cdot p}{(p - q_1 - q_2)^2 + i \epsilon}  \right] 
    %
    %
    \left[ (2\pi) \, \delta\left( q_1^0 - \vec{u}_{i} \cdot \vec{q}_1 \right) \right]
    \left[  (2\pi) \, \delta\left( q_2^0 - \vec{u}_{i} \cdot \vec{q}_2 \right) \right]  
    \notag \\ &=
    \mathcal{C} \sum_i  \:
    \int \frac{d^4 q_1}{(2\pi)^4} \, \frac{d^4 q_2}{(2\pi)^4} \:
    e^{i (q_1 + q_2) \cdot x_i} \: v_i (q_1) \, v_i (q_2)
    \notag \\ &\hspace{1cm} \times
    \left[  \frac{2E (1-u_{i z})}{(p-q_2)^2 + i \epsilon}  \right] 
    \left[  \frac{2E (1-u_{i z})}{(p - q_1 - q_2)^2 + i \epsilon}  \right] 
    J(p - q_1 - q_2) J^* (p) 
    \notag \\ &\hspace{1cm} \times
    \left[ (2\pi) \, \delta\left( q_1^0 - \vec{u}_{i} \cdot \vec{q}_1 \right) \right]
    \left[  (2\pi) \, \delta\left( q_2^0 - \vec{u}_{i} \cdot \vec{q}_2 \right) \right]  
        \left[ 
        1 - 2 \frac{\vec{u}_{i \, \bot} \cdot \vec{p}_\bot}{E (1-u_{i z})}
        \right] \, .
\end{align}
The poles of both propagators $(p-q_2)^2 + i \epsilon$ and $(p-q_1-q_2)^2 + i \epsilon$ are those given in \eqref{e:BroadPoles1} for $q \rightarrow q_2$ and $q \rightarrow q_1 + q_2$, respectively:
\begin{subequations}
\label{e:DblBorn3}
\begin{align}
    (p-q_2)^2 + i \epsilon &= 
    -( 1-u_{i \, z}^2 )  \left[ q_{2 \, z} - Q_{p-q_2}^+ - i \epsilon \right]
    \left[ q_{2 \, z} - Q_{p-q_2}^- + i \epsilon \right] \, ,
    \\
    (p-q_1-q_2)^2 + i \epsilon &= 
    -( 1-u_{i \, z}^2 )  \left[ q_{1 \, z} + q_{2 \, z} - Q_{p-q_1-q_2}^+ - i \epsilon \right] 
    \notag \\ &\hspace{1cm} \times
    \left[ q_{1 \, z} + q_{2 \, z} - Q_{p-q_1-q_2}^- + i \epsilon \right]\,.
\end{align}
\end{subequations}
After taking the $q_0$ integrals, we can immediately collect the residue of $q_{1 \, z}$ since it only enters into $(p - q_1 - q_2)^2$.  The dominant contribution is the pole $q_{1 \, z} = - q_{2 \, z} + Q_{p-q_1-q_2}^- - i \epsilon$, which we can enclose below the real axis for $z_i > 0$ and find
\begin{align}
\label{e:DblBorn4}
    \left\langle M_2 M_0^* \right\rangle &= 
    i\mathcal{C} \sum_i  \:
    \int \frac{d^2 q_{1\bot}}{(2\pi)^2} \, \frac{d^3 q_2}{(2\pi)^3} \:
    e^{i (q_{1 \, i} + q_{2 \, i}) \cdot x_i} \: v_i (q_{1 \, i}) \, v_i (q_{2\,i}) \, J(p - q_{1 \, i} - q_{2 \, i}) J^* (p)
    \notag \\ &\hspace{1cm} \times
    \left[1 - 2 \frac{\vec{u}_{i \, \bot} \cdot \vec{p}_\bot}{E (1-u_{i z})}\right]
    \left[\frac{2 E}{1 + u_{i z}} \:
    \frac{1}{Q_{p-q_1-q_2}^- - Q_{p-q_1-q_2}^+  }
    \right] 
    \notag \\ &\hspace{1cm} \times
    \left[ - \frac{2 E}{1+u_{i z}} \, \frac{1}{\left[ q_{2 \, z} - Q_{p-q_2}^+ - i \epsilon \right]
    \left[ q_{2 \, z} - Q_{p-q_2}^- + i \epsilon \right]}  \right] 
    \notag \\ &
    =
    \mathcal{C} \sum_i  \:
    \int \frac{d^2 q_{1\bot}}{(2\pi)^2} \, \frac{d^3 q_2}{(2\pi)^3} \:
    e^{i (q_{1 \, i} + q_{2 \, i}) \cdot x_i} \: v_i (q_{1 \, i}) \, v_i (q_{2\,i}) \, J(p - q_{1 \, i} - q_{2 \, i}) J^* (p)
    \notag \\ &\hspace{1cm} \times
        (-i)  \left[1 - 2 \frac{\vec{u}_{i \, \bot} \cdot \vec{p}_\bot}{E (1-u_{i z})}\right]
        \left[ 1 - \frac{\vec{u}_{i \, \bot} \cdot (\vec{q}_{1\bot} + \vec{q}_{2\bot})}{E (1 - u_{i \, z})}
        \right]^{-1} 
    \notag \\ &\hspace{1cm} \times
    \left[ - \frac{2 E}{1+u_{i z}} \, \frac{1}{\left[ q_{2 \, z} - Q_{p-q_2}^+ - i \epsilon \right]
    \left[ q_{2 \, z} - Q_{p-q_2}^-  + i \epsilon \right]}  \right]\, ,
\end{align}
where we have expanded $\frac{1}{Q_{p-q_1-q_2}^- - Q_{p-q_1-q_2}^+}$ in powers of $1/E$.  Note that the Fourier exponent has lost the ability to constrain the direction of closure of the $q_{2 \, z}$ integral:
\begin{align}
    e^{i (q_{1 \, i} + q_{2 \, i}) \cdot x_i} &= 
    e^{- i (\vec{q}_{1\bot} + \vec{q}_{2\bot}) \cdot \vec{x}_{i \, \bot}} \,
    \exp\left[-i\:\frac{\vec{u}_{i\,\bot}\cdot(\vec{q}_{1\bot}+\vec{q}_{2\bot})}{1-u_{i z}}z_i-i\:\frac{(p - q_1 - q_2)_\bot^2 - p_\bot^2}{2 E (1-u_{i z})}z_i \right] \, , 
\end{align}
just as in \eqref{e:BBphase1}.  However, including the $q_{2 \, z}$ dependence in the potential, there is more than enough convergence of the $q_{2 \, z}$ integral to close the contour at infinity in either direction and perform the integral by residues.  The various poles of $q_{2 \, z}$ are: 
\begin{subequations}
\begin{align}
    v (q_1): \qquad & \qquad
    \mathcal{P}_1^\pm \equiv 
    Q^-_{p-q_1-q_2} - \frac{u_z}{1-u_z^2} (\vec{u}_\bot \cdot \vec{q}_{1\bot}) 
    \pm \frac{i}{1-u_z^2} R_1  \,  ,
    \\ \label{e:potlpoles1}
    v (q_{2}): \qquad & \qquad
    \mathcal{P}_2^\pm \equiv 
    \frac{u_z}{1-u_z^2} (\vec{u}_\bot \cdot \vec{q}_{2\bot}) 
    \pm \frac{i}{1-u_z^2} R_2 \,  ,
    \\
    \mathrm{Propagators:} \qquad & \qquad
    \mathcal{P}_3 \equiv Q_{p-q_2}^+ + i \epsilon  \,  ,
    \\ & \qquad
    \mathcal{P}_4 \equiv Q_{p-q_2}^- - i \epsilon  \,  ,
\end{align}
\end{subequations}
where we have introduced the shorthand
\begin{align}
    \label{e:Rdef}
    R^2 &\equiv (1-u_z^2)(q_{\perp}^2 + \mu^2) - \left(\vec{u}_\perp\cdot\vec{q}_{\perp}\right)^2 > 0\,,
\end{align}
and suppressed the source index for brevity.
The fact that $R^2 > 0$ can be seen directly by explicitly introducing the polar angle $\theta$ of $\vec{u}$ with respect to the $z-$axis and the azimuthal angle $\phi$ between $\vec{u}_\bot$ and $\vec{q}_\bot$:
\begin{align}
    R^2 &= (1-u_z^2) q_\bot^2 - (\vec{u}_\perp\cdot\vec{q}_\perp)^2 + (1-u_z^2) \mu^2
    \notag \\ &=
    q_\bot^2 \left[
    1 - \vec{u}^2 \left( \cos^2\theta + \sin^2\theta\cos^2\phi \right)
    \right]  + (1-u_z^2) \mu^2 \, .
\end{align}
Since $\cos^2\phi \leq 1$, we have $(\cos^2\theta + \sin^2\theta\cos^2\phi) \leq (\cos^2\theta + \sin^2\theta) = 1$.  Moreover, we also have the relativistic constraint $\vec{u}^2 < 1$, such that $\left[    1 - \vec{u}^2 \left( \cos^2\theta + \sin^2\theta\cos^2\phi \right)   \right] > 0$ and hence $R^2 > 0$.  Knowing the real and imaginary parts of the poles will help considerably in combining the double-Born diagram with its complex conjugate. 

Let us now consider the corresponding $q_{2 \, z}$ integral in (\ref{e:DblBorn4}) which can be written as
\begin{align}
    \mathcal{I} &\equiv \int\frac{dq_{2 \, z}}{2\pi} \, v (q_{1}) \, v (q_{2}) \,
    \frac{1}{q_{2 \, z} - Q_z^+ (q_2) - i \epsilon} \, 
    \frac{1}{q_{2 \, z} - Q_z^- (q_2) + i \epsilon}
    \notag \\ &=
    \frac{g^4}{(1-u_z^2)^2} \int\frac{dq_{2 \, z}}{2\pi} \, 
    \frac{1}{q_{2 \, z} - \mathcal{P}_1^+} \, 
    \frac{1}{q_{2 \, z} - \mathcal{P}_1^-} \, 
    \frac{1}{q_{2 \, z} - \mathcal{P}_2^+} \, 
    \frac{1}{q_{2 \, z} - \mathcal{P}_2^-} \, 
    \frac{1}{q_{2 \, z} - \mathcal{P}_3} \, 
    \frac{1}{q_{2 \, z} - \mathcal{P}_4} \, , 
\end{align}
with
\begin{align}   \label{e:DblBorn5}
    \text{Re}\,\,\mathcal{I} &=\frac{1+u_z}{2 E} v(q_{1\bot}^2) v(q_{2\bot}^2)
    \frac{
    R_1 (\vec{u}_\bot \cdot \vec{q}_{2\bot})
    \Big( 2 R_1 R_2 + (1-u_z^2)(q_{1\bot}^2 + \mu^2) \Big)
    }{
    2 R_1 R_2 \left[ (R_1 + R_2)^2 + \Big(\vec{u}_\bot \cdot (\vec{q}_{1\bot} + \vec{q}_{2\bot}) \Big)^2 \right] ,
    }
    \notag\\
    &\hspace{0.5cm} - (1 \leftrightarrow 2)+\mathcal{O}\left(\frac{\perp^2}{E^2}\right)\, .
\end{align}
and
\begin{align} 
\label{e:DblBorn5prime}
    \text{Im}\,\,\mathcal{I} &=\frac{1+u_z}{4 E} v(q_{1\bot}^2) v(q_{2\bot}^2)+\mathcal{O}\left(\frac{\perp^2}{E^2}\right) \, .
\end{align}
Several comments are in order about Eqs.~\eqref{e:DblBorn5} and \eqref{e:DblBorn5prime}.  As a first check, note that in the limit $\vec{u} \rightarrow 0$, the real part vanishes at the leading order and the contact integral gives its usual value
\begin{align}
    \mathcal{I} \rightarrow \frac{1}{2 E} \left( \frac{i}{2} \right) v(q_{1\bot}^2) v(q_{2\bot}^2)+\mathcal{O}\left(\frac{\perp^2}{E^2}\right) \, .
\end{align}
While the dominant imaginary part is unmodified from the static case, now for the first time a nonzero real part is generated as well. At the leading order in the eikonal expansion, the imaginary part is symmetric under $q_1 \leftrightarrow q_2$ (and hence $R_1 \leftrightarrow R_2$) and the real part is explicitly antisymmetric.  The existence of this term will complicate the interference with the complex conjugate amplitudes $\langle M_2 M_0^* \rangle + \langle M_0 M_2^* \rangle$, and appears to lead to dramatically new phase structures in the LPM interference pattern.  Note also that the particular form of the real part \eqref{e:DblBorn5} is specific to the Gyulassy-Wang potential \eqref{e:potl5}.

With the result \eqref{e:DblBorn5} we return to the contribution to the cross section:
\begin{align}   
\label{e:DblBorn6}
    \left\langle M_2 M_0^* \right\rangle &= 
    \mathcal{C} 
    \int d^3 x \, \rho(\vec{x})
    \int \frac{d^2 q_{1\bot}}{(2\pi)^2} \, \frac{d^2 q_{2\bot}}{(2\pi)^2} \:
    e^{- i (\vec{q}_{1\bot} + \vec{q}_{2\bot}) \cdot \vec{x}_{\bot}} \,
    e^{-i\:\frac{\vec{u}_\bot\cdot(\vec{q}_{1\bot}+\vec{q}_{2\bot})}{1-u_z}z-i\:\frac{(p - q_1 - q_2)_\bot^2 - p_\bot^2}{2 E (1-u_z)}z} \,
    \notag \\ &\hspace{1cm}\times
    J(p - q_{1} - q_{2}) J^* (p)
    \frac{2 E}{1+u_{z}}\left[ 1 - \frac{2\vec{u}_\perp\cdot \left(\vec{p}_\perp -
    \thalf(\vec{q}_{1\perp} + \vec{q}_{2\perp})\right)}
    {E (1 - u_{z})} \right] 
    \,i\,\mathcal{I}\,,
\end{align}
where we have converted from a discrete summation to a continuous integral.  For the source current $J$ we have (in Minkowski coordinates)
\begin{align}
    \hspace{-0.5cm}(p - q_{1} - q_{2})^\mu &=
    \Bigg(
        E - \frac{\vec{u}_\perp\cdot(\vec{q}_{1\perp} + \vec{q}_{2 \perp})}{1-u_z}        
        \:\: , \:\:
        \vec{p}_\bot - \vec{q}_{1\bot} - \vec{q}_{2\bot}
        \:\: , \:\:
        E - \frac{\vec{u}_\perp\cdot(\vec{q}_{1\perp} + \vec{q}_{2\perp})}{1-u_z}
    \Bigg) \, , 
\end{align}
so that we can write $J(p-q_{1} - q_{2}) = J\left(E - \frac{\vec{u}_\perp\cdot(\vec{q}_{1\perp} + \vec{q}_{2 \perp})}{1-u_z} \: , \: \vec{p}_\bot - \vec{q}_{1\bot} - \vec{q}_{2\bot} \right)$. Taylor expanding the shift in 
momentum, 
\begin{align}
    J\left(E - \frac{\vec{u}_\perp\cdot(\vec{q}_{1\perp} + \vec{q}_{2 \perp})}{1-u_z} \: , \: \vec{p}_\bot - \vec{q}_{1\bot} - \vec{q}_{2\bot} \right) & =
    J(E  \: , \: \vec{p}_\bot - \vec{q}_{1\bot} - \vec{q}_{2\bot})
    \notag \\ &
   \times \left[
        1 - \frac{\vec{u}_\perp\cdot(\vec{q}_{1\perp} + \vec{q}_{2\perp})}{1-u_z} \, \frac{1}{J} \frac{\partial J}{\partial E}
    \right] \;, 
\end{align}
gives the complete expression
\begin{align}   
\label{e:DblBorn7}
    \left\langle M_2 M_0^* \right\rangle &= 
    \mathcal{C}  
    \int d^3 x \, \rho(\vec{x})
    \int \frac{d^2 q_{1\bot}}{(2\pi)^2} \, \frac{d^2 q_{2\bot}}{(2\pi)^2} \:
    e^{- i (\vec{q}_{1\bot} + \vec{q}_{2\bot}) \cdot \vec{x}_{\bot}} \,
    e^{-i\:\frac{\vec{u}_\bot\cdot(\vec{q}_{1\bot}+\vec{q}_{2\bot})}{1-u_z}z-i\:\frac{(p - q_1 - q_2)_\bot^2 - p_\bot^2}{2 E (1-u_z)}z} \,
    \notag \\ &\hspace{1cm}\times
    J(E  \: , \: \vec{p}_\bot - \vec{q}_{1\bot} - \vec{q}_{2\bot})
    J^* (E \: , \: \vec{p}_\bot) \frac{2 E}{1+u_{i z}} \:
   \,i\,\mathcal{I}
    \notag \\ &\hspace{1cm}\times
    \left[ 
        1 
        -\frac{2\vec{u}_\perp\cdot\left(\vec{p}_\perp - \thalf(\vec{q}_{1\perp} + \vec{q}_{2\perp} )\right)}{E (1 - u_{z})} 
        - \frac{\vec{u}_\perp\cdot(\vec{q}_{1\perp} + \vec{q}_{2\perp})}{1-u_z} \, \frac{1}{J} \frac{\partial J}{\partial E}
    \right] \, .
\end{align}
Now we may add the complex conjugate term to obtain
\begin{align}   
\label{e:DblBorn8}
    \left\langle M_2 M_0^* \right\rangle + \mathrm{c.c.}&= 
    i\mathcal{C}  
    \int d^3 x \, \rho(\vec{x})
    \int \frac{d^2 q_{1\bot}}{(2\pi)^2} \, \frac{d^2 q_{2\bot}}{(2\pi)^2} \: \frac{2 E}{1+u_{i z}}J^*(E  \: , \: \vec{p}_\bot - \vec{q}_{1\bot} - \vec{q}_{2\bot})
    J (E \: , \: \vec{p}_\bot)
    \notag \\ &\hspace{1cm}\times
    \left[ 
        1 
        -\frac{2\vec{u}_\perp\cdot\left(\vec{p}_\perp - \thalf(\vec{q}_{1\perp} + \vec{q}_{2\perp} )\right)}{E (1 - u_{z})} 
        - \frac{\vec{u}_\perp\cdot(\vec{q}_{1\perp} + \vec{q}_{2\perp} )}{1-u_z} \, \frac{1}{J} \frac{\partial J}{\partial E}
    \right]
    \notag \\ &\hspace{1cm}\times
    \Bigg\{
    e^{- i (\vec{q}_{1\bot} + \vec{q}_{2\bot}) \cdot \vec{x}_{\bot}} \,
    e^{-i\:\frac{\vec{u}_\bot\cdot(\vec{q}_{1\bot}+\vec{q}_{2\bot})}{1-u_z}z-i\:\frac{(p - q_1 - q_2)_\bot^2 - p_\bot^2}{2 E (1-u_z)}z}\,\mathcal{I}-c.c.\Bigg\}\,,
\end{align}
where we have again assumed that $J^*(E  \: , \: \vec{p}_\bot - \vec{q}_{1\bot} - \vec{q}_{2\bot})J (E \: , \: \vec{p}_\bot)=J(E  \: , \: \vec{p}_\bot - \vec{q}_{1\bot} - \vec{q}_{2\bot})J^* (E \: , \: \vec{p}_\bot)$. This is about as far as we can go without next invoking the gradient expansion.

%
\subsection{No Gradients (Translational Invariance)}
%

If transverse gradients of all quantities can be neglected, then we can immediately perform the $d^2 x_\bot$ integral to obtain $\delta^2 (\vec{q}_{1\bot} + \vec{q}_{2\bot})$. We also need to use the explicit form of $\text{Im}\,\,\mathcal{I}$ at $\vec{q}_{1\bot} + \vec{q}_{2\bot}=0$ while $\text{Re}\,\,\mathcal{I}$ cancels between the conjugated parts of the full amplitude. Thus, noticing that
$$
\text{Im}\,\,\mathcal{I}|_{q_{2\bot}=-q_{1\bot}}=\frac{1+u_z}{4 E} \left(v(q_{1\,\perp}^2)\right)^2-\frac{1+u_z}{1-u_z}(\vec{u}_\perp\cdot \vec{q}_{1\perp})\frac{1-2 \left((p+q_1)_\bot^2-p_\bot^2\right)v(q_{1\,\perp}^2)}{4E^2}\left(v(q_{1\,\perp}^2)\right)^2 \, , 
$$
we write
\begin{align}   
\label{e:DblBorn9}
    \left\langle M_2 M_0^* \right\rangle + \mathrm{c.c.}&= 
    - \mathcal{C} 
    \int d z \, \rho(z)
  \int \frac{d^2q_\bot} {(2\pi)^2} \:
    [v( q_{\bot}^2 )]^2  \: \left| J(E  \: , \: \vec{p}_\bot) \right|^2\notag\\
    &\hspace{-0.5cm}\times\Big[ 1 - \frac{2\,\vec{u}_\perp\cdot\vec{p}_\perp }{E (1 - u_{z})}-\frac{(\vec{u}_\perp\cdot \vec{q}_{\perp})\left(1-2\left((p+q)_\bot^2-p_\bot^2\right)v(q_\perp^2)\right)}{E(1-u_z)}\Big]  + \mathcal{O}\left(\pa_\bot\right) \, .
\end{align}
The angular structure of the integrand in the transverse plane can be simplified through the angular averaging. After that one may rewrite this expression in the form of (\ref{e:BroadBorn12}) introducing similar notations
\begin{align}
\label{e:DblBorn10}
    \hspace{-0.5cm}&\left( E \frac{dN^{(1)}}{d^3 p} \right)_\mathrm{DB} = 
    -\int dz \, \rho(z) \int d^2q_\bot  \bar{\s}(q_\bot^2) \left( E \frac{dN^{(0)}}{d^2 p_\perp\,dE} \right) 
    \Big[1 + \vec{u}_\perp\cdot\vec{\Gamma}_{DB}
    \Big]  + \mathcal{O}\left(\pa_\bot\right)\, , 
\end{align}
where $\vec{\Gamma}_{DB}$ is explicitly given in (\ref{e:GamDB}). The effects of the medium motion on the jet broadening are again intrinsically related to the non-zero energy transfer $q^0=\vec{u}_\perp\cdot\vec{q}_\perp$ and the subeikonal corrections to the vertices as in the case of the single Born diagram.

%
\subsection{Linear Gradient Corrections}
%

The gradient corrections to the double Born contribution are considerably more involved since now the transverse coordinate dependence will transform to the momentum derivative $\frac{\pa}{\pa Q}_{i\perp}$ which results in many more terms. While the expression (\ref{e:DblBorn8}) can be used to derive the general answer, here we will omit the corrections which are  simultaneously sub-eikonal and gradient suppressed, and also set velocity to zero. Then, to first order in transverse gradients, we expand the spatial dependence as for the single-Born diagram.

Keeping only the leading eikonal terms in (\ref{e:DblBorn8}) which are not suppressed by powers of $E$ but allowing the hydrodynamic parameters to vary in space and time we start with
\begin{align}   
    &\left\langle M_2 M_0^* \right\rangle + \mathrm{c.c.}= 
    i\mathcal{C} 
    \int d^3 x \, \rho(\vec{x})
    \int \frac{d^2 q_{1\bot}}{(2\pi)^2} \, \frac{d^2 q_{2\bot}}{(2\pi)^2} \: \frac{2 E}{1+u_{z}}\Bigg\{
    e^{- i (\vec{q}_{1\bot} + \vec{q}_{2\bot}) \cdot \vec{x}_{\bot}} \,
    e^{-i \left[ \frac{(p - q_1 - q_2)_\bot^2 - p_\bot^2}{2 E} \right] z}\,\mathcal{I}
    \notag \\ 
    &\hspace{0cm}-e^{ i (\vec{q}_{1\bot} + \vec{q}_{2\bot}) \cdot \vec{x}_{\bot}} \,
    e^{i \left[ \frac{(p - q_1 - q_2)_\bot^2 - p_\bot^2}{2 E} \right] z}\,\mathcal{I}^*     
    \Bigg\}J^*(E  \: , \: \vec{p}_\bot - \vec{q}_{1\bot} - \vec{q}_{2\bot})J (E \: , \: \vec{p}_\bot)+\mathcal{O}\left(\frac{\perp}{E}\right) \,.
\end{align}
The transverse integral appearing from the gradient terms can again be performed through integration by parts:
\begin{align}
    \int d^2 x_\perp \, e^{- 2i\vec{Q}_{\bot} \cdot \vec{x}_\bot} \, x_\bot^j \, f(\vec{Q}_{\bot}, \vec{q}_{12\bot})
    & =
    \int d^2 x_\perp \, 
    \left[ \frac{i}{2} \frac{\partial}{\partial Q_\bot^j} \,
        e^{- 2 i \vec{Q}_{\bot}\cdot \vec{x}_\bot} 
    \right] \,  f(\vec{Q}_{\bot}, \vec{q}_{12\bot})
    \notag \\ & =
    \frac{(2\pi)^2}{2} \delta^{(2)} (2\vec{Q}_\bot)
    \left[ - i \frac{\partial}{\partial Q_\bot^j} \,
         f(\vec{Q}_{\bot}, \vec{q}_{12\bot})
\right]_{\vec{Q}_\perp = 0} ,
\end{align}
with the c.c. term $e^{i (\vec{q}_{1\bot} + \vec{q}_{2\bot}) \cdot \vec{x}_\bot}$ differing by a minus sign.  Let us illustrate that on the example of the gradient of $\r$ which up to $1/E$ corrections  contributes
\begin{align}   
    &\hspace{-0.4cm}\left\langle M_2 M_0^* \right\rangle^{(\text{linear})}_{\r} + \mathrm{c.c.}= 
    \mathcal{C} 
    \int dz\, \int \frac{d^2Q_\bot d^2 q_{12\bot}}{2(2\pi)^2}\,\frac{2\, E}{1+u_{z0}}\,\pa^j\r\,\d(2\vec{Q}_\perp)
    \notag \\
    &\hspace{-0.4cm}\frac{\pa}{\pa Q^j}\left[\Bigg\{e^{-i \left[ \frac{(p -2Q)_\bot^2 - p_\bot^2}{2 E} \right] z}
    \,\mathcal{I}_0+e^{i \left[ \frac{(p - 2Q)_\bot^2 - p_\bot^2}{2 E} \right] z} \,\mathcal{I}^*_0 \Bigg\}J^*(E  \: , \: \vec{p}_\bot - 2\vec{Q}_\bot)J (E \: , \: \vec{p}_\bot)\right]\,,
\end{align}
where $\mathcal{I}_0$ is the value of $\mathcal{I}$ when the gradients are neglected in its definition. One should notice now that the leading eikonal part of $\mathcal{I}_0$ satisfies 
\bea
&\mathcal{I}_0 (q_{12}, Q)=\mathcal{I}_0 (q_{12}, -Q) \, ,   \qquad 
\frac{\pa}{\pa Q^j}\mathcal{I}_0 \Big|_{\vec{Q}_\bot=0}=0   \, .
\eea
The momentum derivative gives a non-zero result only when it acts on the source function or the LPM phase. However $\mathrm{Re} \, \mathcal{I}_0 = 0$ for $\vec{u}=0$; thus, the only non-trivial contribution comes from the term with the momentum derivative acting on the LPM phase giving
\begin{align}   
    &\left\langle M_2 M_0^* \right\rangle^{(\text{linear})}_{\r} + \mathrm{c.c.}\simeq 
    -\mathcal{C}  
    \int dz\, \int \frac{d^2q_\bot}{(2\pi)^2}\,\pa^j\r\,
    \left(\frac{p^j}{E}z\right)\left[v(q_\bot^2)\right]^2|J(E  \: , \: \vec{p}_\bot)|^2\, .
\end{align}
For the contribution due to the gradient of $\m^2$ we write
\begin{align}  
\label{e:DblBorn11}
    &\hspace{-0.5cm}\left\langle M_2 M_0^* \right\rangle^{(\text{linear})}_{\m^2} + \mathrm{c.c.}= 
    \mathcal{C}  
    \int dz\, \int \frac{d^2Q_\bot d^2 q_{12\bot}}{2(2\pi)^2}\,\frac{2\, E}{1+u_{z0}}\,\r\,\pa^j\m^2\,\d(2\vec{Q}_\perp)
    \notag \\
    &\hspace{-0.5cm}\frac{\pa}{\pa Q^j}\left[\Bigg\{e^{-i \left[ \frac{(p -2Q)_\bot^2 - p_\bot^2}{2 E} \right] z}
    \,\frac{\pa\mathcal{I}_0}{\pa\m^2}
    +c.c.
    \Bigg\}J^*(E  \: , \: \vec{p}_\bot - 2\vec{Q}_\bot)J (E \: , \: \vec{p}_\bot)\right]+\mathcal{O}\left(\frac{\perp}{E}\right)\, .
\end{align}
Unless the derivative acts on the LPM phase, the integrand is proportional to $\frac{\pa}{\pa\m^2}\mathrm{Re}\,\mathcal{I}_0$ or its momentum derivative which vanish for $\vec{u} = 0$. Thus, (\ref{e:DblBorn8}) can be simplified to
\begin{align}
    &\left\langle M_2 M_0^* \right\rangle^{(\text{linear})}_{\m^2} + \mathrm{c.c.}\simeq 
    -\mathcal{C} 
    \int dz\, \int \frac{d^2q_\bot} {(2\pi)^2}\,\r\,\pa^j\m^2\,|J(E  \: , \: \vec{p}_\bot)|^2\left[\frac{p_\perp^j}{E}z\right]\frac{\pa}{\pa\m^2}\left(v(q_{T}^2)\right)^2\, .
\end{align}
Similarly, the contributions due to the gradients of $\vec{u}_\perp$ and $u_z$ are zero since $\frac{2\, E}{1+u_{z}}\text{Re}\,\,\mathcal{I}$ gives zero contribution to the result while $\frac{2\, E}{1+u_{z}}\text{Im}\,\,\mathcal{I}$ is independent of $\vec{u}$. Thus, the full leading eikonal gradient correction to the double Born contribution reads
\begin{align}
\label{e:DblBorn12}
    &\hspace{-0.4cm} &\left( E \frac{dN^{(1)}}{d^3 p} \right)_\mathrm{DB}^{(\text{linear})}\simeq 
    \int dz\int d^2 q_\bot{\bar{\s}}_0(q^2_\perp)\left( E \frac{dN^{(0)}}{d^2 p_\perp dE} \right)\left[\frac{p_\perp^j}{E}z\right]
    \left(\pa^j\r+\r_0\,\frac{1}{\bar{\s}_0}\frac{\pa \bar{\s}}{\pa\m^2}\,\pa^j\m^2\right)\, .
\end{align}

\section{In-Medium Branching: Single Born Diagrams}
\label{sec:BRsnglBorn}

At $\ord{g}$ in the external potential, there are three diagrams contributing to the radiative branching of the scalar jet, which we denote $R_1^A$, $R_1^B$, and $R_1^C$ and compute in the sections below.

\subsection{Diagram A}


%
\begin{figure}
    \centering
	\includegraphics[width=0.6\textwidth]{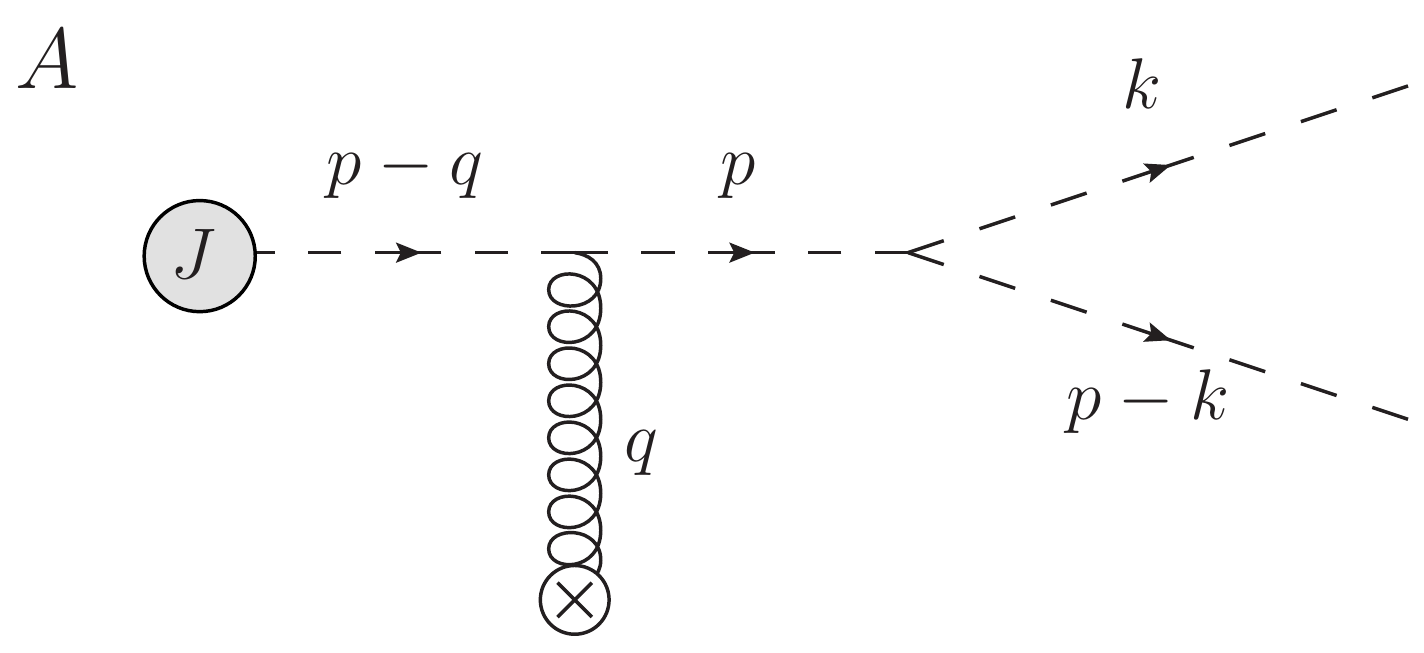}	
	\caption{The initial-state scattering diagram denoted $R_1^A$.} 
	\label{f:Born_Rad_A}
\end{figure}
%

The diagram denoted $R_1^A$ shown in Fig.~\ref{f:Born_Rad_A} corresponds to a final-state branching that occurs after the scattering.  Using the external potential given in Eqs.~\eqref{f:Potential} and \eqref{e:potl4} gives
\bea
R_{1,A}=i \sum_i \mathcal{C}_A^a t^a_i \int\frac{d^4 q}{(2\p)^4}e^{iq\cdot x_i}\,\frac{\l\,u_i^\m(2p-q)_\m}{[p^2+i\e][(p-q)^2+i\e]}\,v_i(q)J(p-q)\,(2\p)\delta\left(q_0-\vec{u}_i \cdot \vec{q}\right) , \notag 
\eea
where $\mathcal{C}_A^a$ is a generic color matrix for the projectile, associated with diagram $A$.  We write the target color matrix $t^a_i$ explicitly and will perform the target color averaging momentarily.  Evaluating the scalar-boson vertex to $\ord{\tfrac{\bot}{E}}$ gives
\bea
R_{1,A}=i \sum_i \mathcal{C}_A^a t^a_i \, 
\int\frac{d^3 q}{(2\p)^3} e^{iq\cdot x_i}\,\frac{2E\l\,(1-u_{i\,z})\left[1-\frac{\vec{u}_{i\,\perp}\cdot \vec{p}_\perp}{E(1-u_{i\,z})}\right]}{[p^2+i\e][(p-q)^2+i\e]}\,v_i(q_i)J(p-q_i) ,
\eea
and for the final-state branching diagram $A$, the kinematics \eqref{e:kin1} of the splitting are unmodified from the vacuum case:
\bea
&&p^2=\frac{\left[k-xp\right]_\perp^2}{(1-x)x}\, , 
\eea 
which can be identified with the scalar branching light-front wave function \eqref{e:scalvac1}.  

Next we evaluate the pole structure of the $q$ integral; the propagator $(p-q)$ can be expressed as
\bea \label{e:prop1}
(p-q)^2=-( 1-u_{i \, z}^2 )  \left[ q_z - Q_{p-q}^+ - i \epsilon \right]
    \left[ q_z - Q_{p-q}^- + i \epsilon \right] \, , 
\eea 
where the large and small poles $Q^+$ and $Q^-$ are given by
\begin{subequations} \label{e:rpoles1}
\begin{align}
Q_{p-q}^{+} &\simeq \frac{2E}{1+u_{i\,z}}\left(1-\frac{\vec{u}_{i\,\perp}\cdot\vec{q}_\perp}{2E}\right) \, , 
\\
Q_{p-q}^{-} &\simeq \frac{\vec{u}_{i\,\perp}\cdot\vec{q}_\perp}{1-u_{i\,z}}-\frac{x(p-k)_\perp^2+(1-x)k_\perp^2-x(1-x)(p-q)_\perp^2}{2x(1-x)(1-u_{i\,z})E}
\notag \\ &=
\frac{\vec{u}_{i\,\bot} \cdot \vec{q}_\bot}{1 - u_{i\,z}} - \frac{1}{1-u_{i\,z}} \left(
\frac{(k - x p)_\bot^2}{2 x (1-x) E} - \frac{(p-q)_\bot^2 - p_\bot^2}{2 E} \right) \, , 
\end{align}
\end{subequations}
with the difference generated from the residue 
\bea \label{e:edenom1}
&&Q_{p-q}^{+}-Q_{p-q}^{-}\simeq \frac{2E}{1+u_{i\,z}}\left(1-\frac{\vec{u}_{i\,\perp}\cdot\vec{q}_\perp}{(1-u_{i\,z})E}\right) ,
\eea
all evaluated to first sub-eikonal accuracy.  Noting that the residue of the large pole $Q^+$ is highly suppressed, we collect the residue of the small pole $Q^-$ for $z_i > 0$ to obtain
\begin{align}
R_{1,A} &= \sum_i \mathcal{C}_A^a t^a_i \: \theta(z_i) \int\frac{d^2 q_\perp}{(2\p)^2}e^{iq\cdot x_i}\,\left[1-\frac{\vec{u}_{i\,\perp}\cdot (\vec{p}-\vec{q})_\perp}{E(1-u_{i\,z})}\right]\,\frac{\lambda\,x(1-x)}{\left[k-xp\right]_\perp^2}\,v_i(q_i)J(p-q_i)
\notag \\ &=
- i \sum_i \mathcal{C}_A^a t^a_i \: \theta(z_i) \, 
\psi(x, \vec{k}_\bot - x \vec{p}_\bot ) \,  \int\frac{d^2 q_\perp}{(2\p)^2}e^{iq\cdot x_i} \, 
\, v_i(q_i) J(p-q_i) \left[1 - \frac{\vec{u}_{i\,\perp}\cdot(\vec{p}-\vec{q})_\perp}{E(1-u_{i\,z})}
\right]\,.
\end{align}
Here the potential $v(q_i)$ and the source current $J(p-q_i)$ depend indirectly on the velocity $\vec{u}_i$ through the pole value of the momentum
\begin{align}
    q_i^\mu = \left( \vec{u}_{i\,\bot} \cdot \vec{q}_\bot + u_{i\,z} Q_{p-q}^- \, , \,
    \vec{q}_\bot \, , \, Q_{p-q}^- \right) .
\end{align}
The Fourier exponent, in turn, reads
\begin{align}
    e^{i q_i \cdot x_i} &= e^{- i \vec{q}_\bot \cdot \vec{x}_{i\,\bot}} \,
    e^{-i Q_{p-q}^- z_i}
    \notag \\ &=
    e^{- i \vec{q}_\bot \cdot \vec{x}_{i\,\bot}} \,
    e^{-i\: \frac{\vec{u}_{i\,\bot}\cdot\vec{q}_\bot}{1-u_{i\,z}} z_i}e^{i \left( \frac{(k - x p)_\bot^2}{2 x (1-x) E(1-u_{i\,z})} \right) z_i} \,
    e^{-i \left( \frac{(p-q)_\bot^2 - p_\bot^2}{2 E(1-u_{i\,z})} \right) z_i } .
\end{align}
We see that the velocity dependence only mildly modifies the usual LPM phase structure with the so-called ``boundary phase'' $e^{i \left( \frac{(k - x p)_\bot^2}{2 x (1-x) E(1-u_{i\,z})} \right) z_i}$ and ``impulse phase'' $e^{-i \left( \frac{(p-q)_\bot^2 - p_\bot^2}{2 E(1-u_{i\,z})} \right) z_i }$ \cite{Sievert:2018imd} clearly identifiable as in the static medium.

The shift in the argument of the source current $J(p - q_i)$ corresponds to the energy shift due to collisional energy transfer with the moving medium:
\begin{align}
    J(p-q_i) &= J(E - \vec{u}_{i\,\bot} \cdot \vec{q}_\bot - u_{i\,z} Q_{p-q}^- \: , \: \vec{p}_\bot - \vec{q}_\bot \: , \: p_z - Q_{p-q}^- )
    \notag \\ &=
    J \left(E - \frac{\vec{u}_{i\,\bot} \cdot \vec{q}_\bot}{1 - u_{i\,z}} \: , \: \vec{p}_\bot - \vec{q}_\bot \: , \: E - \frac{\vec{u}_{i\,\bot} \cdot \vec{q}_\bot}{1-u_{i\,z}} \right)
    \notag \\ &=
    J(E, \vec{p}_\bot - \vec{q}_\bot) \left[ 1 - \frac{\vec{u}_{i\,\bot} \cdot\vec{q}_\bot}{1 - u_{i\,z}}
    \frac{1}{J} \: \frac{\partial J}{\partial E} \right] .
\end{align}
Similarly, the nontrivial shift in the argument of the potential \eq{e:potl5} due to the velocity is 
\begin{align} \label{e:vA}
    v_i (q_i) &= v_i( q_\bot^2 + (Q_{p-q}^-)^2 - q_0^2)
    \notag \\ &=
    v_i\left( q_\bot^2 - \frac{2 (\vec{u}_{i\,\bot} \cdot \vec{q}_\bot)}{1-u_{i\,z}}
    \left( \frac{(k - x p)_\bot^2}{2 x (1-x) E} - \frac{(p-q)_\bot^2 - p_\bot^2}{2 E} \right)   \right)
    \notag \\ &=
    v_i(q_\bot^2) \left[
    1 - \frac{2 (\vec{u}_{i\,\bot} \cdot \vec{q}_\bot)}{1-u_{i\,z}}
    \left( \frac{(k - x p)_\bot^2}{2 x (1-x) E} - \frac{(p-q)_\bot^2 - p_\bot^2}{2 E} \right) \:
    \frac{1}{v_i(q_\bot^2)} \frac{\pa v_i}{\pa q_\bot^2}
    \right] \, .
\end{align}
Altogether, this lets us express the leading velocity corrections to $R_{1,A}$ as
\begin{align}   \label{e:R1Afinal}
R_{1,A}&=
- i \sum_i \mathcal{C}_A^a t^a_i \, \theta(z_i) \:
\psi(x, \vec{k}_\bot - x \vec{p}_\bot ) \int\frac{d^2 q_\perp}{(2\p)^2} \, v_i(q_\bot^2) \, J(E, \vec{p}_\bot - \vec{q}_\bot) \,
e^{- i \vec{q}_\bot \cdot \vec{x}_{i\,\bot}} \,
\notag \\ &\times
e^{-i\: \frac{\vec{u}_{i\,\bot}\cdot\vec{q}_\bot}{1-u_{i\,z}} z_i}\,e^{i \left( \frac{(k - x p)_\bot^2}{2 x (1-x) E(1-u_{i\,z})} \right) z_i} \,
e^{-i \left( \frac{(p-q)_\bot^2 - p_\bot^2}{2 E(1-u_{i\,z})} \right) z_i }
\left[1-\frac{\vec{u}_{i\,\bot}\cdot (\vec{p}-\vec{q})_\perp}{E(1-u_{i\,z})}\right] \,
\notag \\ &\times  \left[ 1 - \frac{\vec{u}_{i\,\bot} \cdot\vec{q}_\bot}{1 - u_{i\,z}}
\frac{1}{J} \: \frac{\partial J}{\partial E} \right] \; 
\left[
1 - \frac{2 (\vec{u}_{i\,\bot} \cdot \vec{q}_\bot)}{1-u_{i\,z}}
\left( \frac{(k - x p)_\bot^2}{2 x (1-x) E} - \frac{(p-q)_\bot^2 - p_\bot^2}{2 E} \right) \:
\frac{1}{v_i(q_\bot^2)} \frac{\pa v_i}{\pa q_\bot^2}
\right] 
\notag \\ &=
- i \sum_i \mathcal{C}_A^a t^a_i \, \theta(z_i) \:
\psi(x, \vec{k}_\bot - x \vec{p}_\bot ) \int\frac{d^2 q_\perp}{(2\p)^2} \, v_i(q_\bot^2) \, J(E, \vec{p}_\bot - \vec{q}_\bot) \,
e^{- i \vec{q}_\bot \cdot \vec{x}_{i\,\bot}} \,
\notag \\ & \hspace{1cm} \times
e^{-i\: \frac{\vec{u}_{i\,\bot}\cdot\vec{q}_\bot}{1-u_{i\,z}} z_i}\,e^{i \left( \frac{(k - x p)_\bot^2}{2 x (1-x) E(1-u_{i\,z})} \right) z_i} \,
e^{-i \left( \frac{(p-q)_\bot^2 - p_\bot^2}{2 E(1-u_{i\,z})} \right) z_i }
\, \Big[ 1 + \vec{u}_{i\,\bot} \cdot \vec{\Omega}_A \Big],
\end{align}
where the velocity correction $\vec{\Omega}_A$ is given by
\begin{align}
\label{e:OA}
    \vec{\Omega}_A = 
    - \frac{\vec{p}_\bot - \vec{q}_\bot}{(1-u_{i\,z}) E} -
    \frac{\vec{q}_\bot}{(1-u_{i\,z})} \frac{1}{J}\frac{\partial J}{\partial E}
    - \frac{2 \vec{q}_\bot}{1-u_{i\,z}}
\left( \frac{(k - x p)_\bot^2}{2 x (1-x) E} - \frac{(p-q)_\bot^2 - p_\bot^2}{2 E} \right) \:
\frac{1}{v_i(q_\bot^2)} \frac{\pa v_i}{\pa q_\bot^2}\,,
\end{align}
with the LPM phase shifts directly coupling to the medium velocity.

\subsection{Diagrams B and C}


%
\begin{figure}
    \centering
	\includegraphics[width=\textwidth]{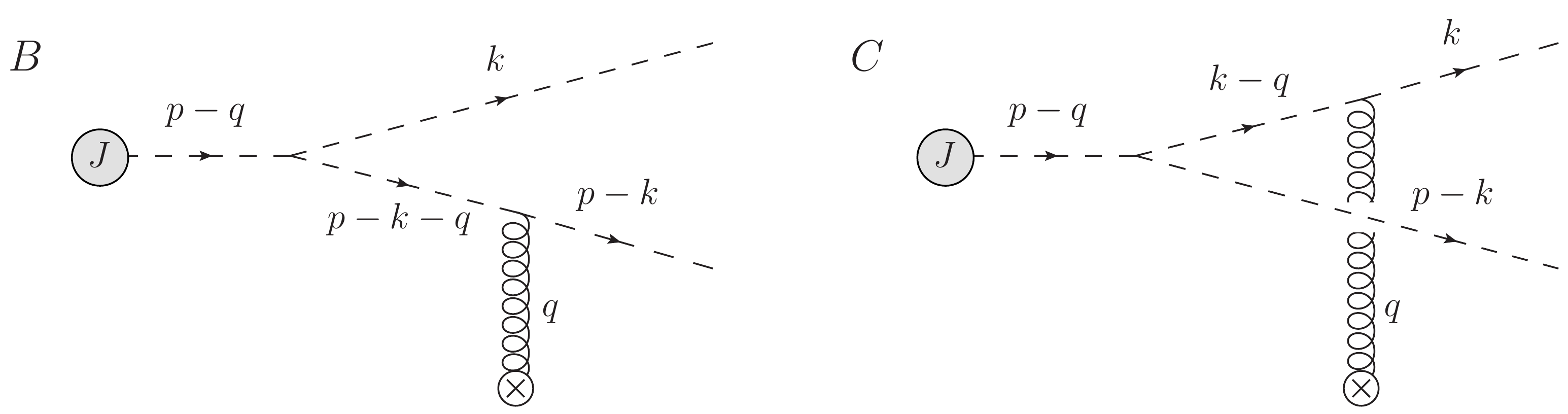}	
	\caption{The final-state scattering diagrams denoted $R_1^B$ and $R_1^C$.} 
	\label{f:Born_Rad_BC}
\end{figure}
%

In the same way we begin the evaluation of the final-state scattering diagrams $R_{1,B}$ and $R_{1,C}$ shown in Fig.~\ref{f:Born_Rad_BC}.  The diagram $R_1^B$ denotes the case of an initial-state branching followed by rescattering on the parton which emerges with momentum $(p-k)$.  The diagram $R_1^C$, in which the rescattering takes place on the parton with final-state momentum $k$ is obtained from $R_1^B$ by the replacement $k \rightarrow (p-k)$.  For the simplified case of scalar jets considered here, this replacement is sufficient to generate the third diagram, but in full QCD one has also the conversion between fundamental and adjoint partons.  For diagram $R_{1,B}$ we have
\bea \label{e:R1B}
R_{1,B} = i \sum_i \mathcal{C}_B^a t^a_i
\int\frac{d^4 q}{(2\p)^4}e^{iq\cdot x_i}\,\frac{\lambda\,u^\m_i(2(p-k)-q)_\m}{[(p-q)^2+i\e][(p-k-q)^2+i\e]}\,v_i(q)J(p-q)(2\p)\delta\left(q_0-\vec{u}_i\cdot\vec{q}\right) .\notag
\eea
Computing the numerator to $\ord{\frac{\perp}{E}}$ gives
\bea
R_{1,B}=i \sum_i \mathcal{C}_B^a t^a_i \int\frac{d^3 q}{(2\p)^3}e^{iq\cdot x_i}\,\frac{2\l(1-u_{i\,z})(1-x)E\left[1-\frac{\vec{u}_{i\,\perp}\cdot (\vec{p}_\perp-\vec{k}_\perp)}{(1-u_{i\,z})(1-x)E}\right]}{[(p-q)^2+i\e][(p-k-q)^2+i\e]}\,v_i(q_i)J(p-q_i). \notag\\
\eea
Next we collect the poles of $q_z$ from the propagators; in addition to the propagator $(p-q)$ computed in \eqref{e:prop1} we also have
\bea
(p-k-q)^2+i\epsilon=-(1-u_{i\,z}^2)\left(q_z-Q_{p-k-q}^{+}-i\e\right)\left(q_z-Q_{p-k-q}^{-}+i\e\right) \, , 
\eea 
which contributes new large and small poles given by
\begin{subequations}
\begin{align}
Q_{p-k-q}^{+} &\simeq\frac{2E(1-x)}{1+u_{i\,z}}\left(1-\frac{\vec{u}_{i\,\perp}\cdot\vec{q}_\perp}{2(1-x)E}\right) \, , 
\\
Q_{p-k-q}^{-} &\simeq \frac{\vec{u}_{i\,\perp}\cdot\vec{q}_\perp}{1-u_{i\,z}} + \frac{(p-k-q)_\perp^2-(p-k)_\perp^2}{2(1-x)(1-u_{i\,z})E} \, .
\end{align}
\end{subequations}
As with the diagram $R_{1,A}$, we close the $q_z$ contour below to generate a factor of $(-2\pi i) \theta(z_i)$ to encircle the small poles $Q_{p-k-q}^-$ and $Q_{p-q}^-$; the residue of the large poles $Q^+$ we neglect as (further) power-suppressed.

Unlike with the diagram $R_{1,A}$, now the pole structure couples to the splitting wave function as well through the new propagator $(p-k-q)$ and generates two distinct contributions to the residue:
\begin{align}
    \int\frac{dq_z}{2\pi} & \, e^{- i q_z z_i} \, v_i (q_i) \, J(p-q_i)
    \frac{1}{(p-k-q)^2 + i \epsilon}
    \frac{1}{(p-q)^2 + i \epsilon}
   \notag \\ &=
    \frac{1}{(1 - u_{i\,z}^2)^2} \int\frac{dq_z}{2\pi} \, 
    e^{- i q_z z_i} \, v_i (q_i) \, J(p-q_i)
    \frac{1}{(q_z - Q_{p-k-q}^+ - i \epsilon)(q_z - Q_{p-k-q}^- + i \epsilon)}
    \notag \\ &\hspace{1cm}\times
    \frac{1}{(q_z - Q_{p-q}^+ - i \epsilon)(q_z - Q_{p-q}^- + i \epsilon)}
\notag \\ &\simeq
    \frac{-i \: \theta(z_i)}{(1 - u_{i\,z}^2)^2} 
    \frac{1}{Q_{p-k-q}^- - Q_{p-q}^-}
    \Bigg[
    \frac{e^{- i (Q_{p-k-q}^-) z_i} \, v_i (Q_{p-k-q}^-) \,
    J\left(p - q_i(Q^-_{p-k-q})\right) }
    {(Q_{p-k-q}^+ - Q_{p-k-q}^-)(Q_{p-q}^+ - Q_{p-k-q}^-)
    }
    \notag \\ &\hspace{1cm}-
    \frac{e^{- i (Q_{p-q}^-) z_i} \, v_i (Q_{p-q}^-) \, 
    J\left(p - q_i(Q^-_{p-q})\right) }
    {(Q_{p-k-q}^+ - Q_{p-q}^-)(Q_{p-q}^+ - Q_{p-q}^-)}
    \Bigg] .
\end{align}
The various contributions to the residue denominators all need to be evaluated, keeping their leading (eikonal) contributions and first sub-leading correction in $1/E$ to maintain control over the coupling to the medium velocity.  In addition to the difference $Q_{p-q}^+ - Q_{p-q}^-$ already computed in \eqref{e:edenom1}, we now have
\begin{subequations} \label{e:edenom2}
\begin{align}
Q_{p-k-q}^{+} - Q_{p-k-q}^{-} & \simeq \frac{2E(1-x)}{1+u_{i\,z}}\left(1-\frac{\vec{u}_{i\,\perp}\cdot\vec{q}_\perp}{(1-u_{i\,z})(1-x)E}\right) \; , 
\\
Q_{p-q}^{+}-Q_{p-k-q}^{-} & \simeq \frac{2E}{1+u_{i\,z}}\left(1-\frac{\vec{u}_{i\,\perp}\cdot\vec{q}_\perp}{(1-u_{i\,z})E}\right) \; ,
\\
Q_{p-k-q}^{+}-Q_{p-q}^{-}&\simeq \frac{2E(1-x)}{1+u_{i\,z}}\left(1-\frac{\vec{u}_{i\,\perp}\cdot\vec{q}_\perp}{(1-u_{i\,z})(1-x)E}\right) \; .
\end{align}
\end{subequations}
Note that, in order to maintain control of the overall $\ord{\frac{\bot}{E}}$ corrections, we must also keep the small poles $Q_{p-q}^-$ and $Q_{p-k-q}^-$ to $\ord{\frac{\bot^2}{E^2}}$ accuracy in
\bea
\label{e:dblsub}
Q_{p-k-q}^{-}-Q_{p-q}^{-} &\simeq  \frac{(k-xp+xq)_\perp^2}{2Ex(1-x)(1-u_{i\,z})}+\frac{\left[(p-k-q)_\perp^2-(1-x)^2(p-q)_\perp^2\right](\vec{u}_{i\,\perp}\cdot\vec{q}_\perp)}{2E^2(1-x)^2(1-u_{i\,z})^2}\,. 
\eea
This is because the product of three factors in the denominator with leading terms $E \times E \times 1/E$ includes an overall correction of $\ord{\frac{\bot}{E}}$ from both the $\ord{\frac{\bot}{E}}$ corrections to the large poles $Q_{p-q}^+, Q_{p-k-q}^+$ and from the $\ord{\frac{\bot^2}{E^2}}$ correction to the small poles.

Luckily, since the leading term in both poles is the same, $Q_{p-q}^- \approx Q_{p-k-q}^- \approx \frac{\vec{u}_{i\,\perp}\cdot\vec{q}_\perp}{1-u_{i\,z}}$, the $\ord{\frac{\bot}{E}}$ shift in the source current is the same for both terms:
\begin{align}
    J\left(p-q_i(Q^-_{p-k-q})\right) \simeq J\left(p-q_i(Q^-_{p-q})\right) &\simeq
    J\left(E - \frac{\vec{u}_{i\,\perp} \cdot \vec{q}_\bot}{1-u_{i\,z}}, \vec{p}_\bot - \vec{q}_\bot\right)
    \notag \\ &=
    J(E, \vec{p}_\bot - \vec{q}_\bot) 
    \left[ 1 - \frac{\vec{u}_{i\,\perp} \cdot \vec{q}_\bot}{1-u_{i\,z}} \:
    \frac{1}{J} \frac{\partial J}{\partial E} \right] .
\end{align}
We also see from \eqref{e:edenom1} and \eqref{e:edenom2} that at this accuracy, we have $(Q_{p-k-q}^+ - Q_{p-k-q}^-) \simeq (Q_{p-k-q}^+ - Q_{p-k}^-)$ and $(Q_{p-q}^+ - Q_{p-k-q}^-) \simeq (Q_{p-q}^+ - Q_{p-k}^-)$: compared to the values of the large poles, the difference between the small poles is negligible.  This allows us to factor out the denominators from both of the residue terms, giving
\begin{align}
R_{1,B} &= \sum_i \mathcal{C}_B^a t^a_i \, \theta(z_i) \int\frac{d^2 q_\perp}{(2\p)^2} \, 
e^{-i\vec{q}_\perp\cdot \vec{x}_{i\perp}} \, \frac{2\l(1-u_{i\,z})(1-x)E}{(1-u_{i\,z}^2)^2\left(Q_{p-k-q}^{-}-Q_{p-q}^{-}\right)}
J(E, \vec{p}_\bot - \vec{q}_\bot) 
\notag\\ &\times
\left[
\frac{
v_i(Q_{p-k-q}^{-})
e^{-iQ^-_{p-k-q}z_i }
-
v_i(Q_{p-q}^{-})
e^{-iQ^-_{p-q}z_i }
}{(Q_{p-q}^{+}-Q_{p-q}^{-})(Q_{p-k-q}^{+}-Q_{p-q}^{-})}
\right]
\notag\\ &\times
\left[1-\frac{\vec{u}_{i\,\perp}\cdot (\vec{p}_\perp-\vec{k}_\perp)}{(1-u_{i\,z})(1-x)E}\right]
\left[ 1 - \frac{\vec{u}_{i\,\perp} \cdot \vec{q}_\bot}{1-u_{i\,z}} \:
\frac{1}{J} \frac{\partial J}{\partial E} \right] .
\end{align} 
Then inserting the explicit correction factors from \eqref{e:edenom1} and \eqref{e:edenom2} we obtain
\begin{align}
R_{1,B} &= \sum_i \mathcal{C}_B^a t^a_i \, \theta(z_i)
\int\frac{d^2 q_\perp}{(2\p)^2} \, 
e^{-i\vec{q}_\perp\cdot \vec{x}_{i\perp}} \, 
\frac{\l x (1-x)}{ (k - x p + x q)_\bot^2 } \:
J(E, \vec{p}_\bot - \vec{q}_\bot) 
\notag\\ &\times
\left[
v_i(Q_{p-k-q}^{-})
e^{-iQ^-_{p-k-q} z_i }
-
v_i(Q_{p-q}^{-})
e^{-iQ^-_{p-q} z_i }
\right]
\notag\\ &\times
\bigg[
1 - 
\frac{\vec{u}_{i\,\perp}\cdot(\vec{p}_\perp-\vec{k}_\perp-\vec{q}_\bot)}
{(1-u_{i\,z})(1-x)E} 
- \frac{\vec{u}_{i\,\perp} \cdot \vec{q}_\bot}{1-u_{i\,z}} \:
\frac{1}{J} \frac{\partial J}{\partial E} 
+\frac{\vec{u}_{i\,\perp}\cdot\vec{q}_\perp}{(1-u_{i\,z})E}
\notag\\ &\hspace{4cm}-\frac{x(\vec{u}_{i\,\perp}\cdot\vec{q}_\perp)\left[(p-k-q)_\perp^2-(1-x)^2(p-q)_\perp^2\right]}{(1-u_{i\,z})(1-x)E\,(k-xp+xq)_\perp^2}
\bigg]\,.
\end{align} 
We identify the light-front wave function $\psi(x, \vec{k}_\bot - x \vec{p}_\bot + x \vec{q}_\bot)$ and LPM phases in the exponents to write
\begin{align}
R_{1,B} &= -i \sum_i \mathcal{C}_B^a t^a_i \, \theta(z_i)
\int\frac{d^2 q_\perp}{(2\p)^2} \, 
e^{-i\vec{q}_\perp\cdot \vec{x}_{i\perp}}e^{-i\:\frac{\vec{u}_{i\,\bot}\cdot\vec{q}_\bot}{1-u_{i\,z}}z_i} \, 
\psi(x, \vec{k}_\bot - x \vec{p}_\bot + x \vec{q}_\bot) \:
J(E, \vec{p}_\bot - \vec{q}_\bot) 
\notag\\ &\times
\left[
v_i(Q_{p-k-q}^{-})
e^{-i 
\left( \frac{(p-k-q)_\perp^2-(p-k)_\perp^2}{2(1-x)E(1-u_{i\,z})} \right) 
z_i }
-
v_i(Q_{p-q}^{-})
e^{+i \left(
\frac{(k - x p)_\bot^2}{2 x (1-x) E(1-u_{i\,z})} - \frac{(p-q)_\bot^2 - p_\bot^2}{2 E(1-u_{i\,z})} \right) z_i }
\right]
\notag\\ &\times
\bigg[
1 - 
\frac{\vec{u}_{i\,\perp}\cdot(\vec{p}_\perp-\vec{k}_\perp-\vec{q}_\bot)}
{(1-u_{i\,z})(1-x)E} 
- \frac{\vec{u}_{i\,\perp} \cdot \vec{q}_\bot}{1-u_{i\,z}} \:
\frac{1}{J} \frac{\partial J}{\partial E} 
+\frac{\vec{u}_{i\,\perp}\cdot\vec{q}_\perp}{(1-u_{i\,z})E}
\notag\\ &\hspace{4cm}-\frac{x(\vec{u}_{i\,\perp}\cdot\vec{q}_\perp)\left[(p-k-q)_\perp^2-(1-x)^2(p-q)_\perp^2\right]}{(1-u_{i\,z})(1-x)E\,(k-xp+xq)_\perp^2}
\bigg]\,.
\end{align} 
We note that there are two sets of velocity-dependent corrections to $R_{1,B}$.  First there are the common corrections to the amplitude coming from the changes in the energy denominators and the correction to the external potential vertex.  Second, there are the shifts in the virtuality of the potential which differ between the two terms in brackets.  These term-specific corrections are
\begin{align}
    v_i(Q_{p-k-q}^{-}) &= v_i\left(q_\bot^2 + (Q_{p-k-q}^{-})^2 - q_0^2
    \right)
    \notag \\ &=
    v_i(q_\bot^2) \left[
    1 + \frac{2 (\vec{u}_{i\,\perp} \cdot \vec{q}_\bot) }{1-u_{i\,z}}
    \left( \frac{(p-k-q)_\bot^2 - (p-k)_\bot^2}{2 (1-x) E} \right)
    \frac{1}{v_i} \frac{\partial v_i}{\partial q_\bot^2}
    \right] \, , 
   \notag \\
    v_i(Q_{p-q}^{-}) &= v_i\left(q_\bot^2 + (Q_{p-q}^{-})^2 - q_0^2
    \right)
    \notag \\ &=
    v_i(q_\bot^2) \left[
    1 - \frac{ 2 (\vec{u}_{i\,\perp} \cdot \vec{q}_\bot) }{1-u_{i\,z}}
    \left( \frac{(k - x p)_\bot^2}{2 x (1-x) E} 
    - \frac{(p-q)_\bot^2 - p_\bot^2}{2 E} \right)
    \frac{1}{v_i} \frac{\partial v_i}{\partial q_\bot^2}
    \right]\,,
\end{align}
which explicitly couple the LPM phases which drive the radiative energy loss to the medium velocity.  Together, all these corrections are written as
\begin{align}   \label{e:RBfinal}
R_{1,B} &= -i \sum_i \mathcal{C}_B^a t^a_i \, \theta(z_i)
\int\frac{d^2 q_\perp}{(2\p)^2} \, 
e^{-i\vec{q}_\perp\cdot \vec{x}_{i\perp}}\,e^{-i\:\frac{\vec{u}_{i\bot}\cdot\vec{q}_\bot}{1-u_{i\,z}}z_i} \, v(q_\bot^2) \,
\psi(x, \vec{k}_\bot - x \vec{p}_\bot + x \vec{q}_\bot) \:
J(E, \vec{p}_\bot - \vec{q}_\bot) 
\notag\\ 
&\times
\Bigg\{
e^{-i 
\left( \frac{(p-k-q)_\perp^2-(p-k)_\perp^2}{2(1-x)E(1-u_{i\,z})} \right) 
z_i }
\left[
    1 + \vec{u}_{i\,\perp} \cdot \vec{\O}_{I\,B}
    \right]
-e^{+i \left(
\frac{(k - x p)_\bot^2}{2 x (1-x) E(1-u_{i\,z})} - \frac{(p-q)_\bot^2 - p_\bot^2}{2 E(1-u_{i\,z})} \right) z_i }
\left[
    1 + \vec{u}_{i\,\perp} \cdot \vec{\O}_{II\,B}
    \right]
\Bigg\}\,,
\end{align}
where we have introduced shorthand notations
\begin{subequations}
\label{e:OB}
\begin{align}
    \vec{\Omega}_{I\,B} &=
    -\frac{\vec{p}_\perp-\vec{k}_\perp-\vec{q}_\bot}
{(1-u_{i\,z})(1-x)E} 
- \frac{\vec{q}_\bot}{1-u_{i\,z}} \:
\frac{1}{J} \frac{\partial J}{\partial E} 
+\frac{2 \vec{q}_\bot}{1-u_{i\,z}}
    \left( \frac{(p-k-q)_\bot^2 - (p-k)_\bot^2}{2 (1-x) E} \right)
    \frac{1}{v} \frac{\partial v}{\partial q_\bot^2}\notag\\
    &\hspace{1cm}+\frac{\vec{q}_\perp}{(1-u_{i\,z})E}
-\frac{x\vec{q}_\perp\left[(p-k-q)_\perp^2-(1-x)^2(p-q)_\perp^2\right]}{(1-u_{i\,z})(1-x)E\,(k-xp+xq)_\perp^2}\\
   \vec{\Omega}_{II\,B} &=
    -\frac{\vec{p}_\perp-\vec{k}_\perp-\vec{q}_\bot}
{(1-u_{i\,z})(1-x)E} 
- \frac{\vec{q}_\bot}{1-u_{i\,z}} \:
\frac{1}{J} \frac{\partial J}{\partial E} 
-\frac{2 \vec{q}_\bot}{1-u_{i\,z}}
    \left( \frac{(k-xp)_\bot^2}{2x(1-x)E}-\frac{(p-q)_\bot^2 - p_\bot^2}{2 E} \right)
    \frac{1}{v} \frac{\partial v}{\partial q_\bot^2}\notag\\
    &\hspace{1cm}+\frac{\vec{q}_\perp}{(1-u_{i\,z})E}
-\frac{x\vec{q}_\perp\left[(p-k-q)_\perp^2-(1-x)^2(p-q)_\perp^2\right]}{(1-u_{i\,z})(1-x)E\,(k-xp+xq)_\perp^2}\,.
\end{align}
\end{subequations}
Notice that $\vec{\O}_{II\,B}$ is similar to $\vec{\O}_A$ in (\ref{e:OA}) while both factors appear in front of the same LPM phase exponent:
$\vec{\Omega}_{II\,B} =
   \vec{\O}_A+\frac{\vec{k}_\perp-x\vec{p}_\perp}
{(1-u_{i\,z})(1-x)E}+\frac{\vec{q}_\perp}{(1-u_{i\,z})E}\frac{k_\bot^2-x^2(p-q)^2_\bot}{(k-xp+xq)^2_\bot}\,.$

Finally, we consider the diagram $R_{1,C}$, in which the final-state rescattering occurs on the parton with momentum $k$ rather than the one with momentum $(p-k)$.  The amplitude is given by
\bea
R_{1,C}= i \sum_i \mathcal{C}_C^a t^a_i \int\frac{d^4 q}{(2\p)^4}e^{iq\cdot x_i}\,\frac{  \lambda\,u^\m_i(2k-q)_\m}{[(p-q)^2+i\e][(k-q)^2+i\e]}\,v(q)J(p-q)(2\p)\delta\left(q_0-\vec{u}\cdot\vec{q}\right)\notag\\
\eea
Comparison with \eqref{e:R1B} confirms that they are related by $k \leftrightarrow (p-k)$ (including the longitudinal momentum fraction $x \leftrightarrow (1-x)$) and by replacement of the color factor.  This allows us to immediately write the final expression
\begin{align}   \label{e:RCfinal}
R_{1,C} &= -i \sum_i \mathcal{C}_C^a t^a_i \, \theta(z_i)
\int\frac{d^2 q_\perp}{(2\p)^2} \, 
e^{-i\vec{q}_\perp\cdot \vec{x}_{i\perp}} \,e^{-i\:\frac{\vec{u}_{i\bot}\cdot\vec{q}_\bot}{1-u_{i\,z}}z_i}\, v(q_\bot^2) \,
\psi(x, \vec{k}_\bot - x \vec{p}_\bot - (1-x) \vec{q}_\bot) \:
J(E, \vec{p}_\bot - \vec{q}_\bot) 
\notag\\ &\times
\Bigg\{
e^{-i 
    \left( \frac{(k-q)_\perp^2-k_\perp^2}{2 x E(1-u_{i\,z})} \right) 
z_i }
\left[
    1 + \vec{u}_{i\,\perp} \cdot \vec{\O}_{I\,C}
    \right]
-e^{+i \left(
\frac{(k - x p)_\bot^2}{2 x (1-x) E(1-u_{i\,z})} - \frac{(p-q)_\bot^2 - p_\bot^2}{2 E(1-u_{i\,z})} \right) z_i }
\left[
    1 + \vec{u}_{i\,\perp}\cdot\vec{\O}_{II\,C}\right]
\Bigg\}\,,
\end{align}
where $\vec{\O}_{I\,C}$ and $\vec{\O}_{II\,C}$ follow from $\vec{\O}_{I\,B}$ and $\vec{\O}_{II\,B}$ under the substitution $k\leftrightarrow (p-k)$ and $x\leftrightarrow 1-x$.

\section{In-Medium Branching: Double Born Diagrams}
\label{sec:BRdblBorn}

%
\subsection{Diagram D}
%

%
\begin{figure}[t]
    \centering
	\includegraphics[width=0.5\textwidth]{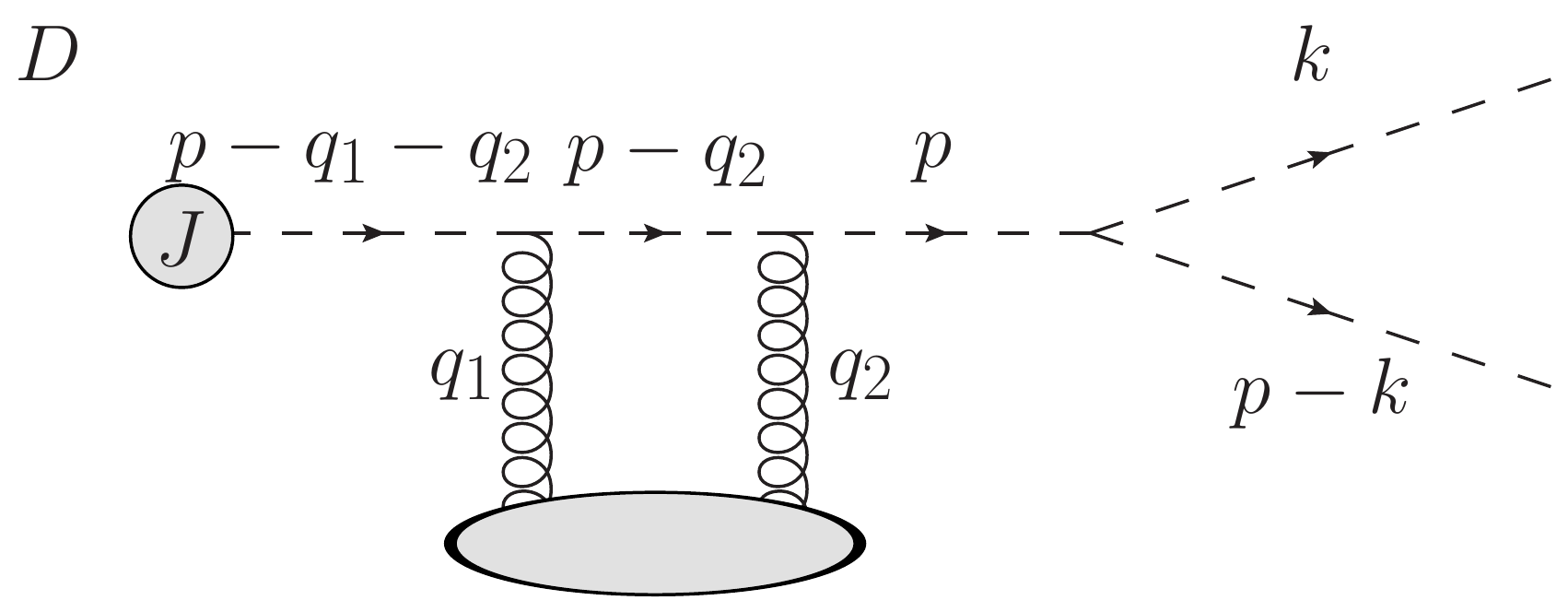}
	\caption{Double-Born diagram denoted $R_{2, D}$ corresponding to initial-state scattering followed by final-state branching.}
	\label{f:DblBorn_D}
\end{figure}
%
%

We begin with the double-Born diagram D shown in Fig.~\ref{f:DblBorn_D}.  As before, using the external potential expressed in \eqref{f:Potential} and \eqref{e:potl4} gives
\begin{align}
    R_{2, D} &=
    - i \lambda \sum_{i \, j} \mathcal{C}_D^{b a} t^b_{j} t^a_i 
    \int\frac{d^4 q_1}{(2\pi)^4} \frac{d^4 q_2}{(2\pi)^4} \,
    e^{ i q_1 \cdot x_i} \, e^{i q_2 \cdot x_{j}}
    \frac{2 p \cdot u_{j}}{\left[p^2 + i \epsilon\right]\left[(p-q_1-q_2)^2 + i \epsilon\right]} 
    \notag \\ & \times
    \frac{2 (p-q_2) \cdot u_i}{(p-q_2)^2 + i \epsilon}
    v_i (q_1) \, v_{j} (q_2) \: J(p-q_1-q_2) \:
    (2\pi)^2 \: \delta(q_1^0 - \vec{u}_i \cdot \vec{q}_1) \,
    \delta(q_2^0 - \vec{u}_{j} \cdot \vec{q}_2) ,
\end{align}
where $\mathcal{C}_D^{b a}$ is a generic color matrix depending on the representation of the projectile partons and we have used the constraints $u_{i} \cdot q_1 = u_{j} \cdot q_2 = 0$ imposed by the delta functions.  Identifying $\frac{i \lambda}{p^2 + i \epsilon}$ as the light-front wave function $\psi(x, \vec{k}_\bot - x \vec{p}_\bot)$ from \eqref{e:scalvac1}, we observe that $R_{2, D}$ can be written in the suggestive form $R_{2, D} = - i \psi(x, \vec{k}_\bot - x \vec{p}_\bot) M_2 (p)$ with the double-Born amplitude $M_2 (p)$ given by \eqref{e:DblBorn1}.  However, despite this simple relation, one cannot jump directly to the final expression for $R_{2, D}$ because here the kinematics are slightly different than for the jet broadening diagram $M_2$.  In the case of jet broadening without branching, $p^2 = 0$ is on-shell, whereas in the case with final-state branching it is off-shell.  In particular, this leads to different values of $p_z$ which propagate and affect the resulting poles of the propagators $(p-q_2)$ and $(p-q_1-q_2)$.  

Proceeding, we multiply by the complex conjugate $R_0^* = - \psi^* (x, \vec{k}_\bot - x \vec{p}_\bot) \, J^*(p)$ of \eqref{e:scalvac1} and perform the averaging over the color states of the target to obtain
\begin{align}
    \left\langle R_{2, D} R_0^* \right\rangle &=
    \mathcal{C}_{(D, 0)} \left| \psi(x, \vec{k}_\bot - x \vec{p}_\bot) \right|^2 \sum_i         
    \int\frac{d^4 q_1}{(2\pi)^4} \frac{d^4 q_2}{(2\pi)^4} \,
    e^{ i (q_1 + q_2) \cdot x_i}  J(p-q_1-q_2) \, J^*(p) 
    \notag \\ & \times
 \frac{ [2 p \cdot u_i]^2 \: v_i (q_1) \, v_i (q_2) } {[(p-q_2)^2 + i \epsilon][(p-q_1-q_2)^2 + i \epsilon]}     
    (2\pi)^2 \: \delta(q_1^0 - \vec{u}_i \cdot \vec{q}_1) \,
    \delta(q_2^0 - \vec{u}_i \cdot \vec{q}_2) ,
\end{align}
where setting $j = i$ has simplified $(p-q_2) \cdot u_ i = p \cdot u_i$.  We have defined the overall color factor as
\begin{align}
    \langle \mathcal{C}_D^{b a} t^b_{j} t^a_i \,
    \mathcal{C}_0^\dagger \rangle
    &\equiv 
    \mathcal{C}_{(D, 0)} \delta_{i j}  =
    \frac{1}{2 C_{\bar{R}}} 
    \langle \mathcal{C}_D^{a a} \,
    \mathcal{C}_0^\dagger \rangle \delta_{i j} \, .
\end{align}
The correction to the eikonal vertex computed in \eqref{e:BroadBorn3} gives
\begin{align}
    \left\langle R_{2, D} R_0^* \right\rangle &=
    \mathcal{C}_{(D, 0)}  \left| \psi(x, \vec{k}_\bot - x \vec{p}_\bot) \right|^2 \int d^3 x \, \rho
    \int\frac{d^4 q_1}{(2\pi)^4} \frac{d^4 q_2}{(2\pi)^4} \,
    e^{ i (q_1 + q_2) \cdot x}
    \notag \\ & \times
    \frac{ [2 E (1-u_z)]^2 \: v(q_1) \, v (q_2) } {[(p-q_2)^2 + i \epsilon][(p-q_1-q_2)^2 + i \epsilon]}
    \left[ 1 - 2 \frac{\vec{u}_\bot \cdot \vec{p}_\bot}{(1-u_z) E} \right] \:
    J(p-q_1-q_2) \, J^*(p)
    \notag \\ & \times
    (2\pi)^2 \: \delta(q_1^0 - \vec{u} \cdot \vec{q}_1) \,
    \delta(q_2^0 - \vec{u} \cdot \vec{q}_2) ,
\end{align}
where we have also converted the discrete sum over scattering centers to a continuous integral.  Note that the position dependence of $\rho$, $\vec{u}$, and $v(q)$ is left implicit for brevity.

The pole analysis now proceeds as before.  We start by computing the integral over $q_{1z}$ by residues, where the only poles are given by the propagator
\begin{align}
    (p-q_1-q_2)^2+i\e=-(1-u_z^2) \left(q_{1z}+q_{2z}-Q^+_{p-q_1-q_2}-i\e\right)\left(q_{1z}+q_{2z}-Q^-_{p-q_1-q_2}+i\e\right) .\notag
\end{align}
The pole values $Q^\pm_{p-q_1-q_2}$ are the same as for the single-Born radiation diagrams \eqref{e:rpoles1} but with total momentum transfer $q = q_1 + q_2$:
\begin{subequations} \label{e:rpoles2}
\begin{align}
Q_{p-q_1-q_2}^{+} &\simeq \frac{2E}{1+u_z}\left(1-\frac{\vec{u}_\perp \cdot (\vec{q}_{1 \, \perp} + \vec{q}_{2\bot})}{2E}\right),
\\
Q_{p-q_1-q_2}^{-} &\simeq
\frac{\vec{u}_\bot \cdot (\vec{q}_{1\bot} + \vec{q}_{2\bot}) }{1 - u_z} - \frac{1}{1-u_z} \left(
\frac{(k - x p)_\bot^2}{2 x (1-x) E} - \frac{(p-q_1-q_2)_\bot^2 - p_\bot^2}{2 E} \right),
\\
Q_{p-q_1-q_2}^{+} &- Q_{p-q_1-q_2}^{-} \simeq \frac{2E}{1+u_z}\left(1-\frac{\vec{u}_\perp\cdot(\vec{q}_{1 \, \perp} + \vec{q}_{2\bot})} {(1-u_z)E}\right).
\end{align}
\end{subequations}
Since the large pole $q_{1z} + q_{2z} = Q^+_{p - q_1 - q_2} \sim \ord{E}$ is parametrically large, its residue is highly suppressed and can be neglected.  Thus the unsuppressed contribution comes from closing the contour below the real axis, generating $- 2\pi i \, \theta(z)$ and picking up $q_{1z} + q_{2z} = Q^-_{p - q_1 - q_2}$:
\begin{align}
    \left\langle R_{2, D} R_0^* \right\rangle &=
    -i \, \mathcal{C}_{(D, 0)} \left| \psi(x, \vec{k}_\bot - x \vec{p}_\bot) \right|^2 \int d^3 x \, \rho \,
    \theta(z)
    \int\frac{d^2 q_{1\bot}}{(2\pi)^2} 
    \frac{d^4 q_2}{(2\pi)^4} \,
    e^{ -iQ^-_{p-q_1-q_2} z}
    \notag \\ & \times
    e^{-i (\vec{q}_{1\bot} + \vec{q}_{2\bot}) \cdot \vec{x}_\bot}
    \left[
    4 E^2 (1-u_z)^2 \,
    \frac{-1}{1-u_z^2} \,
    \frac{1}{Q^-_{p-q_1-q_2} - Q^+_{p-q_1-q_2}}
    \right]
    \frac{ v(q_1) \, v (q_2) } {(p-q_2)^2 + i \epsilon}
    \notag \\ & \times
    \left[ 1 - 2 \frac{\vec{u}_\bot \cdot \vec{p}_\bot}{(1-u_z) E} \right] \:
    J(p-q_1-q_2) \, J^*(p)
    (2\pi) \: \delta(q_2^0 - \vec{u} \cdot \vec{q}_2) .
\end{align}
Using the explicit form of the pole $Q^-_{p-q_1-q_2}$ we can simplify the expressions for the Fourier phase, the source current $J(p-q_1-q_2)$, and the explicit residue to obtain
\begin{align}
    \left\langle R_{2, D} R_0^* \right\rangle &=
    -i \, \mathcal{C}_{(D, 0)}  \left| \psi(x, \vec{k}_\bot - x \vec{p}_\bot) \right|^2 \int d^3 x \, \rho \,
    \theta(z)
    \int\frac{d^2 q_{1\bot}}{(2\pi)^2} 
    \frac{d^4 q_2}{(2\pi)^4} \,
    e^{ i \left[ \frac{(k - x p)_\bot^2}{2 x (1-x) E(1-u_{z})} - \frac{(p-q_1-q_2)_\bot^2 - p_\bot^2}{2 E(1-u_{z})} \right] z}
    \notag \\ &  \hspace*{-1cm}\times
    e^{-i (\vec{q}_{1\bot} + \vec{q}_{2\bot}) \cdot \vec{x}_\bot}\,e^{-i\:\frac{\vec{u}_\bot\cdot(\vec{q}_{1\bot}+\vec{q}_{2\bot})}{1-u_{z}}z}
    \left[
    2 (1-u_z) E \,
    \frac{ v(q_1) \, v (q_2) } {(p-q_2)^2 + i \epsilon}
    \right] \:
    J(E, \vec{p}_\bot - \vec{q}_{1\bot} - \vec{q}_{2\bot}) J^* (E, \vec{p}_\bot)
    \notag \\ &  \hspace*{-1cm}
    \times
    \left[ 1 - 2 \frac{\vec{u}_\bot \cdot \vec{p}_\bot}{(1-u_z) E} 
    + \frac{\vec{u}_\perp \cdot (\vec{q}_{1 \, \perp} + \vec{q}_{2\bot})} {(1-u_z)E}
    - \frac{\vec{u}_\perp \cdot (\vec{q}_{1 \, \perp} + \vec{q}_{2\bot})} {(1-u_z)}
    \frac{1}{J} \frac{\partial J}{\partial E}
    \right] 
    \:
    (2\pi) \: \delta(q_2^0 - \vec{u} \cdot \vec{q}_2) .
\end{align}

The next step will be to evaluate the $q_{2z}$ integral by residues, including particularly the poles of $q_{2z}$ present in \textit{both} potentials $v(q_2)$ and $v(q_1)$, where the pole in the latter is induced by the previous pole value $q_{1 \,z} = Q^-_{p-q_1-q_2} - q_{2 \, z}$.  These expressions can become quite unwieldy very quickly, so let us note from the outset that we limit ourselves to the leading corrections from the velocity at $\ord{\frac{\bot}{E}}$ and from transverse gradients at $\ord{ \vec{x}_\bot \cdot \vec{\nabla}_\bot}$ but not both at the same time.  For our present purposes in these double-Born radiation diagrams, we are interested primarily in the velocity corrections, so we will employ the assumption of transverse translational invariance (neglecting transverse gradients) in this section.  This significantly simplifies the (already complicated) expressions by allowing us to compute the $d^2 x$ integration immediately, obtaining $(2\pi)^2 \, \delta^2 (\vec{q}_{1\bot} + \vec{q}_{2\bot})$ and giving
\begin{align}
    \left\langle R_{2, D} R_0^* \right\rangle &=
    -i \, \mathcal{C}_{(D, 0)} \left| \psi(x, \vec{k}_\bot - x \vec{p}_\bot) \right|^2 \int_0^L dz \, \rho \:
    \int \frac{d^4 q_2}{(2\pi)^4} \,
    e^{ i \left[ \frac{(k - x p)_\bot^2}{2 x (1-x) E(1-u_{z})} \right] z}
    \left| J(E, \vec{p}_\bot) \right|^2
    \notag \\ & \times
    \left[
    2 (1-u_z) E \,
    \frac{ v(q_1) \, v (q_2) } {(p-q_2)^2 + i \epsilon}
    \right]
    \left[ 1 - 2 \frac{\vec{u}_\bot \cdot \vec{p}_\bot}{(1-u_z) E} 
    \right] 
    \:
    (2\pi) \: \delta(q_2^0 - \vec{u} \cdot \vec{q}_2) \; .
\end{align}

Writing $\vec{q}_\bot \equiv \vec{q}_{2\bot} = - \vec{q}_{1\bot}$ and writing the second pole integral as $\mathcal{I}_D$, the expression is
\begin{align} \label{e:RD1}
    \left\langle R_{2, D} R_0^* \right\rangle &=
    -i \, \mathcal{C}_{(D, 0)}  \left| \psi(x, \vec{k}_\bot - x \vec{p}_\bot) \right|^2 \int_0^L dz \, \rho \:
    \int \frac{d^2q_\bot} {(2\pi)^2} \,
    e^{ i \left[ \frac{(k - x p)_\bot^2}{2 x (1-x) E(1-u_{z})} \right] z}
    \left| J(E, \vec{p}_\bot) \right|^2
    \notag \\ & \times
    \mathcal{I}_D
    \left[ 1 - 2 \frac{\vec{u}_\bot \cdot \vec{p}_\bot}{(1-u_z) E} 
    \right]  .
\end{align}
where for the explicit form of the Gyulassy-Wang potential \eqref{e:potl5} the pole integral is
\begin{align}
    \mathcal{I}_D &\equiv
    2 (1-u_z) E \,
    \int \frac{dq_z}{2\pi}
    \frac{ v(q_1) \, v (q_2) } {(p-q_2)^2 + i \epsilon}
    \notag \\ &=
    - \frac{2 (1-u_z) E g^4}{(1-u_z^2)^3} \,
    \int \frac{dq_z}{2\pi}
    \frac{1}{
    [q_{2z} - Q^+_{p-q_2} - i \epsilon]
    [q_{2z} - Q^-_{p-q_2} + i \epsilon]
    [q_{2z} - \mathcal{P}^+]
    [q_{2z} - \mathcal{P}^-]}
    \notag \\ &\hspace{1cm}\times
    \frac{1}{
    [q_{2z} - \mathcal{P}^+ - Q^-_p]
    [q_{2z} - \mathcal{P}^- - Q^-_p]
    } \; ,
\end{align}
where $Q^\pm_{p-q_2}$ are defined analogously to \eqref{e:rpoles1} and \eqref{e:rpoles2}, $Q^-_p$ is similarly defined with $q=0$, that is,
\begin{align} \label{e:Qpminus1}
    Q_p^- &= \frac{-1}{1-u_z} \: \frac{(k - x p)_\bot^2}{2 x (1-x) E} ,
\end{align}
and the poles $\mathcal{P}^\pm$ associated with the potential were first defined in \eqref{e:potlpoles1}:
\begin{align}   \label{e:potlpoles2}
    \mathcal{P}^\pm \equiv \frac{u_z (\vec{u}_\bot \cdot \vec{q}_\bot) \pm i R }{1-u_z^2}
\end{align}
with finite imaginary part $R$ first defined in \eqref{e:Rdef}:
\begin{align}
    R^2 \equiv (1-u_z^2) (q_\bot^2 + \mu^2) - (\vec{u}_\bot \cdot \vec{q}_\bot)^2 > 0 .
\end{align}
The integral $\mathcal{I}_D$ is calculable but unwieldy, so before we present the result, let us add the complex conjugate term to \eqref{e:RD1} and simplify the structure of the result:
\begin{align} \label{e:RD2_first}
    \left\langle R_{2, D} R_0^* \right\rangle + \mathrm{c.c.} &=
    \mathcal{C}_{(D, 0)} \left| \psi(x, \vec{k}_\bot - x \vec{p}_\bot) \right|^2 \int_0^L dz \, \rho \:
    \left| J(E, \vec{p}_\bot) \right|^2
    \left[ 1 - 2 \frac{\vec{u}_\bot \cdot \vec{p}_\bot}{(1-u_z) E} 
    \right]
    \notag \\ & \times
    \int \frac{d^2q_\bot} {(2\pi)^2} \:
    \left[
    (-i) \: e^{ i \left[ \frac{(k - x p)_\bot^2}{2 x (1-x) E(1-u_{z})} 
    \right] z}  \:
    \mathcal{I}_D \:
    + \mathrm{c.c.}
    \right]
    \notag \\ &
   \hspace*{-2cm}  =  2\mathcal{C}_{(D, 0)}  \left| \psi(x, \vec{k}_\bot - x \vec{p}_\bot) \right|^2 \int_0^L dz \, \rho \:
    \left| J(E, \vec{p}_\bot) \right|^2 
    \left[ 1 - 2 \frac{\vec{u}_\bot \cdot \vec{p}_\bot}{(1-u_z) E} 
    \right] 
        \notag \\ &
 \hspace*{-2cm}    \times  \int \frac{d^2q_\bot} {(2\pi)^2} \:
    \left[
    \sin\left( \frac{(k - x p)_\bot^2}{2 x (1-x) E(1-u_{z})} z
    \right) \: \mathrm{Re} \, \mathcal{I}_D  
    + 
    \cos\left( \frac{(k - x p)_\bot^2}{2 x (1-x) E(1-u_{z})} z
    \right) \: \mathrm{Im} \, \mathcal{I}_D
    \right] \;.
\end{align}

Although $\mathcal{I}_D$ has both real and imaginary parts at $\ord{1}$ and $\ord{\frac{\bot}{E}}$ in eikonal power counting, not all of these terms will contribute to the final answer.  At the leading order (eikonal approximation) the expressions are
\begin{subequations} \label{e:ID1}
\begin{align}
    \mathrm{Re} \, \mathcal{I}_D &=
    - \frac{\vec{u}_\bot \cdot \vec{q}_\bot}{4 R^3}
    [v(q_\bot^2)]^2
    \left( 
    2 (\vec{u}_\bot \cdot \vec{q}_\bot)^2 - 3 (1-u_z^2) (q_\bot^2 + \mu^2) \right) + \ord{\frac{\bot}{E}} \, ,
    \\
    \mathrm{Im} \, \mathcal{I}_D &=
    - \frac{1}{2} [v(q_\bot^2)]^2
    + \ord{\frac{\bot}{E}} \, .
\end{align}
\end{subequations}
Clearly the eikonal real part $\mathrm{Re} \, \mathcal{I}_D$, which is an odd function of $\vec{q}_\bot$, integrates to zero after angular averaging and does not contribute to the final answer.  At first sub-eikonal accuracy $\ord{\frac{\bot}{E}}$, both $\mathrm{Re} \, \mathcal{I}_D$ and $\mathrm{Im} \, \mathcal{I}_D$ contain nonvanishing corrections due to the velocity.  The real part, in particular, generates qualitatively new phase structures proportional to $\sin\phi$ in the LPM interference pattern, as clearly seen in \eqref{e:RD2_first}.    Propagating these new $\sin\phi$ phase structures significantly increases the length and complexity of the final answer, so we defer these new terms for future analysis.

Instead, we will tune the parametric regime of interest to allow us to neglect them.  We previously noted that while genuine subeikonal terms of $\ord{\frac{\bot}{E}}$ are generally small, the LPM phases receive an enhancement because the separation $z$ between scattering centers can be large, $z \sim \ord{L}$ if the opacity is not too large.  In this case the LPM phases $\phi \sim \ord{\frac{\bot^2 \, z}{E}}$ are much larger than the genuine sub-eikonal corrections.  To allow us to justify neglecting the $\sin\phi \, \mathrm{Re} \, \mathcal{I}_D$ terms at present, we consider the parameteric regime
\begin{align} \label{e:regime}
    \frac{\bot^2}{E^2} \ll \frac{\bot^3 \, z}{E^2} \ll
    \frac{\bot}{E} \ll \left( \frac{\bot^2 z}{E} \right)^n \lesssim 1
\end{align}
and drop corrections of $\ord{\frac{\bot^3 \, z}{E^2}}$ or smaller.  In this regime, we would keep the full $\sin\left(\frac{\bot^2 \, z}{E}\right)$ phase structure multiplying the eikonal part of $\mathrm{Re} \, \mathcal{I}_D$ written in \eqref{e:ID1}; however, this term integrates to zero and does not contribute.  We will drop, however, the $\sin\left(\frac{\bot^2 \, z}{E}\right)$ phase multiplying the $\ord{\frac{\bot}{E}}$ sub-eikonal part of $\mathrm{Re} \, \mathcal{I}_D$ because the combination of the sine and the sub-eikonal suppression factor is beyond the working accuracy.  In contrast, for the cosine phase structure multiplying $\mathrm{Im} \, \mathcal{I}_D$, we must keep the full phase structure together with the leading eikonal part written in \eqref{e:ID1}, and at this accuracy we may either keep the phase structure or set $\cos\phi \approx 1$ in conjunction with the $\ord{\frac{\bot}{E}}$ sub-eikonal correction to $\mathrm{Im} \, \mathcal{I}_D$.

Within this approximation we need not explicitly consider the sub-eikonal correction to $\mathrm{Re} \, \mathcal{I}_D$, and because the eikonal part vanishes after angular averaging, we can neglect $\mathrm{Re} \, \mathcal{I}_D$ entirely.  Thus we write
\begin{align} \label{e:RD2}
    \left\langle R_{2, D} R_0^* \right\rangle + \mathrm{c.c.} &=
    \mathcal{C}_{(D, 0)} \left| \psi(x, \vec{k}_\bot - x \vec{p}_\bot) \right|^2 \int_0^L dz \, \rho \:
    \left| J(E, \vec{p}_\bot) \right|^2
    \left[ 1 - 2 \frac{\vec{u}_\bot \cdot \vec{p}_\bot}{(1-u_z) E} 
    \right]
    \notag \\ &\times
    \int \frac{d^2q_\bot} {(2\pi)^2} \:
    2 \cos\left( \frac{(k - x p)_\bot^2}{2 x (1-x) E(1-u_{z})} z
    \right) \: \mathrm{Im} \, \mathcal{I}_D  \, , 
\end{align}
with
\begin{align}   \label{e:ID2}
    \mathrm{Im} \, \mathcal{I}_D &=
    - \frac{1}{2} [v(q_\bot^2)]^2 -
    \frac{(\vec{u}_\bot \cdot \vec{q}_\bot)}{q_\bot^2 + \mu^2} [v(q_\bot^2)]^2
    \frac{
    (k - x p)_\bot^2 + x (1-x) (4 (\vec{p}_\bot \cdot \vec{q}_\bot) - q_\bot^2 + \mu^2)
    }{2E (1-u_z) x (1-x)}  \; .
\end{align}
Angular averaging again significantly simplifies the answer, allowing us to replace
\begin{align}   \label{e:ID3}
    \mathrm{Im} \, \mathcal{I}_D &\rightarrow
    - \frac{1}{2} [v(q_\bot^2)]^2 -
    \frac{\vec{u}_\bot \cdot \vec{p}_\bot}{
    (1-u_z) E} \: [v(q_\bot^2)]^2
    \frac{ q_\bot^2 }{q_\bot^2 + \mu^2}
 \notag \\ &=
    - \frac{1}{2} [v(q_\bot^2)]^2 
    \left(
    1 - 
    \frac{\vec{u}_\bot \cdot \vec{p}_\bot}{
    (1-u_z) E} \: 
    \frac{ q_\bot^2 }{\bar\sigma(q_\bot^2)} \frac{\partial\bar\sigma}{\partial q_\bot^2}
    \right) 
\end{align}
for the specific form of the Gyulassy-Wang potential.  Substituting this result back into \eqref{e:RD2} gives
\begin{align} \label{e:RD3}
    \left\langle R_{2, D} R_0^* \right\rangle + \mathrm{c.c.} &=
    - \mathcal{C}_{(D, 0)} \left| \psi(x, \vec{k}_\bot - x \vec{p}_\bot) \right|^2 \int_0^L dz \, \rho \:
    \left| J(E, \vec{p}_\bot) \right|^2
    \left[ 1 - 2 \frac{\vec{u}_\bot \cdot \vec{p}_\bot}{(1-u_z) E} 
    \right]
    \notag \\ & \hspace{-0.5cm} \times
    \int \frac{d^2q_\bot} {(2\pi)^2} \:
    [v(q_\bot^2)]^2
    \cos\left( \frac{(k - x p)_\bot^2}{2 x (1-x) E(1-u_{z})} z
    \right)
    \left( 1 - 
    \frac{\vec{u}_\bot \cdot \vec{p}_\bot}{
    (1-u_z) E} \: 
    \frac{ q_\bot^2 }{\bar\sigma(q_\bot^2)} \frac{\partial\bar\sigma}{\partial q_\bot^2} \right).
\end{align}
Simplifying the result by converting $[v(q_\bot^2)]^2$ into the elastic scattering cross section $\bar\sigma(q_\bot^2)$, multiplying by the final-state phase space factors, and combining the correction factors we obtain
\begin{align} \label{e:RD4}
    &\left(E\frac{dN^{(1)}}{d^2 k \, dx \: d^2 p \, dE} \right)_D  =
    - \frac{1}{2 (2\pi)^3\,x (1-x)}\frac{\mathcal{C}_{(D, 0)}}{\mathcal{C}}
    \left| \psi(x, \vec{k}_\bot - x \vec{p}_\bot) \right|^2 \int_0^L dz \, \rho \:
    \left( E \frac{dN^{(0)}}{d^2p_\perp \, dE} \right)
    \notag \\ &\hspace{0.65cm}\times
    \cos\left( \frac{(k - x p)_\bot^2}{2 x (1-x) E(1-u_{z})} z
    \right)
    \int d^2q_\bot \: \bar\s(q_\bot^2)
    \left[ 1 - 2 \frac{\vec{u}_\bot \cdot         \vec{p}_\bot}{(1-u_z) E} 
    - \frac{\vec{u}_\bot \cdot \vec{p}_\bot}{
    (1-u_z) E} \: 
    \frac{ q_\bot^2 }{\bar\sigma(q_\bot^2)} \frac{\partial\bar\sigma}{\partial q_\bot^2}
    \right] .
\end{align}
Recognizing the velocity corrections as the same factor $\vec{u}_\bot \cdot \vec{\Gamma}_{DB} (\vec{q}_\bot)$ for the double-Born broadening diagram $M_2$ from \eqref{e:GamDB}, we see that \eqref{e:RD4} can be compactly summarized as
\begin{align}   \label{e:RD5}
    &\left(E \frac{dN^{(1)}}{d^2 k \, dx \: d^2 p \, dE} \right)_D  =
    - \frac{1}{2(2\pi)^3\,x (1-x)}\frac{\mathcal{C}_{(D, 0)}}{\mathcal{C}}
    \left| \psi(x, \vec{k}_\bot - x \vec{p}_\bot) \right|^2 \int_0^L dz \, \rho \:
    \left( E \frac{dN^{(0)}}{d^2p_\perp \, dE} \right)
    \notag \\ & \hspace{1.5cm} \times
    \cos\left( \frac{(k - x p)_\bot^2}{2 x (1-x) E(1-u_{z})} z
    \right)
    \int d^2q_\bot \: \bar\sigma(q_\bot^2)
    \left[ 1 + \vec{u}_\bot \cdot \vec{\Gamma}_{DB} (\vec{q}_\bot)
    \right]
\notag \\ &
=
    \frac{1}{2 (2\pi)^3\,x (1-x)}\frac{\mathcal{C}_{(D, 0)}}{\mathcal{C}}
    \left| \psi(x, \vec{k}_\bot - x \vec{p}_\bot) \right|^2
    \cos\left( \frac{(k - x p)_\bot^2}{2 x (1-x) E(1-u_{z})} z
    \right)
    \left( E \frac{dN^{(0)}}{d^2p_\perp \, dE} \right)_\mathrm{DB}.
\end{align}

Eq.~\eqref{e:RD5} should be compared with the vacuum-like production of jets in Eq.~\eqref{e:scalvac2}, which are unmodified by the medium.  For the particularly simple case of the double-Born scattering followed by final-state branching shown in Fig.~\ref{f:DblBorn_D}, we see that the phase space distribution cleanly factorizes into a product involving the similarly-broadened jet distribution \eqref{e:DblBorn10}, followed by final-state branching as in vacuum \eqref{e:scalvac2}.  This is the manifestation at the level of the final distribution of the simple factorization $R_{2, D} = - i \psi(x, \vec{k}_\bot - x \vec{p}_\bot) \, M_2 (p)$ at the amplitude level noted previously.  However, as anticipated, the final distribution \eqref{e:RD5} is {\em not} solely a product of the broadening and vacuum-like branching factors; there is an essential additional feature generated by the interplay of these two ingredients.  This is the LPM phase $\cos\left(\frac{(k-xp)_\bot^2}{2 x (1-x) E (1-u_z)} z\right)$ which is fundamental to radiative energy loss due to stimulated emission in the medium.  As the calculation above shows, this structure is preserved under the influence of velocity corrections at first sub-eikonal accuracy.  It is also important to emphasize the emergence of \textit{modified} LPM phase structures sensitive to $\sin\left(\frac{(k-xp)_\bot^2}{2 x (1-x) E(1-u_z)} z\right)$ as well, as seen in \eqref{e:RD2}, which can couple to the velocity in new and interesting ways.  While we do not explore in full detail these structures here, we note that they do emerge in the general answer and will qualitatively change the LPM interference pattern.

%
\subsection{Diagrams E and F}
%

%
\begin{figure}
    \centering
	\includegraphics[width=\textwidth]{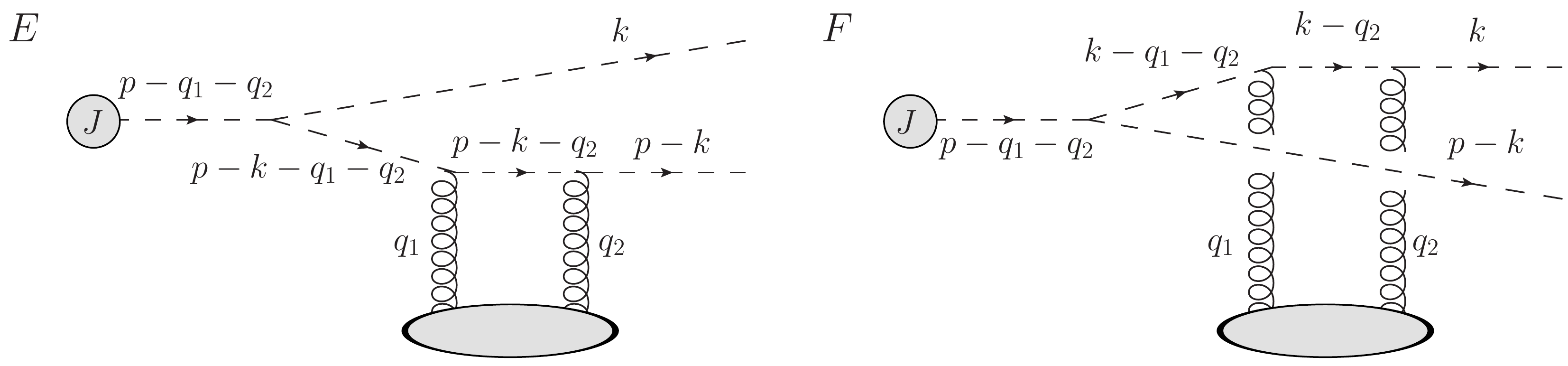}
	\caption{Left: Double-Born diagram denoted $R_{2, E}$ corresponding to initial-state branching followed by final-state scattering on one of the daughter partons.  Right: The diagram denoted $R_{2, F}$ with scattering on the other daughter parton is obtained by the substitution $k \leftrightarrow (p-k)$ from $R_{2, E}$.}
	\label{f:DblBorn_E}
\end{figure}
%

Next we consider the double-Born diagram shown in Fig.~\ref{f:DblBorn_E}.  The starting point for the amplitude is
\begin{align}
    R_{2, E} &= -i \lambda\sum_{i \, j} 
    \mathcal{C}_E^{b a} t^b_{j} t^a_i 
    \int\frac{d^4 q_1}{(2\pi)^4} \frac{d^4 q_2}{(2\pi)^4}
    e^{i q_1 \cdot x_i} \, e^{i q_2 \cdot x_{j}} \:
    v_i (q_1) \, v_{j} (q_2) \: J(p-q_1-q_2)
    \notag \\ &\times
    \frac{
    [2 (p-k)\cdot u_{j}]
    [2 (p-k-q_2) \cdot u_{i}]
    }{
    [(p-k-q_2)^2 + i \epsilon]
    [(p-k-q_1-q_2)^2 + i \epsilon]
    [(p-q_1-q_2)^2 + i \epsilon]
    }
    \notag \\ &\times
    \, (2\pi)^2 
    \delta(q_1^0 - \vec{u}_i \cdot \vec{q}_1) \,
    \delta(q_2^0 - \vec{u}_{j} \cdot \vec{q}_2) \;  .
\end{align}
Multiplying by the vacuum splitting amplitude $R_0^* = - \psi^* (x, \vec{k}_\bot - x \vec{p}_\bot) J^*(p)$ and performing the color averaging, we obtain
\begin{align}
    \left\langle R_{2, E} R_0^* \right\rangle &= 
    i \lambda \mathcal{C}_{(E,0)} 
    \psi^* (x, \vec{k}_\bot - x \vec{p}_\bot)
    \int d^3 x \, \rho \,
    \int\frac{d^4 q_1}{(2\pi)^4} \frac{d^4 q_2}{(2\pi)^4}
    e^{i (q_1+q_2) \cdot x} \:
    J(p-q_1-q_2) \, J^*(p)
    \notag \\ &\times
    \frac{
    [2 (p-k)\cdot u]^2
    }{
    [(p-k-q_2)^2 + i \epsilon]
    [(p-k-q_1-q_2)^2 + i \epsilon]
    [(p-q_1-q_2)^2 + i \epsilon]
    } v (q_1) \, v (q_2) \:
    \notag \\ &\times
    \, (2\pi)^2 
    \delta(q_1^0 - \vec{u} \cdot \vec{q}_1) \,
    \delta(q_2^0 - \vec{u} \cdot \vec{q}_2) \; ,
\end{align}
where we have converted the discrete summation into a continuous integral over the density and left the $\vec{x}$ dependence of quantities such as $\rho$ and $\vec{u}$ implicit.  Evaluating the correction to the eikonal vertex \eqref{e:BroadBorn3} gives
\begin{align}
    \left\langle R_{2, E} R_0^* \right\rangle &= 
    i \lambda \mathcal{C}_{(E,0)} 
    \psi^* (x, \vec{k}_\bot - x \vec{p}_\bot)
    \int d^3 x \, \rho \,
    \int\frac{d^4 q_1}{(2\pi)^4} \frac{d^4 q_2}{(2\pi)^4}
    e^{i (q_1+q_2) \cdot x} \:
    J(p-q_1-q_2) \, J^*(p)
    \notag \\ &\times
    \frac{
    [2 (1-u_z) (1-x) E]^2
    }{
    [(p-k-q_2)^2 + i \epsilon]
    [(p-k-q_1-q_2)^2 + i \epsilon]
    [(p-q_1-q_2)^2 + i \epsilon]
    } v (q_1) \, v (q_2) \:
    \notag \\ &\times
    \left[ 1 - 2 \frac{\vec{u}_\bot \cdot (\vec{p}_\bot - \vec{k}_\bot)}{(1-u_z) (1-x) E} \right]
    \, (2\pi)^2 
    \delta(q_1^0 - \vec{u} \cdot \vec{q}_1) \,
    \delta(q_2^0 - \vec{u} \cdot \vec{q}_2) \;  .
\end{align}

Now we proceed to perform the $dq_{1z}$ integral by residues of the poles of the propagators
\begin{align}
    [(p-k-q_1-q_2)^2 & + i \epsilon]
    [(p-q_1-q_2)^2 + i \epsilon]
    \notag \\ &=
    (1-u^2_z)^2 
    \left(q_{1z} + q_{2z} - Q^+_{p-q_1-q_2} - i \epsilon \right) 
    \left(q_{1z} + q_{2z} - Q^-_{p-q_1-q_2} + i \epsilon \right) 
    \notag \\ &\times
    \left(q_{1z} + q_{2z} - Q^+_{p-k-q_1-q_2} - i \epsilon \right)
    \left(q_{1z} + q_{2z} - Q^-_{p-k-q_1-q_2} + i \epsilon \right)
\end{align}
with $Q^\pm_{p-q_1-q_2}$ given by Eqs.~\eqref{e:rpoles2} from diagram $R_{2, D}$ and
\begin{subequations}    \label{e:polespkq1q2}
\begin{align}
    Q^+_{p-k-q_1-q_2} &= \frac{2 (1-x) E}{1+u_z} -
    \frac{\vec{u}_\bot \cdot (\vec{q}_{1 \bot} + \vec{q}_{2 \bot})}{1+u_z} \; , 
    \\
    Q^-_{p-k-q_1-q_2} &= \frac{\vec{u}_\bot \cdot (\vec{q}_{1 \bot} + \vec{q}_{2 \bot})}{1-u_z} +
    \frac{(p-k-q_1-q_2)_\bot^2 - (p-k)_\bot^2}{2(1-u_z) (1-x) E} \; .
\end{align}
\end{subequations}
As before we neglect the highly suppressed contributions from $Q^+$ which necessitates closing the contour below the real axis, generating a factor of $- 2\pi \, i \, \theta(z)$ and picking up the residues at $Q^-_{p-q_1-q_2} \, , \, Q^-_{p-k-q_1-q_2}$.  Unlike for the amplitude $R_{2, D}$, this produces a sum of two terms from the two enclosed poles of the propagators:
\begin{align}
    &\left\langle R_{2, E} R_0^* \right\rangle = 
    \lambda \mathcal{C}_{(E,0)} 
    \psi^* (x, \vec{k}_\bot - x \vec{p}_\bot)
    \int d^3 x \, \rho \, \theta(z) \,
    \int\frac{d^2 q_{1\bot}}{(2\pi)^2} \frac{d^4 q_2}{(2\pi)^4}
    e^{-i (\vec{q}_{1\bot}+\vec{q}_{2\bot}) \cdot \vec{x}_\bot} \:
    \, J^*(p)
    \notag \\ &\times
    \frac{
    [2 (1-u_z) (1-x) E]^2
    }{
    [(p-k-q_2)^2 + i \epsilon]
    } v (q_2) \:
    \left[ 1 - 2 \frac{\vec{u}_\bot \cdot (\vec{p}_\bot - \vec{k}_\bot)}{(1-u_z) (1-x) E} \right]
    \, (2\pi) \,
    \delta(q_2^0 - \vec{u} \cdot \vec{q}_2)
    \notag \\ & \times
    \Bigg\{
    \mathrm{Res}_{(Q_{p-q_1-q_2}^-)}
    \left(
    e^{-i (q_{1z} + q_{2z}) z}
    \frac{ v (q_1) \, J(p-q_1-q_2) }
    { [(p-k-q_1-q_2)^2 + i \epsilon][(p-q_1-q_2)^2 + i \epsilon] }
    \right)
    \notag \\ &  +
    \mathrm{Res}_{(Q_{p-k-q_1-q_2}^-)}
    \left(
    e^{-i (q_{1z} + q_{2z}) z}
    \frac{ v (q_1) \, J(p-q_1-q_2) }
    { [(p-k-q_1-q_2)^2 + i \epsilon][(p-q_1-q_2)^2 + i \epsilon] }
    \right)  \Bigg\} .
\end{align}

While the two residues are in general quite different, we note that the leading part of the two small poles are equal: $Q^-_{p-q_1-q_2} \simeq Q^-_{p-k-q_1-q_2} + \ord{\frac{\bot}{E}}$.  Thus when either pole value $Q^-$ is combined with a large scale of $\ord{E}$, the two small poles can be considered equal with a relative accuracy of $\ord{\frac{\bot}{E}}$.  Working at this accuracy allows us to simplify and combine several parts of the two residues:
\begin{subequations}
\begin{align}
    \mathrm{Res}_{(Q_{p-q_1-q_2}^-)} &\, J(p-q_1-q_2) \: \simeq \:
    \mathrm{Res}_{(Q_{p-k-q_1-q_2}^-)} \, J(p-q_1-q_2) 
    \notag \\ & =
    J(E, \vec{p}_\bot - \vec{q}_{1\bot} - \vec{q}_{2\bot})
    \left[ 1 -\frac{\vec{u}_\bot \cdot (\vec{q}_{1\bot} + \vec{q}_{2\bot})}{1-u_z} \frac{1}{J} \frac{\partial J}{\partial E} \right] \, ,
  \\ \label{e:resemph1}
    \mathrm{Res}_{(Q_{p-q_1-q_2}^-)} & \, 
    \frac{1}{[(p-k-q_1-q_2)^2 + i \epsilon][(p-q_1-q_2)^2 + i \epsilon]} 
    \notag \\ & 
    \simeq \:
    - \: \mathrm{Res}_{(Q_{p-k-q_1-q_2}^-)} \, 
    \frac{1}{[(p-k-q_1-q_2)^2 + i \epsilon][(p-q_1-q_2)^2 + i \epsilon]}
    \notag \\ &
    =
    \frac{1}{(1-u_z^2)^2}
    \frac{1}{[Q^+_{p-q_1-q_2} - Q^-_{p-q_1-q_2}] 
    [Q^+_{p-k-q_1-q_2} - Q^-_{p-q_1-q_2}]
    [Q^-_{p-q_1-q_2} - Q^-_{p-k-q_1-q_2}]} \, .
\end{align}
\end{subequations}
Note the minus sign in the second line of \eqref{e:resemph1} arising from substituting one pole into the other.  The different pole values of $q_{1z} + q_{2z}$ entering the LPM phases $e^{- i (q_{1z} + q_{2z}) z}$ and the potential $v(q_1)$ are not compared with a large scale of $\ord{E}$ and thus remain distinct, giving
\begin{align}
    &\left\langle R_{2, E} R_0^* \right\rangle = 
    \lambda \mathcal{C}_{(E,0)} 
    \psi^* (x, \vec{k}_\bot - x \vec{p}_\bot)
    \int d^3 x \, \rho \, \theta(z) \,
    \int\frac{d^2 q_{1\bot}}{(2\pi)^2} \frac{d^4 q_2}{(2\pi)^4}
    e^{-i (\vec{q}_{1\bot}+\vec{q}_{2\bot}) \cdot \vec{x}_\bot}
    \notag \\ &\times
    \frac{ v (q_2) }{ (p-k-q_2)^2 + i \epsilon }  \:
    \left[ 1 - 2 \frac{\vec{u}_\bot \cdot (\vec{p}_\bot - \vec{k}_\bot)}{(1-u_z) (1-x) E} 
    - \frac{\vec{u}_\bot \cdot (\vec{q}_{1\bot} + \vec{q}_{2\bot})}{1-u_z} \frac{1}{J} \frac{\partial J}{\partial E} 
    \right]
    \notag \\ & \times
    e^{-i\:\frac{\vec{u}_\bot\cdot(\vec{q}_{1\bot}+\vec{q}_{2\bot})}{1-u_{z}}z}J(E, \vec{p}_\bot - \vec{q}_{1\bot} - \vec{q}_{2\bot}) \,
    J^* (E, \vec{p}_\bot)
    \, (2\pi) \, \delta(q_2^0 - \vec{u} \cdot \vec{q}_2)
    \notag \\ & \times
    \frac{4 (1-x)^2 E^2}{(1 + u_z)^2}
    \frac{1}{[Q^+_{p-q_1-q_2} - Q^-_{p-q_1-q_2}] 
    [Q^+_{p-k-q_1-q_2} - Q^-_{p-q_1-q_2}]
    [Q^-_{p-q_1-q_2} - Q^-_{p-k-q_1-q_2}]}
    \notag \\ & \times
    \Bigg\{
    e^{i \left( \frac{(k-xp)_\bot^2}{2x(1-x)E(1-u_{z})} -
    \frac{(p-q_1-q_2)_\bot^2 - p_\bot^2}{2 E(1-u_{z})} \right) z} 
    \: v (\vec{q}_{1\bot} \, ; \, q_{1z} = -q_{2z} + Q^-_{p-q_1-q_2})
    \notag \\ &\hspace{1cm} -
    e^{- i \left( \frac{(p-k-q_1-q_2)_\bot^2 - (p-k)_\bot^2}{2 E(1-u_{z})}
    \right) z} 
    \: v (\vec{q}_{1\bot} \, ; \, q_{1z} = -q_{2z} + Q^-_{p-k-q_1-q_2}) \Bigg\} \; .
\end{align}
Inserting the explicit form of the difference between the large poles $Q^+$ and small poles $Q^-$ with a relative accuracy of $\ord{\frac{\bot}{E}}$ simplifies the expression somewhat:
\begin{align}
    &\left\langle R_{2, E} R_0^* \right\rangle = 
    \lambda \mathcal{C}_{(E,0)} 
    \psi^* (x, \vec{k}_\bot - x \vec{p}_\bot)
    \int d^3 x \, \rho \, \theta(z) \,
    \int\frac{d^2 q_{1\bot}}{(2\pi)^2} \frac{d^4 q_2}{(2\pi)^4}
    e^{-i (\vec{q}_{1\bot}+\vec{q}_{2\bot}) \cdot \vec{x}_\bot}e^{-i\:\frac{\vec{u}_\bot\cdot(\vec{q}_{1\bot}+\vec{q}_{2\bot})}{1-u_{z}}z}
    \notag \\ &\times
    \frac{ v (q_2) }{ (p-k-q_2)^2 + i \epsilon }  \:
    \frac{(1-x)}{Q^-_{p-q_1-q_2} - Q^-_{p-k-q_1-q_2}} \:
    J(E, \vec{p}_\bot - \vec{q}_{1\bot} - \vec{q}_{2\bot}) \,
    J^* (E, \vec{p}_\bot)
    \, (2\pi) \, \delta(q_2^0 - \vec{u} \cdot \vec{q}_2)
    \notag \\ & \times
    \left[ 1 
    - 2 \frac{\vec{u}_\bot \cdot (\vec{p}_\bot - \vec{k}_\bot)}{(1-u_z) (1-x) E} 
    - \frac{\vec{u}_\bot \cdot (\vec{q}_{1\bot} + \vec{q}_{2\bot})}{1-u_z} \frac{1}{J} \frac{\partial J}{\partial E}  
    + \frac{\vec{u}_\bot \cdot (\vec{q}_{1\bot} + \vec{q}_{2\bot})}{(1-u_z) E} 
    + \frac{\vec{u}_\bot \cdot (\vec{q}_{1\bot} + \vec{q}_{2\bot})}{(1-u_z) (1-x) E}
    \right]
    \notag \\ & \times
    \Bigg\{
    e^{i \left( \frac{(k-xp)_\bot^2}{2x(1-x)E(1-u_{z})} -
    \frac{(p-q_1-q_2)_\bot^2 - p_\bot^2}{2 E(1-u_{z})} \right) z} 
    \: v (\vec{q}_{1\bot} \, ; \, q_{1z} = -q_{2z} + Q^-_{p-q_1-q_2})
    \notag \\ &\hspace{1cm} -
    e^{- i \left( \frac{(p-k-q_1-q_2)_\bot^2 - (p-k)_\bot^2}{2 E(1-u_{z})}
    \right) z} 
    \: v (\vec{q}_{1\bot} \, ; \, q_{1z} = -q_{2z} + Q^-_{p-k-q_1-q_2}) \Bigg\} \, .
\end{align}

To proceed, we further simplify the expression by neglecting transverse gradients to perform the transverse coordinate averaging; the $d^2 x$ integral generates a delta function setting $\vec{q}_{1\bot} = - \vec{q}_{2\bot}$, giving
\begin{align}
    \left\langle R_{2, E} R_0^* \right\rangle &= 
    \lambda \mathcal{C}_{(E,0)} 
    \psi^* (x, \vec{k}_\bot - x \vec{p}_\bot)
    \int_0^L dz \, \rho \, \int \frac{d^4 q_2}{(2\pi)^4} \:
    \left| J(E, \vec{p}_\bot) \right|^2
    \notag \\ &\times
    \frac{ v (q_2) }{ (p-k-q_2)^2 + i \epsilon }  \:
    \frac{(1-x)}{Q^-_{p} - Q^-_{p-k}}
    \left[ 1 - 2 \frac{\vec{u}_\bot \cdot (\vec{p}_\bot - \vec{k}_\bot)}{(1-u_z) (1-x) E} \right]
    \, (2\pi) \, \delta(q_2^0 - \vec{u} \cdot \vec{q}_2)
    \notag \\ & \times
    \Bigg\{
    v (\vec{q}_{1\bot} = - \vec{q}_{2\bot} \, ; \, q_{1z} = -q_{2z} + Q^-_{p}) \:
    e^{i \frac{(k-xp)_\bot^2}{2x(1-x)E(1-u_{z})} z} 
    \notag \\ & \hspace{6cm} -
    v (\vec{q}_{1\bot} = - \vec{q}_{2\bot} \, ; \, q_{1z} = -q_{2z} + Q^-_{p-k}) \Bigg\} \; .
\end{align}
After setting $\vec{q}_{1\bot} + \vec{q}_{2\bot} = 0$, we have $Q_p^-$ given already by \eqref{e:Qpminus1}, whereas $Q^-_{p-k} \sim \ord{\frac{\bot^3}{E^3}}$ can be neglected entirely at this accuracy.  Together, this allows us to write the difference from the small poles in terms of the light-front wave function through $\lambda / (Q_p^- - Q_{p-k}^-) \simeq 2 i (1-u_z) E \: \psi(x, \vec{k}_\bot - x \vec{p}_\bot)$.  Moreover, we can use the fact that $v(q) = v(\vec{q}^2 - (\vec{u}\cdot\vec{q})^2)$ is an even function of the three-vector $\vec{q}$ (see Eq.~\eqref{e:potl5}) to simplify the arguments of the potential in braces.  These simplifications allow us to write
\begin{align}
    \left\langle R_{2, E} R_0^* \right\rangle &= 
    i \mathcal{C}_{(E,0)} 
    \left| \psi(x, \vec{k}_\bot - x \vec{p}_\bot) \right|^2
    \int_0^L dz \, \rho \, \int \frac{d^4 q_2}{(2\pi)^4} \:
    \left| J(E, \vec{p}_\bot) \right|^2
    \, (2\pi) \, \delta(q_2^0 - \vec{u} \cdot \vec{q}_2)
    \notag \\ &\times
    \frac{ 2 (1-u_z) (1-x) E }{ (p-k-q_2)^2 + i \epsilon }  \:
    \left[ 1 - 2 \frac{\vec{u}_\bot \cdot (\vec{p}_\bot - \vec{k}_\bot)}{(1-u_z) (1-x) E} \right]
    \notag \\ & \times
    \Bigg\{
    v (\vec{q}_{2\bot} \, ; q_{2z} - Q^-_{p}) \, v(q_2) \:
    e^{i \frac{(k-xp)_\bot^2}{2x(1-x)E(1-u_{z})} z} 
    - [ v (q_2) ]^2 \Bigg\} 
  \notag \\ &=
    i \mathcal{C}_{(E,0)} 
    \left| \psi(x, \vec{k}_\bot - x \vec{p}_\bot) \right|^2
    \int_0^L dz \, \rho \, \left| J(E, \vec{p}_\bot) \right|^2
    \left[ 1 - 2 \frac{\vec{u}_\bot \cdot (\vec{p}_\bot - \vec{k}_\bot)}{(1-u_z) (1-x) E} \right]
    \notag \\ &\times
    \int \frac{d^2 q_{2\bot}}{(2\pi)^2} \:
    \left[
    \mathcal{I}_E^{(1)} \, e^{i \frac{(k-xp)_\bot^2}{2x(1-x)E(1-u_{z})} z}
    - \mathcal{I}_E^{(2)}
    \right] \; ,
\end{align}
where we have written the remaining pole integrals as
\begin{subequations}
\begin{align}
    \mathcal{I}_{E}^{(1)} &\equiv
    2 (1-u_z) (1-x) E \int\frac{dq_{2z}}{2\pi}
    \frac{v (\vec{q}_{2\bot} \, ; q_{2z} - Q^-_{p}) \, v(q_2)}
    {(p-k-q_2)^2 + i \epsilon}
  \notag \\ &=
    \frac{- 2 (1-u_z) (1-x) E}{(1-u_z^2)^3}
    \int\frac{dq_{2z}}{2\pi}
    \frac{1}{( q_{2z} - Q^+_{p-k-q_2} - i \epsilon )
    ( q_{2z} - Q^-_{p-k-q_2} + i \epsilon )}
    \notag \\ &\hspace{1cm} \times
    \frac{1}{( q_{2z} - \mathcal{P}^+)
    ( q_{2z} - \mathcal{P}^- )}
    \frac{1}{
    ( q_{2z} - \mathcal{P}^+ - Q_p^- ) 
    ( q_{2z} - \mathcal{P}^- - Q_p^- ) } \; ,
    \\ \notag 
    \mathcal{I}_{E}^{(2)} &\equiv
    2 (1-u_z) (1-x) E \int\frac{dq_{2z}}{2\pi}
    \frac{[ v(q_2) ]^2}
    {(p-k-q_2)^2 + i \epsilon}
 \notag \\ &=
    \frac{- 2 (1-u_z) (1-x) E}{(1-u_z^2)^3}
    \int\frac{dq_{2z}}{2\pi}
    \frac{1}{( q_{2z} - Q^+_{p-k-q_2} - i \epsilon )
    ( q_{2z} - Q^-_{p-k-q_2} + i \epsilon )}
    \notag \\ &\hspace{1cm} \times
    \left[
    \frac{1}{
    ( q_{2z} - \mathcal{P}^+ )
    ( q_{2z} - \mathcal{P}^- ) }
    \right]^2  \; ,
\end{align}
\end{subequations}
with $\mathcal{P}^\pm$ defined as in \eqref{e:potlpoles2}.

Similar to the what we did for diagram $R_{2, D}$, we consider only the parts of the integrals $\mathcal{I}_E$ which contribute after adding the complex conjugate amplitude in the limit \eqref{e:regime}.  We have
\begin{align}   \label{e:R2E_1}
    &\left\langle R_{2, E} R_0^* \right\rangle + \mathrm{c.c.} = 
    \mathcal{C}_{(E,0)} 
    \left| \psi(x, \vec{k}_\bot - x \vec{p}_\bot) \right|^2
    \int_0^L dz \, \rho \, \left| J(E, \vec{p}_\bot) \right|^2
    \left[ 1 - 2 \frac{\vec{u}_\bot \cdot (\vec{p}_\bot - \vec{k}_\bot)}{(1-u_z) (1-x) E} \right]
    \notag \\ &\hspace{1cm}\times
    \int \frac{d^2 q_{2\bot}}{(2\pi)^2} \:
    \bigg[
    - 2 \, \mathrm{Re} \, \mathcal{I}_E^{(1)} \, 
    \sin\left( \frac{(k-xp)_\bot^2}{2x(1-x)E(1-u_{z})} z \right)
    \notag\\
    &\hspace{5cm}-2 \, \mathrm{Im} \, \mathcal{I}_E^{(1)} \, 
    \cos\left( \frac{(k-xp)_\bot^2}{2x(1-x)E(1-u_{z})} z \right) +
    2 \, \mathrm{Im} \, \mathcal{I}_E^{(2)}    
    \bigg] \, ,
\end{align}
and working in the limit \eqref{e:regime}, we see that we only need to evaluate $\mathrm{Re} \, \mathcal{I}_E^{(1)}$ to $\ord{1}$ but need to keep $\mathrm{Im} \, \mathcal{I}_E^{(1)}$ and $\mathrm{Im} \, \mathcal{I}_E^{(2)}$ to $\ord{\frac{\perp}{E}}$.  The relevant integrals are
\begin{subequations}
\begin{align}
    \mathrm{Re} \, \mathcal{I}_E^{(1)} &=
    \frac{- g^4 (\vec{u}_\bot \cdot \vec{q}_{2\bot}) \, 
    [3 R^2 + (\vec{u}_\bot \cdot \vec{q}_{2\bot})^2 ]}
    {4 R^3 (q_\bot^2 + \mu^2)^2
    } + \ord{\frac{\perp}{E}} \,  , 
\notag \\
    \mathrm{Im} \, \mathcal{I}_E^{(1)} &=
    - \frac{g^4}{2 (q_\bot^2 + \mu^2)^2} 
    \notag \\ & -
    \frac{g^4 (\vec{u}_\bot \cdot \vec{q}_{2\bot})}{2 (1-u_z) x (1-x) E} 
    \frac{ -2 x (p-k-q_2)_\bot^2 + 2 x (p-k)_\bot^2 + x (q_\bot^2 + \mu^2) - (k - x p)_\bot^2 }{ (q_{2\bot}^2 + \mu^2)^3 }\;  ,
   \notag \\
    \mathrm{Im} \, \mathcal{I}_E^{(2)} &=
    - \frac{g^4}{2 (q_\bot^2 + \mu^2)^2} 
    \notag \\ & -
    \frac{g^4 (\vec{u}_\bot \cdot \vec{q}_{2\bot})}{2 (1-u_z) (1-x) E}
    \frac{ -2(p-k-q_2)_\bot^2 +2(p-k)_\bot^2 + q_{2\bot}^2 + \mu^2 }{
    (q_{2\bot}^2 + \mu^2)^3 } \; ,
\end{align}
\end{subequations}
and we can further simplify the expressions by utilizing angular averaging over the $d^2 q_{2\bot}$ integral.  Doing so gives
\begin{subequations}
\begin{align}
    \mathrm{Re} \, \mathcal{I}_E^{(1)} &\rightarrow
    0 + \ord{\frac{\perp}{E}} \, , 
  \notag \\
    \mathrm{Im} \, \mathcal{I}_E^{(1)} =
    \mathrm{Im} \, \mathcal{I}_E^{(2)}
    &\rightarrow
    - \frac{1}{2} \left[ v(q_{2\bot}^2) \right]^2
    \left( 1 - \frac{\vec{u}_\bot \cdot (\vec{p}_\bot - \vec{k}_\bot)}{ (1-u_z) (1-x) E } \: 
    \frac{ q_{2\bot}^2 }{\bar\sigma} \frac{\partial\bar\sigma}
    {\partial q_{2\bot}^2} \right) \, ,
\end{align}
\end{subequations}
where, after angular averaging, the imaginary parts of the two integrals are equal.  

Substituting this expression back into \eqref{e:R2E_1}, we obtain
\begin{align}
    \left\langle R_{2, E} R_0^* \right\rangle + \mathrm{c.c.} &= 
    \mathcal{C}_{(E,0)} 
    \left| \psi(x, \vec{k}_\bot - x \vec{p}_\bot) \right|^2
    \int_0^L dz \, \rho \, \left| J(E, \vec{p}_\bot) \right|^2
    \left[ 1 - 2 \frac{\vec{u}_\bot \cdot (\vec{p}_\bot - \vec{k}_\bot)}{(1-u_z) (1-x) E} \right]
    \notag \\ &\times
    \int \frac{d^2 q_{2\bot}}{(2\pi)^2} \:
    \left[ v(q_{2\bot}^2) \right]^2
    \left[ \cos\left( \frac{(k-xp)_\bot^2}{2x(1-x)E(1-u_{z})} z \right) - 1
    \right]
    \notag \\ & \hspace{0cm} \times
    \left( 1 - \frac{\vec{u}_\bot \cdot (\vec{p}_\bot - \vec{k}_\bot)}{ (1-u_z) (1-x) E } \: 
    \frac{ q_{2\bot}^2 }{\bar\sigma} \frac{\partial\bar\sigma}
    {\partial q_{2\bot}^2} \right) \, ,
\end{align}
and inserting the relevant phase space factors, we obtain the corresponding phase space distribution
\begin{align}
    \left( E \frac{dN^{(0)}}{d^2 k \, dx \: d^2 p \, dE} \right)_{E} &=
    - \frac{1}{2(2\pi)^3\,x (1-x)} \frac{\mathcal{C}_{(E,0)}}{\mathcal{C}}\,
    \left| \psi(x, \vec{k}_\bot - x \vec{p}_\bot) \right|^2
    \int_0^L dz \, \rho \, 
    \left( E \frac{dN^{(0)}}{d^2 p \, dE} \right)
    \notag \\ &\times
    \left[ 1 - \cos\left( \frac{(k-xp)_\bot^2}{2x(1-x)E(1-u_{z})} z \right)
    \right]
    \int d^2q_\bot \, \bar\sigma(q_\bot^2)
    \notag \\ & \hspace{0cm} \times
    \left[ 1 
    - 2 \frac{\vec{u}_\bot \cdot (\vec{p}_\bot - \vec{k}_\bot)}{(1-u_z) (1-x) E}
    - \frac{\vec{u}_\bot \cdot (\vec{p}_\bot - \vec{k}_\bot)}{ (1-u_z) (1-x) E } \: 
    \frac{ q_\bot^2 }{\bar\sigma} \frac{\partial\bar\sigma}
    {\partial q_\bot^2} \right] \, .
\end{align}
Recognizing the velocity correction factor in brackets as possessing the same form as $\vec{\Gamma}_{DB}$ from \eqref{e:GamDB}, but for scattering on the parton with momentum $p-k$ instead (that is, having energy $E \rightarrow (1-x) E$ and transverse momentum $\vec{p}_\bot \rightarrow (\vec{p}_\bot - \vec{k}_\bot)$), we define
\begin{align}   \label{e:GamDB2}
    \vec{\Gamma}_{DB}^{(p-k)} = 
    \frac{- 2 (\vec{p}_\bot - \vec{k}_\bot)}{(1-u_z) (1-x) E} -
    \frac{(\vec{p}_\bot - \vec{k}_\bot)}{(1-u_z) (1-x) E}
    \frac{q_\bot^2}{\bar\sigma (q_\bot^2)} \frac{\partial \bar\sigma}{\partial q_\bot^2} \, .
\end{align}
to write the final answer as
\begin{align}   \label{e:R2E_2}
    \left( E \frac{dN^{(1)}}{d^2 k \, dx \: d^2 p \, dE} \right)_E &=
    - \frac{1}{2 (2\pi)^3\,x(1-x)}\frac{\mathcal{C}_{(E,0)}}{\mathcal{C}}\left| \psi(x, \vec{k}_\bot - x \vec{p}_\bot) \right|^2
    \int_0^L dz \, \rho \, 
    \left( E \frac{dN^{(0)}}{d^2 p \, dE} \right)
    \notag \\ &\times
    \left[ 1 - \cos\left(\frac{(k-xp)_\bot^2}{2 x (1-x) E(1-u_{z})} z \right) \right]
    \int d^2q_\bot \, \bar\sigma(q_\bot^2) \left[ 1 + \vec{u}_\bot \cdot \vec{\Gamma}_{DB}^{(p-k)} \right] \, .
\end{align}
The structure of \eqref{e:R2E_2} is intuitively easy to understand.  We see that the LPM phase structure $1 - \cos(\phi)$ is unmodified from the static case\footnote{At least in the limit \eqref{e:regime} where the $\sin\phi$ terms are negligible.}, that the double-Born contribution enters with a relative minus sign reflecting the depletion of jets with center-of-mass transverse momentum $\vec{p}_\bot$, and that the velocity corrections enter in the standard form \eqref{e:GamDB2} associated with double-Born scattering on the associated parton.

From Eq.~\eqref{e:R2E_2} we can immediately obtain the counterpart diagram $R_{2, F}$ with final-state double-Born scattering on the parton with momentum $k$ by interchanging $x \leftrightarrow (1-x)$, $\vec{k}_\bot \leftrightarrow (\vec{p}_\bot - \vec{k}_\bot)$, and $\mathcal{C}_{(E,0)}\leftrightarrow\mathcal{C}_{(F,0)}$ to obtain
\begin{align}  \label{e:R2F} 
    \left( E \frac{dN^{(1)}}{d^2 k \, dx \: d^2 p \, dE} \right)_F &=
    - \frac{1}{2 (2\pi)^3\,x(1-x)} \frac{\mathcal{C}_{(F,0)}}{\mathcal{C}}
    \left| \psi(x, \vec{k}_\bot - x \vec{p}_\bot) \right|^2
    \int_0^L dz \, \rho \, 
    \left( E \frac{dN^{(0)}}{d^2 p \, dE} \right)
    \notag \\ &\times
    \left[ 1 - \cos\left(\frac{(k-xp)_\bot^2}{2 x (1-x) E(1-u_{z})} z \right) \right]
    \int d^2q_\bot \, \bar\sigma(q_\bot^2) \left[ 1 + \vec{u}_\bot \cdot \vec{\Gamma}_{DB}^{(k)} \right]
\end{align}
with the associated velocity correction
\begin{align}  
\label{e:GamDB3}
    \vec{\Gamma}_{DB}^{(k)} = 
    \frac{- 2 \vec{k}_\bot}{(1-u_z) x E} -
    \frac{\vec{k}_\bot}{(1-u_z) x E}
    \frac{q_\bot^2}{\bar\sigma (q_\bot^2)} \frac{\partial \bar\sigma}{\partial q_\bot^2} \; .
\end{align}

%
\begin{figure}[t]
    \centering
	\includegraphics[width=0.5\textwidth]{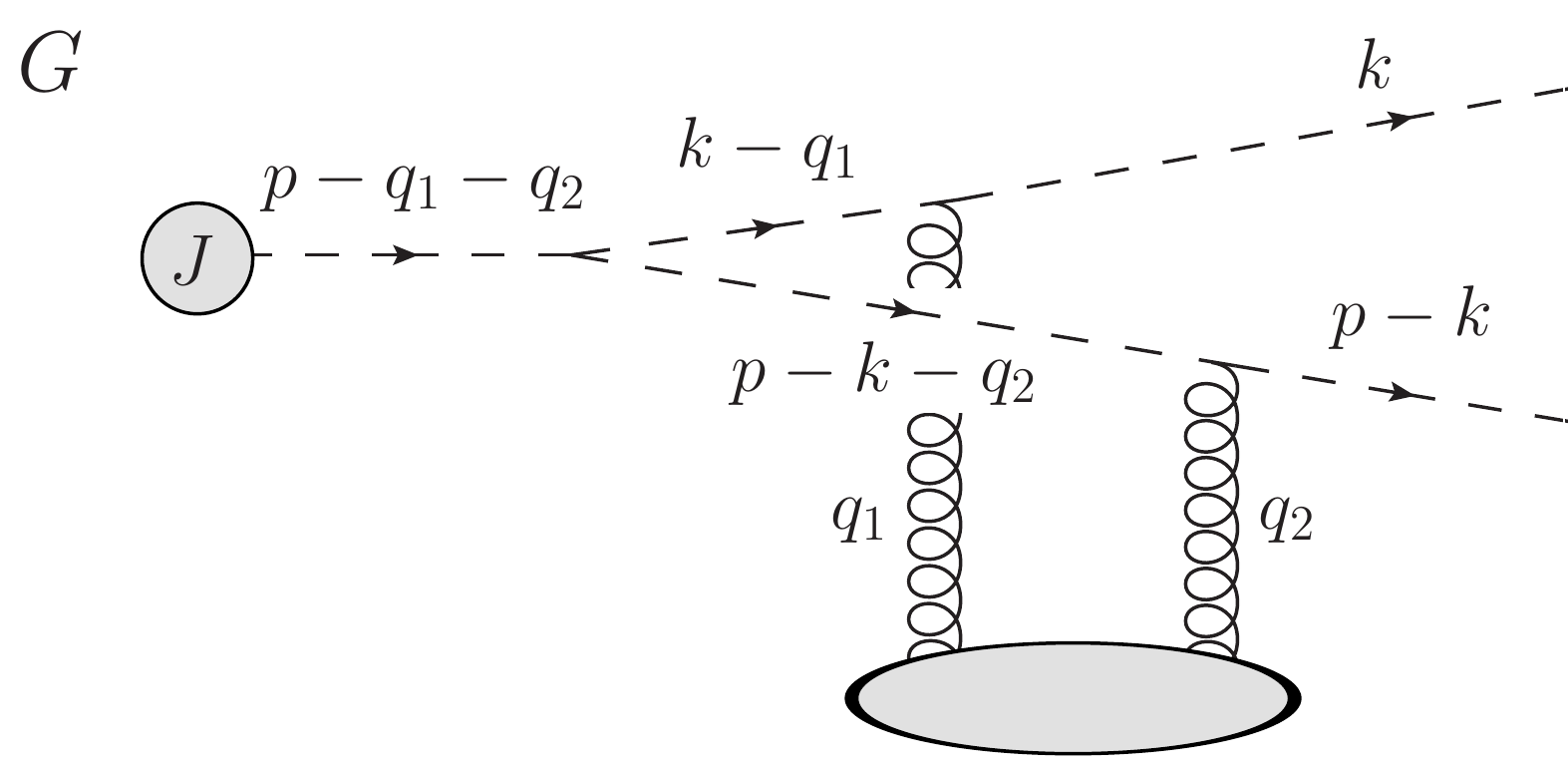}
	\caption{Double-Born diagram denoted $R_{2, G}$ corresponding to initial-state branching followed by final-state scattering on \textit{both} of the daughter partons.}
	\label{f:DblBorn_G}
\end{figure}
%

%
\subsection{Diagram G}
%

Next we continue with the final-state scattering diagram G shown in Fig.~\ref{f:DblBorn_G} with a single double-Born interaction coupling \textit{both} of the daughter partons.  The amplitude is given by
\begin{align}
    R_{2, G} &= - i \lambda \sum_{i \, j} 
    \mathcal{C}_G^{b a} t^b_{j} t^a_i
    \int \frac{d^4 q_1}{(2\pi)^4} \frac{d^4 q_2}{(2\pi)^4} \,
    e^{i q_1 \cdot x_i} \, e^{i q_2 \cdot x_{j}} \, v_i (q_1) \, v_{j} (q_2) \, 
    J(p-q_1-q_2)
    \notag \\ &\times
    \frac{
        [ 2 k \cdot u_i ] [2 (p-k) \cdot u_{j}] 
    }{
        [(p-k-q_2)^2 + i \epsilon] \, [(k-q_1)^2 + i\epsilon] \,
        [(p-q_1-q_2)^2 + i\epsilon]
    }
    \notag \\ &\times
    (2\pi)^2 \, \delta(q_1^0 - \vec{u}_i \cdot \vec{q}_1) \,
    \delta(q_2^0 - \vec{u}_{j} \cdot \vec{q}_2) \; .
\end{align}
Multiplying by the vacuum splitting amplitude $R_0^* = - \psi^* (x, \vec{k}_\bot - x \vec{p}_\bot) J^*(p)$ and performing the color averaging as before, we obtain
\begin{align}
    \left\langle R_{2, G} R_0^* \right\rangle &= 
    i \lambda \mathcal{C}_{(G,0)} \, \psi^* (x, \vec{k}_\bot - x \vec{p}_\bot)
    \int d^3 x \, \rho \,
    \int \frac{d^4 q_1}{(2\pi)^4} \frac{d^4 q_2}{(2\pi)^4} \,
    e^{i (q_1 + q_2) \cdot x} \, v (q_1) \, v (q_2) \, 
    \notag \\ &\times
    \frac{
        [ 2 k \cdot u ] [2 (p-k) \cdot u] 
    }{
        [(p-k-q_2)^2 + i \epsilon] \, [(k-q_1)^2 + i\epsilon] \,
        [(p-q_1-q_2)^2 + i\epsilon]
    }
    \, J(p-q_1-q_2) \, J^* (p)
    \notag \\ &\times
    (2\pi)^2 \, \delta(q_1^0 - \vec{u} \cdot \vec{q}_1) \,
    \delta(q_2^0 - \vec{u} \cdot \vec{q}_2) \; , 
\end{align}
and evaluating the correction to the eikonal vertex \eqref{e:BroadBorn2} gives
\begin{align}
    \left\langle R_{2, G} R_0^* \right\rangle &= 
    i \lambda \mathcal{C}_{(G,0)} \, \psi^* (x, \vec{k}_\bot - x \vec{p}_\bot)
    \int d^3 x \, \rho \,
    \int \frac{d^4 q_1}{(2\pi)^4} \frac{d^4 q_2}{(2\pi)^4} \,
    e^{i (q_1 + q_2) \cdot x} \, v (q_1) \, v (q_2) \, 
    \notag \\ &\times
    \frac{
        [ 2(1-u_z) x E ] [2 (1-u_z) (1-x) E ] 
    }{
        [(p-k-q_2)^2 + i \epsilon] \, [(k-q_1)^2 + i\epsilon] \,
        [(p-q_1-q_2)^2 + i\epsilon]
    }
    \, J(p-q_1-q_2) \, J^* (p)
    \notag \\ &\times
    \left[ 1 - \frac{\vec{u}_\bot \cdot \vec{k}_\bot}{(1-u_z) x E} 
    - \frac{\vec{u}_\bot \cdot (\vec{p}_\bot - \vec{k}_\bot)}{(1-u_z) (1-x) E} \right] \; 
    (2\pi)^2 \, \delta(q_1^0 - \vec{u} \cdot \vec{q}_1) \,
    \delta(q_2^0 - \vec{u} \cdot \vec{q}_2) \; .
\end{align}
Next we collect the poles of $q_{1z}$
\begin{align}
    \big[ (k-q_1)^2 + i \epsilon \big]
    \big[ (p-q_1-q_2)^2 + i \epsilon \big]
    &= (1-u_z^2)^2
    \big[ q_{1z} - Q_{k-q_1}^+ - i \epsilon\big]
    \big[ q_{1z} - Q_{k-q_1}^- + i \epsilon\big]
    \notag \\ & \times
    \big[ q_{1z} + q_{2z} - Q_{p-q_1-q_2}^+ - i \epsilon\big]
    \big[ q_{1z} + q_{2z} - Q_{p-q_1-q_2}^- + i \epsilon\big]\notag
\end{align}
with
\begin{subequations}
\begin{align}
    Q_{k-q_1}^+ &= \frac{2 x E}{1+u_z} - \frac{\vec{u}_\bot \cdot \vec{q}_{1 \bot}}{1+u_z} \; , 
    \\
    Q_{k-q_1}^- &= \frac{\vec{u}_\bot \cdot \vec{q}_{1 \bot}}{1-u_z} + 
    \frac{(k-q_1)_\bot^2 - k_\bot^2}{2 (1-u_z) x E} \; , 
\end{align}
\end{subequations}
defined analogously to \eqref{e:polespkq1q2} and $Q_{p-q_1-q_2}^\pm$ given in \eqref{e:rpoles2}.  Keeping only the unsuppressed contributions $Q^-$, we close the contour below and obtain
\begin{align}   \label{e:RG1}
    &\left\langle R_{2, G} R_0^* \right\rangle = 
    \mathcal{C}_{(G,0)} \, \lambda \,
    \psi^* (x, \vec{k}_\bot - x \vec{p}_\bot)
    \int d^3 x \, \rho \, \theta(z)
    \int \frac{d^2 q_{1\bot}}{(2\pi)^2} \frac{d^4 q_2}{(2\pi)^4} \,
    e^{-i (\vec{q}_{1\bot} + \vec{q}_{2\bot}) \cdot \vec{x}_\bot} \,  
    \notag \\ &\times
    \frac{4(1-u_z)^2 x (1-x) E^2}{(p-k-q_2)^2 + i \epsilon}
    \, v (q_2) \, J^* (p)
    \left[ 1 - \frac{\vec{u}_\bot \cdot \vec{k}_\bot}{(1-u_z) x E} 
    - \frac{\vec{u}_\bot \cdot (\vec{p}_\bot - \vec{k}_\bot)}{(1-u_z) (1-x) E} \right]
    (2\pi) \,\delta(q_2^0 - \vec{u} \cdot \vec{q}_2)
    \notag \\ &\times
    \left(
    \mathrm{Res}_{Q_{k-q_1}^-} + 
    \mathrm{Res}_{- q_{2z} + Q_{p-q_1-q_2}^-}
    \right)
    \left(
    \frac{
         v (q_1)
    \, J(p-q_1-q_2)  
    }{
        [(k-q_1)^2 + i\epsilon] \,
        [(p-q_1-q_2)^2 + i\epsilon]
    }
    \, e^{-i (q_{1z} + q_{2z}) z}
    \right) .
\end{align}
The two residues are
\begin{subequations}    \label{e:Gres1}
\begin{align}
    &\mathrm{Res}_{Q_{k-q_1}^-}
    \left(
    \frac{
         v (q_1)
    \, J(p-q_1-q_2)  
    }{
        [(k-q_1)^2 + i\epsilon] \,
        [(p-q_1-q_2)^2 + i\epsilon]
    }
    \, e^{-i (q_{1z} + q_{2z}) z}
    \right)
 \notag \\ &=
    \frac{1}{(1-u_z^2)^2}
    \frac{
    v (q_{1\bot}^2) \: e^{-i q_{2z} z} \:
    e^{-i\:\frac{\vec{u}_\bot\cdot\vec{q}_{1\bot}}{1-u_{z}}z-i \left[ \frac{(k-q_1)_\bot^2 - k_\bot^2}{2 x E (1-u_z)} \right] z}
    }{
        [ Q_{k-q_1}^- - Q_{k-q_1}^+ ] \,
        [ q_{2z} - Q_{p-q_1-q_2}^+ + Q_{k-q_1}^- - i \epsilon ] 
        [ q_{2z} - Q_{p-q_1-q_2}^- + Q_{k-q_1}^-]
    }
    \notag \\ &\times
    J \left(E - \frac{\vec{u}_\bot \cdot \vec{q}_{1\bot}}{1-u_z} - \vec{u}_\bot \cdot \vec{q}_{2\bot} - u_z q_{2z} \: ,  \: \vec{p}_\bot - \vec{q}_{1\bot} - \vec{q}_{2\bot} \: ,  \: E - \frac{\vec{u}_\bot \cdot \vec{q}_{1\bot}}{1-u_z} - q_{2z} \right)
    \notag \\ &\times
    \left[
    1 + \frac{\vec{u}_\bot \cdot \vec{q}_{1\bot}}{(1-u_z)  x E} \frac{(k-q_1)_\bot^2 - k_\bot^2}{v(q_{1\bot}^2)} \frac{\partial v}{\partial q_{1\bot}^2} 
    \right] \, ,
    \\ \notag 
    &\mathrm{Res}_{- q_{2z} + Q_{p-q_1-q_2}^-}
    \left(
    \frac{
         v (q_1)
    \, J(p-q_1-q_2)  
    }{
        [(k-q_1)^2 + i\epsilon] \,
        [(p-q_1-q_2)^2 + i\epsilon]
    }
    \, e^{-i (q_{1z} + q_{2z}) z}
    \right)
 \notag \\ &=
    \frac{1}{(1-u_z^2)^2}
    \frac{
          e^{-i\frac{\vec{u}_\bot\cdot(\vec{q}_{1\bot}+\vec{q}_{2\bot})}{1-u_{z}}z+i
\frac{(k - x p)_\bot^2}{2 x (1-x) E(1-u_z)}-i \frac{(p-q_1-q_2)_\bot^2 - p_\bot^2}{2(1-x)E(1-u_z)} z} 
    }{
        [Q_{p-q_1-q_2}^- - Q_{p-q_1-q_2}^+] \,
        [ q_{2z} - Q^-_{p-q_1-q_2} + Q^+_{k-q_1} + i\epsilon ]
        [ q_{2z} - Q^-_{p-q_1-q_2} + Q^-_{k-q_1}]
    }
    \notag \\ &\times
    \left[ 1 - \frac{\vec{u}_\bot \cdot (\vec{q}_{1\bot} + \vec{q}_{2\bot})}{1-u_z} \frac{1}{J} \frac{\partial J}{\partial E}
    \right]  \;  v (\vec{q}_{1\bot}; \, - q_{2z} + Q^-_{p-q_1-q_2})
    \, J(E, \, \vec{p}_\bot - \vec{q}_{1\bot} - \vec{q}_{2\bot})    \, .
\end{align}
\end{subequations}

Next we must consider the poles associated with the $q_{2z}$ integral.  On the face of it, the calculation appears to be quite challenging: for each of the two residues \eqref{e:Gres1} arising from the $q_{1z}$ integral, there are several potential poles of $q_{2z}$.  In addition to the two poles of the propagator $(p-k-q_2)^2$ and the potential $v(q_2)$, the term $\mathrm{Res}_{Q_{k-q_1}^-}$ contains 2 explicit poles, and the term $\mathrm{Res}_{- q_{2z} + Q_{p-q_1-q_2}^-}$ contains 2 more explicit poles plus 2 induced poles from the potential $v (\vec{q}_{1\bot} \, ; \, - q_{2z} + Q^-_{p-q_1-q_2})$.  However, in practice, the situation is not so bad.

First we note that as usual, the poles corresponding to the large scales $Q^+ \sim \ord{E}$ are highly suppressed and need not be considered.  Second, we observe that residue of the pole $q_{2z} = Q_{p-q_1-q_2}^- - Q_{k-q_1}^-$, which is common to both terms in \eqref{e:Gres1}, cancels exactly between the two terms.  Finally, we note that the first term $\mathrm{Res}_{Q_{k-q_1}^-}$ is not a contact term: it retains the Fourier factor dictating the direction of closure of the contour; as such, the poles arising from the potential which have finite imaginary parts lead to exponentially decaying functions of $z$ and can be neglected.  

After taking into account these considerations, we find that in the term $\mathrm{Res}_{Q_{k-q_1}^-}$, only the single residue of the pole $q_{2z} = Q_{p-k-q_2}^-$ contributes.  In this term, we must close the $q_{2z}$ contour below the real axis because of the presence of the Fourier factor $e^{-i q_{2z} z}$ and the constraint $\theta(z)$.  On the other hand, the contact term $\mathrm{Res}_{- q_{2z} + Q_{p-q_1-q_2}^-}$ may have its integration contour closed either above or below the real axis; we choose to close it below.  In doing so, we collect the three residues $q_{2z} = Q^-_{p-k-q_2}$ as well as the poles $q_{2z} = \mathcal{P}^-$ and $q_{2z} = Q_{p-q_1-q_2}^- + \mathcal{P}^-$ from the potentials, since in this contact term they do not lead to decaying exponentials.  As a result, we only need to consider a single residue coming from the $\mathrm{Res}_{Q_{k-q_1}^-}$ term and the sum of three residues from the $\mathrm{Res}_{- q_{2z} + Q_{p-q_1-q_2}^-}$ term.

If we first neglect transverse gradients and perform the $d^2 x$ integral to set $\vec{q}_{1\bot} + \vec{q}_{2\bot} = 0$, then the above discussion becomes compact enough to write as follows:
\begin{align}   \label{e:RG2}
    &\left\langle R_{2, G} R_0^* \right\rangle = 
    \mathcal{C}_{(G,0)} \, \lambda \,
    \psi^* (x, \vec{k}_\bot - x \vec{p}_\bot)
    \int_0^L dz \, \rho
    \int \frac{d^2 q_{2\bot}}{(2\pi)^2} \: J^* (p) \:
    \left[ \frac{- 2 E x(1-x)}{1+u_z} \right]
    \notag \\ &\times
    \left[ 1 - \frac{\vec{u}_\bot \cdot \vec{k}_\bot}{(1-u_z) x E} 
    - \frac{\vec{u}_\bot \cdot (\vec{p}_\bot - \vec{k}_\bot)}{(1-u_z) (1-x) E} \right]
    \notag \\ &\times
    \bigg\{
    \frac{v(q_{2\bot}^2)}{x} \left[ 1 - \frac{\vec{u}_\bot \cdot \vec{q}_{2\bot}}{(1-u_z) x E} -
    \frac{\vec{u}_\bot \cdot \vec{q}_{2\bot}}{(1-u_z) x E} 
    \frac{(k+q_2)_\bot^2 - k_\bot^2}{v(q_{2\bot}^2)} \frac{dv}{dq_{2\bot}^2} \right]   \:
    \mathcal{I}_G^{(1)}
    %
    %
    +e^{-i Q_p^- z} \, J(E,\vec{p}_\bot) \:
    \mathcal{I}_G^{(2)}
    \bigg\}
\end{align}
with
\begin{subequations}    \label{e:Gint1}
\begin{align}   \label{e:Gint1_1}
    \mathcal{I}_G^{(1)} &=
    \int\frac{dq_{2z}}{2\pi} e^{-i(Q_{k+q_2}^- + q_{2z}) z}
    \frac{v (q_2)}{(p-k-q_2)^2 + i \epsilon} 
    \notag \\ &\times
    \frac{
        J \left(E + \frac{\vec{u}_\bot \cdot \vec{q}_{2\bot}}{1-u_z} - \vec{u}_\bot \cdot \vec{q}_{2\bot} - u_z q_{2z} \: ,  \: \vec{p}_\bot\: ,  \: E + \frac{\vec{u}_\bot \cdot \vec{q}_{2\bot}}{1-u_z} - q_{2z} \right)
    }{
        [ q_{2z} - Q_p^- + Q_{k+q_2}^- + i \epsilon]
        [ q_{2z} - Q_p^+ + Q_{k+q_2}^- - i \epsilon]
    }  \; ,
\\    \label{e:Gint1_2}
    \mathcal{I}_G^{(2)} &=
    \int\frac{dq_{2z}}{2\pi} 
    \frac{v (q_2)}{(p-k-q_2)^2 + i \epsilon} 
    \frac{
        v(\vec{q}_{2\bot}; \, q_{2z} - Q_p^-)
    }{
        [ q_{2z} - Q_p^- + Q_{k+q_2}^- + i \epsilon]
        [ q_{2z} - Q_p^- + Q_{k+q_2}^+ + i \epsilon]
    } \; .
\end{align}
\end{subequations}
In arriving at \eqref{e:RG2} from \eqref{e:RG1}, we have used the residues \eqref{e:Gres1} of the $q_{1z}$ integral, substituted explicitly the results for the pole values $Q_{k-q_1}^- - Q_{k-q_1}^+$ and $Q_{p-q_1-q_2}^- - Q_{p-q_1-q_2}^+$.  We have also used parity symmetry to write $v(-\vec{q}_{1\bot} ; \, -q_{2z} + Q_p^-) = v(\vec{q}_{2\bot} ; \, q_{2z} - Q_p^-)$ in \eqref{e:Gint1_2}.

As discussed above, while the integrals $\mathcal{I}_G^{(1)} \, , \, \mathcal{I}_G^{(2)}$ in \eqref{e:Gint1} contain many possible poles, in the end only one pole of $\mathcal{I}_G^{(1)}$ and three poles of $\mathcal{I}_G^{(2)}$ will contribute.  Keeping just the surviving residue $q_{2z} = Q_{p-k-q_2}^-$ of $\mathcal{I}_G^{(1)}$ gives
\begin{align}
    \mathcal{I}_G^{(1)} &\rightarrow  -i \: \mathrm{Res}_{Q_{p-k-q_2}^-}
    \Bigg\{
    \frac{e^{-i(Q_{k+q_2}^- + q_{2z}) z} \: v (q_2)}{(p-k-q_2)^2 + i \epsilon} 
    \notag \\ &\times
    \frac{
        J \left(E - \frac{\vec{u}_\bot \cdot \vec{q}_{1\bot}}{1-u_z} - \vec{u}_\bot \cdot \vec{q}_{2\bot} - u_z q_{2z} \: ,  \: \vec{p}_\bot - \vec{q}_{1\bot} - \vec{q}_{2\bot} \: ,  \: E - \frac{\vec{u}_\bot \cdot \vec{q}_{1\bot}}{1-u_z} - q_{2z} \right)
    }{
        [ q_{2z} - Q_p^- + Q_{k+q_2}^- + i \epsilon]
        [ q_{2z} - Q_p^+ + Q_{k+q_2}^- - i \epsilon]
    }
    \bigg\} \;  .
\end{align}
All of these factors can be straightforwardly evaluated, although we will need to keep the energy denominator $Q_{p-k-q_2}^- - Q_p^- + Q_{k+q_2}^-$ to an overall accuracy of $\ord{\frac{\perp^2}{E^2}}$ to obtain all of the relative $\ord{\frac{\perp}{E}}$ corrections:
\begin{align}
    &Q_{p-k-q_2}^- - Q_p^- + Q_{k+q_2}^- =
    \frac{(k-xp+q_2)_\bot^2}{2x(1-x)(1-u_z)E} - 
    \frac{(\vec{u}_\bot \cdot\vec{q}_{2\bot}) [(1-x)^2 (k+q_2)_\bot^2 - x^2 (p-k-q_2)_\bot^2]
    }{2 x^2 (1-x)^2 (1-u_z)^2 E^2}
    \notag \\ &\hspace{1.5cm}=
    \frac{(k-xp+q_2)_\bot^2}{2x(1-x)(1-u_z)E}
    \left[
    1 - \frac{\vec{u}_\bot \cdot\vec{q}_{2\bot}}{x (1-x) (1-u_z) E}
    \frac{(1-x)^2 (k+q_2)_\bot^2 - x^2 (p-k-q_2)_\bot^2}{(k-xp+q_2)_\bot^2}
    \right].
\end{align}
Including the prefactors which multiply $\mathcal{I}_G^{(1)}$, the result is
\begin{align}
    \frac{\lambda}{x} v(q_{2\bot}^2) &\left[ \frac{-2 E x(1-x)}{1+u_z} \right] 
    \mathcal{I}_G^{(1)} \rightarrow
    \notag \\ &\rightarrow
    - \psi(x, \vec{k}_\bot - x \vec{p}_\bot + \vec{q}_{2\bot}) \:
    e^{-i \left[ \frac{ (k-xp + q_2)_\bot^2 - (k-xp)_\bot^2}{2 x (1-x) E (1-u_z)} \right] z} \:
    [v (q_{2\bot}^2)]^2 \: J(E, \vec{p}_\bot)
    \notag \\ &\times
    \bigg[
    1 + \frac{\vec{u}_\bot \cdot \vec{q}_{2\bot}}{(1-u_z) (1-x) E}
    + \frac{\vec{u}_\bot \cdot \vec{q}_{2\bot}}{1-u_z}
    \frac{(p-k-q_2)_\bot^2 - (p-k)_\bot^2}{(1-x)E} \frac{1}{v(q_{2\bot}^2)} \frac{dv}{dq_{2\bot}^2}
    \notag \\ &\hspace{1cm} +
    \frac{\vec{u}_\bot \cdot \vec{q}_{2\bot}}{x (1-x) (1-u_z) E}
    \frac{(1-x)^2 (k+q_2)_\bot^2 - x^2 (p-k-q_2)_\bot^2}{(k-xp+q_2)_\bot^2}
    \bigg] .
\end{align}
Similarly, keeping the three relevant residues $q_{2z} = Q_{p-k-q_2}^-, \mathcal{P}^-, \mathcal{P}^- + Q_p^-$ of $\mathcal{I}_G^{(2)}$
\begin{align}
    \mathcal{I}_G^{(2)} &\rightarrow -i 
    \left(
    \mathrm{Res}_{Q_{p-k-q_2}^-} + \mathrm{Res}_{\mathcal{P}^-} + \mathrm{Res}_{\mathcal{P}^- + Q_p^-}
    \right)
    \notag \\ &
    \times \Bigg\{
    \frac{v(q_2) \, v(\vec{q}_{2\bot} \, ; \, q_{2z} - Q_p^-)}
    {
        [(p-k-q_2)^2 + i \epsilon]
        [q_{2z} - Q_p^- - Q_{k+q_2}^- + i \epsilon]
        [q_{2z} - Q_p^- + Q_{k+q_2}^+ + i \epsilon]
    }
    \Bigg\} \; , 
\end{align}
and keeping only the imaginary part $\mathrm{Im} \,\mathcal{I}_G^{(2)}$ which leads to cosines rather than sines of the LPM phases, we obtain
\begin{align}
    \lambda \Bigg[ & \frac{-2 E x(1-x)}{1+u_z} \Bigg] 
     i \, \mathrm{Im}\, \mathcal{I}_G^{(2)} \rightarrow
    \psi(x, \vec{k}_\bot - x \vec{p}_\bot + \vec{q}_{2\bot}) \,
    [ v(q_{2\bot}^2) ]^2 
    \Bigg[
        1
        - \frac{\vec{u}_\bot \cdot \vec{q}_{2\bot}}{(1-u_z) x E}
    \notag \\ &
        + \frac{\vec{u}_\bot \cdot \vec{q}_{2\bot}}{(1-u_z) (1-x) E}
        + \frac{\vec{u}_\bot \cdot \vec{q}_{2\bot}}{x (1-x) (1-u_z) E} \frac{(1-x)^2 (k+q_2)_\bot^2 - x^2 (p-k-q_2)_\bot^2}{(k - x p + q_2)_\bot^2}
    \notag \\ &+
        \frac{\vec{u}_\bot \cdot \vec{q}_{2\bot}}{x(1-x)(1-u_z) E}
        \frac{x (p-k-q_2)_\bot^2 - x(p-k)_\bot^2 - (1-x) (k+q_2)_\bot^2 + (1-x) k_\bot^2}{v(q_{2\bot}^2)}
    \frac{\partial v}{\partial q_{2\bot}^2}
    \Bigg] \; .
\end{align}

Inserting these results back into \eqref{e:RG2} we find
\begin{align}   \label{e:RG3}
    \left\langle R_{2, G} R_0^* \right\rangle &= 
    \mathcal{C}_{(G,0)} \, 
    \int_0^L dz \, \rho \,
    \int \frac{d^2 q_{2\bot}}{(2\pi)^2} \: [ v(q_{2\bot}^2) ]^2 \: 
    \left| J(E, \vec{p}_\bot) \right|^2
    \psi(x, \vec{k}_\bot - x \vec{p}_\bot + \vec{q}_{2\bot})
    \psi^* (x, \vec{k}_\bot - x \vec{p}_\bot)
    \notag \\ & \hspace*{-1cm} \times
    \left(
    e^{i \left[ \frac{(k - x p)_\bot^2}{2 x (1-x) E(1-u_z)} \right] z} 
    - e^{-i \left[ \frac{ (k-xp + q_2)_\bot^2 - (k-xp)_\bot^2}{2 x (1-x) E(1-u_z)} \right] z} 
    \right)
    \Bigg[ 
        1 
        - \frac{\vec{u}_\bot \cdot (\vec{k}_\bot + \vec{q}_{2\bot})}{(1-u_z) x E} 
    \notag \\ &\hspace*{-1cm} 
        - \frac{\vec{u}_\bot \cdot (\vec{p}_\bot - \vec{k}_\bot - \vec{q}_{2\bot})}{(1-u_z) (1-x) E} 
        + \frac{\vec{u}_\bot \cdot \vec{q}_{2\bot}}{x (1-x) (1-u_z) E}   \frac{(1-x)^2 (k+q_2)_\bot^2 - x^2 (p-k-q_2)_\bot^2}{(k-xp+q_2)_\bot^2}
    \notag \\ & \hspace*{-1cm} 
        + \frac{\vec{u}_\bot \cdot \vec{q}_{2\bot}}{x(1-x)(1-u_z) E}
        \frac{x (p-k-q_2)_\bot^2 - x(p-k)_\bot^2 - (1-x) (k+q_2)_\bot^2 + (1-x) k_\bot^2}{v(q_{2\bot}^2)}
    \frac{\partial v}{\partial q_{2\bot}^2}
    \Bigg].
\end{align}
Adding the complex conjugate and the appropriate phase space factors, we obtain 
\begin{align}   \label{e:RGfinal}
    \hspace{-0.5cm}\left( E \frac{dN^{(1)}}{d^2k \, dx \: d^2p_\perp \, dE} \right)_G &=
    \frac{1}{2(2\pi)^3\,x(1-x)}\frac{\mathcal{C}_{(G,0)}}{\mathcal{C}}
    \int_0^L dz \, \rho \: \int d^2 q_{2\bot} \,
   \bar\sigma(q_{2\bot}^2)   
    \notag \\ \hspace{-1.5cm}&\times 
    \psi(x, \vec{k}_\bot - x \vec{p}_\bot) \,
    \psi^* (x, \vec{k}_\bot - x \vec{p}_\bot + \vec{q}_{2\bot})
\left( E \frac{dN^{(0)}}{d^2 p \, dE} \right)
    \bigg[ 1 + \vec{u}_\bot \cdot \vec{\Gamma}_G \bigg]    
    \notag \\ \hspace{-1.5cm}&\times
    \bigg[
    2 \cos\left(\frac{(k-xp)_\bot^2}{2 x(1-x) E(1-u_z)} z \right)
    -
    2 \cos\left( \frac{ (k - x p + q_2)_\bot^2 - (k - x p)_\bot^2 }{ 2 x (1-x) E(1-u_z)} z \right)
    \bigg]
\end{align}
with
\begin{align}
\label{e:GamDBG}
    \vec{\Gamma}_G &\equiv 
    - \frac{\vec{k}_\bot + \vec{q}_{2\bot}}{(1-u_z) x E} 
    - \frac{\vec{p}_\bot - \vec{k}_\bot - \vec{q}_{2\bot}}{(1-u_z) (1-x) E} 
    \notag \\ &
    + \frac{\vec{q}_{2\bot}}{x (1-x) (1-u_z) E}
    \frac{x (p-k-q_2)_\bot^2 - x(p-k)_\bot^2 - (1-x) (k+q_2)_\bot^2 + (1-x) k_\bot^2}{v(q_{2\bot}^2)}
    \frac{\partial v}{\partial q_{2\bot}^2}
    \notag \\ &
    + \frac{\vec{q}_{2 \bot}}{x (1-x) (1-u_z) E}
    \frac{(1-x)^2 (k + q_2)_\bot^2 - x^2 (p - k - q_2)_\bot^2}{(k - x p + q_2)_\bot^2} \; .
\end{align}
Moreover, if we identify the first correction factor from diagrams $R_{1,B}$ and $R_{1,C}$ from \eqref{e:RBfinal} and \eqref{e:RCfinal} we see that $\vec{\Gamma}_G$ is similar to the sum of $\vec{\O}_{I\,B}$ and $\vec{\O}_{I\,C}$. This provides an intuitive relation between the velocity corrections to the double-Born diagram $G$ and two single-Born corrections of the type seen in diagrams $B$ and $C$ associated with each of the two final-state partons. However, the similarity is not complete and one cannot express $\vec{\Gamma}_G$ through other leading corrections in a simple way.

\subsection{Diagrams H and I}

%
\begin{figure}[ht]
    \centering
	\includegraphics[width=\textwidth]{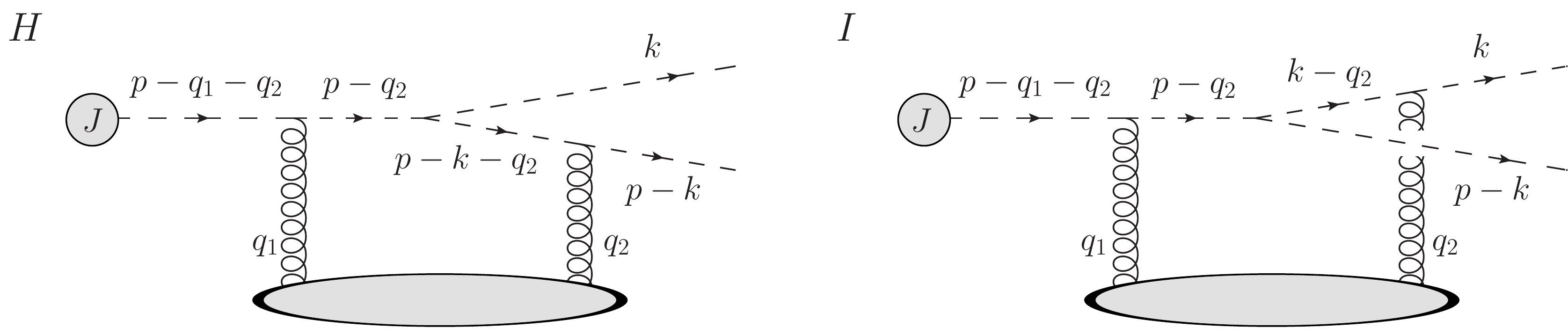}
	\caption{Double-Born diagrams denoted $R_{2, H} \, , \, R_{2, I}$ in which the branching occurs \textit{between} the double-scattering.}
	\label{f:DblBorn_H}
\end{figure}
%

Finally, we consider the two last double-Born diagrams shown in Fig.\ref{f:DblBorn_H}. These diagrams normally do not contribute in the eikonal limit, because the double-scattering is essentially instantaneous, leaving no time in between the two scatterings for the branching to occur.  However, at sub-eikonal accuracy, where the energy is large but finite, these contributions cannot be automatically neglected, and they should begin to contribute at some sub-eikonal order.  Therefore, we consider these diagrams here to examine whether they contribute at the $\ord{\frac{\perp}{E}}$ accuracy at which the velocity corrections enter.

The amplitude corresponding to the diagram H reads 
\begin{align}
    R_{2, H} &= -i \lambda \sum_{i \, j} 
    \mathcal{C}_H^{b a} t^b_{j} t^a_i
    \int\frac{d^4 q_1}{(2\pi)^4} \frac{d^4 q_2}{(2\pi)^4}
    e^{i q_1 \cdot x_i} \, e^{i q_2 \cdot x_{j}} \:
    v_i (q_1) \, v_{j} (q_2) \: J(p-q_1-q_2)
    \notag \\ &\times
    \frac{
    [2 (p-k)\cdot u_{j}]
    [2 (p-q_2) \cdot u_{i}]
    }{
    [(p-q_2)^2 + i \epsilon]
    [(p-k-q_2)^2 + i \epsilon]
    [(p-q_1-q_2)^2 + i \epsilon]
    }
    \notag \\ &\times
    \, (2\pi)^2 
    \delta(q_1^0 - \vec{u}_i \cdot \vec{q}_1) \,
    \delta(q_2^0 - \vec{u}_{j} \cdot \vec{q}_2) \; .
\end{align}
We again multiply by $R_0^* = - \psi^* (x, \vec{k}_\bot - x \vec{p}_\bot) J^*(p)$ and perform the color averaging, obtaining
\begin{align}
    \left\langle R_{2, H} R_0^* \right\rangle &= 
    i \lambda \mathcal{C}_{(H,0)}
    \psi^* (x, \vec{k}_\bot - x \vec{p}_\bot)
    \int d^3 x \, \rho \,
    \int\frac{d^4 q_1}{(2\pi)^4} \frac{d^4 q_2}{(2\pi)^4}
    e^{i (q_1+q_2) \cdot x} \:
    J(p-q_1-q_2) \, J^*(p)
    \notag \\ &\times
    \frac{
    [2 (p-k)\cdot u][2 p\cdot u]
    }{
    [(p-q_2)^2 + i \epsilon]
    [(p-k-q_2)^2 + i \epsilon]
    [(p-q_1-q_2)^2 + i \epsilon]
    } v (q_1) \, v (q_2) \:
    \notag \\ &\times
    \, (2\pi)^2 
    \delta(q_1^0 - \vec{u} \cdot \vec{q}_1) \,
    \delta(q_2^0 - \vec{u} \cdot \vec{q}_2)\; .
\end{align}
The correction to the eikonal vertex can be simplified and the full expression reads
\begin{align}
    \left\langle R_{2, H} R_0^* \right\rangle &= 
    i \lambda \mathcal{C}_{(H,0)}
    \psi^* (x, \vec{k}_\bot - x \vec{p}_\bot)
    \int d^3 x \, \rho \,
    \int\frac{d^4 q_1}{(2\pi)^4} \frac{d^4 q_2}{(2\pi)^4}
    e^{i (q_1+q_2) \cdot x} \:
    J(p-q_1-q_2) \, J^*(p)
    \notag \\ &\times
    \frac{
    [2 (1-u_z) (1-x) E] [2 (1-u_z) E]
    }{
    [(p-q_2)^2 + i \epsilon]
    [(p-k-q_2)^2 + i \epsilon]
    [(p-q_1-q_2)^2 + i \epsilon]
    } v (q_1) \, v (q_2) \:
    \notag \\ &\times
    \left[ 1 - \frac{\vec{u}_\bot \cdot (\vec{p}_\bot - \vec{k}_\bot)}{(1-u_z) (1-x) E} 
    - \frac{\vec{u}_\bot \cdot \vec{p}_\bot}{(1-u_z) E} \right]
    \, (2\pi)^2 
    \delta(q_1^0 - \vec{u} \cdot \vec{q}_1) \,
    \delta(q_2^0 - \vec{u} \cdot \vec{q}_2) .
\end{align}

The residues of the poles in $q_{1z}$ are $Q^\pm_{p-q_1-q_2}$ given by Eqs.~\eqref{e:rpoles2}. We close the contour below the real axis, generating a factor of $- 2\pi \, i \, \theta(z)$ and picking up the single residue at $Q^-_{p-q_1-q_2}-q_{2z}$ as in the case of $R_{2,D}$. Then,
\begin{align}
    \left\langle R_{2, H} R_0^* \right\rangle &= 
    \lambda \mathcal{C}_{(H,0)}
    \psi^* (x, \vec{k}_\bot - x \vec{p}_\bot)
    \int d^3 x \, \rho \, \theta(z) \,
    \int\frac{d^2 q_{1\bot}}{(2\pi)^2} \frac{d^4 q_2}{(2\pi)^4}
    e^{-i Q^-_{p-q_1-q_2} z} e^{-i (\vec{q}_{1\perp}+\vec{q}_{2\perp})\cdot \vec{x}_\perp}\:
    \notag \\ &\times
    \frac{
    4(1-x)(1-u_z)E^2
    }{(1+u_z)\left(Q^+_{p-q_1-q_2}-Q^-_{p-q_1-q_2}\right)}
    \frac{v (-q_{2z} + Q_{p-q_1-q_2}^- , \vec{q}_{1\bot}) \, v (q_2)}{
    [(p-q_2)^2 + i \epsilon]
    [(p-k-q_2)^2 + i \epsilon]
    }  \:
    \notag \\ &\times
    \left[ 1 - \frac{\vec{u}_\bot \cdot (\vec{p}_\bot - \vec{k}_\bot)}{(1-u_z) (1-x) E}  - \frac{\vec{u}_\bot \cdot \vec{p}_\bot}{(1-u_z) E}
    - \frac{\vec{u}_\bot \cdot (\vec{q}_{1 \bot} + \vec{q}_{2 \bot})}{1-u_z} \frac{\partial j}{\partial E}
    \right]
    \notag \\ &\times
    J(E, \vec{p}_\bot - \vec{q}_{1 \bot} - \vec{q}_{2 \bot}) \, J^*(E, \vec{p}_\bot)
    \, (2\pi)\delta(q_2^0 - \vec{u} \cdot \vec{q}_2)\,,
\end{align}
and, averaging over the transverse directions and setting $\vec{q}_\perp\equiv\vec{q}_{2\perp}=-\vec{q}_{1\perp}$, we obtain 
\begin{align}
    \left\langle R_{2, H} R_0^* \right\rangle &= 
    \lambda \mathcal{C}_{(H,0)}
    \psi^* (x, \vec{k}_\bot - x \vec{p}_\bot)
    \int_0^L dz \, \rho \,
    \int\frac{d^4 q_2}{(2\pi)^4}e^{i\frac{(k-xp)_\bot^2}{2x(1-x)E(1-u_z)}z} 
    |J(E, \vec{p}_\bot)|^2
    \notag \\ &\times
    2(1-x)(1-u_z)E
    \frac{v (q_{2z} - Q_p^- \, , \, \vec{q}_{2\bot}) \, v (q_2)}{
    [(p-q_2)^2 + i \epsilon]
    [(p-k-q_2)^2 + i \epsilon]
    }  \:
    \notag \\ &\times
    \left[ 1 - \frac{\vec{u}_\bot \cdot (\vec{p}_\bot - \vec{k}_\bot)}{(1-u_z) (1-x) E} - \frac{\vec{u}_\bot \cdot \vec{p}_\bot}{(1-u_z) E} \right]
    \, (2\pi)\delta(q_2^0 - \vec{u} \cdot \vec{q}_2)\; .
\end{align}
Isolating the $q_{2z}$ dependence, we have
\begin{align}
    \left\langle R_{2, H} R_0^* \right\rangle &= 
    \lambda \mathcal{C}_{(H,0)}
    \psi^* (x, \vec{k}_\bot - x \vec{p}_\bot)
    \int dz \, \rho \,
    \int\frac{d^2q_\bot}{(2\pi)^2}e^{i\frac{(k-xp)_\bot^2}{2x(1-x)E(1-u_z)}z}\: |J(E, \vec{p}_\bot)|^2
    \notag \\ &\times
    \mathcal{I}_H\left[ 1 - \frac{\vec{u}_\bot \cdot (\vec{p}_\bot - \vec{k}_\bot)}{(1-u_z) (1-x) E} - \frac{\vec{u}_\bot \cdot \vec{p}_\bot}{(1-u_z) E} \right]\,,
\end{align}
where
\begin{align}
\mathcal{I}_H &\equiv
2(1-x)(1-u_z)E
\int\frac{dq_z}{2\pi}\frac{v(q_{2z} - Q_p^- , \vec{q}_{2\bot}) v(q_2)} {[(p-q_2)^2 + i \epsilon][(p-k-q_2)^2 + i \epsilon]}%
    \notag \\ 
    &=\frac{2(1-x)(1-u_z)E g^4}{(1-u_z^2)^4}
    \int \frac{dq_z}{2\pi}
    \frac{1}{
    [q_{2z} - Q^+_{p-q_2} - i \epsilon]
    [q_{2z} - Q^-_{p-q_2} + i \epsilon]
    [q_{2z} - \mathcal{P}^+]
    [q_{2z} - \mathcal{P}^-]
    }
    \notag \\ &\hspace{0.5cm}\times
    \frac{1}{
    [q_{2z} - Q^+_{p-k-q_2} - i \epsilon]
    [q_{2z} - Q^-_{p-k-q_2} + i \epsilon]
    [q_{2z} - \mathcal{P}^+ - Q^-_p]
    [q_{2z} - \mathcal{P}^- - Q^-_p]
    }\,.
\end{align}
After evaluating the integral, we find
\begin{align}
\text{Im}\,\mathcal{I}_H
    &=\frac{\vec{u}_\perp\cdot\vec{q}_\perp}{(1-u_z)E}
    \frac{\left[v(q^2_\bot)\right]^2}{q_\bot^2 + \mu^2}
    +\mathcal{O}\left(\frac{\perp^2}{E^2}\right) \, ,  \notag\\
\text{Re}\,\mathcal{I}_H
    &=\mathcal{O}\left(\frac{\perp}{E}\right)\,.
\end{align}
Clearly the imaginary part integrates to zero after angular averaging, and the real part leads to terms of order $\frac{\bot}{E} \sin\frac{(k-xp)_\bot^2}{2 x (1-x)E(1-u_z)} z \sim \frac{\bot^3 z}{E^2}$ which we neglect (remembering the factor of $i$ included in the wave function).

Thus we see that even at $\ord{\frac{\bot}{E}}$ sub-eikonal accuracy, diagram $H$ still does not contribute.  From the diagram $H$ we can also obtain the contribution of the diagram $I$ by substituting $x\leftrightarrow(1-x)$ and $k\leftrightarrow(p-k)$ into the diagram $H$, and replacing the color factor. Both contributions are accordingly zero.

\end{document}